\documentclass[twocolumn]{aastex63}

\usepackage{textcomp}
\usepackage{comment}
\usepackage{changepage}
\usepackage{amsmath}

\shorttitle{A multiwavelength study of CLAGN}
\shortauthors{Birmingham et al.}  

\graphicspath{{./}{figures/}}

\begin{document}

\title{The birth of young radio jets in changing-look AGN: a population study}

\correspondingauthor{Sufia Birmingham}
\email{sb8685@princeton.edu}

\author[0009-0003-7044-9751]{Sufia Birmingham}
\affil{Department of Astrophysical Sciences, Princeton University, Princeton, NJ 08544, USA} 
\author[0000-0002-4557-6682]{Charlotte Ward}
\affil{Department of Astrophysical Sciences, Princeton University, Princeton, NJ 08544, USA} 
\author[0000-0003-1991-370X]{Kristina Nyland}
\affil{U.S. Naval Research Laboratory, 4555 Overlook Ave SW, Washington, DC 20375, USA}
\author[0000-0003-0699-7019]{Dougal Dobie}
\affil{Sydney Institute for Astronomy, School of Physics, The University of Sydney, NSW 2006, Australia}
\affil{ARC Centre of Excellence for Gravitational Wave Discovery (OzGrav), Hawthorn, 3122, Victoria, Australia}
\author[0000-0002-3168-0139]{Matthew J. Graham}
\affil{Cahill Center for Astronomy and Astrophysics, California Institute of Technology, Pasadena, CA 91125, USA}
\author[0000-0001-6295-2881]{David L. Kaplan}
\affil{Center for Gravitation, Cosmology, and Astrophysics, Department of Physics, University of Wisconsin-Milwaukee, P.O. Box 413, Milwaukee, WI 53201, USA}
\author[0000-0002-2686-438X]{Tara Murphy}
\affil{Sydney Institute for Astronomy, School of Physics, The University of Sydney, NSW 2006, Australia}

\begin{abstract}
Changing-Look Active Galactic Nuclei (CLAGN) are a rare subset of AGN that show significant changes to the flux of broad Balmer emission lines. Recent studies of CLAGN, such as 1ES 1927+654 and Mrk 590, have revealed that changes in the optically observed accretion rate are accompanied by changes in radio activity. We present a time-domain population study of 474 spectroscopically confirmed CLAGN at radio wavelengths using the Australia SKA Pathfinder Variable and Slow Transients Survey and the Very Large Array Sky Survey. We compare the radio properties of this CLAGN sample to a control sample of AGN that have not had recent changing-look events, and to AGN that were found to have transitioned from radio-quiet to radio-loud over 10-year timescales in VLASS. For 20 newly studied CLAGN detected in ASKAP VAST, we do not detect Mrk 590 or 1ES 1927+654-like fading of the radio flux in the 10 years following changing-look events. For 6 CLAGN with a sufficiently low redshift and high enough mass, we rule out a Mrk 590-like flare. We find that at the population level, CLAGN have higher VAST/VLASS detection rates, lower fractions of radio loudness, and higher variability rates in the 1 GHz frequency compared to the control AGN. Through VLA observations of radio SEDs and Magellan spectroscopic observations, we do not find evidence of a link between CLAGN and AGN that transitioned from radio-loud to radio-quiet in VLASS. We discuss the implications of this study for the physical mechanisms that drive enhanced accretion episodes.

\vspace{1cm}
\end{abstract}

\section{Introduction} \label{Introduction}
Supermassive black holes reside in the center of most galaxies \citep{Kormendy1995InwardNuclei,Ferrarese2005}. Determining the duty cycle, efficiency, and triggering mechanism of supermassive black hole (SMBH) accretion episodes is key to our understanding of SMBH growth and host galaxy co-evolution \citep{Pacucci2021TheGalaxies}. Important probes of episodic AGN accretion are changing-look AGN \citep[see][for a review]{Ricci2022Changing-lookNuclei}. CLAGN exhibit changes in optical emission from high velocity gas close to the SMBH (the broad line region; BLR) in response to changes in the accretion state \citep{Elitzur2014EvolutionNuclei,Schawinski2015ActiveYr}. The changes may be the appearance or disappearance of the BLR corresponding to a switch between Seyfert 1 or Seyfert 2 classifications, or a change in flux that corresponds to a transition between intermediate types. Recently, time-domain surveys have expanded the number of known CLAGN from a tiny population to around 500 objects using either multi-epoch spectroscopy \citep{Green2022TheSDSS-IV,Zeltyn2024ExploringResults,Guo2024Changing-lookData} or spectroscopic follow-up of AGN that showed variability changes or flares in optical photometry \citep{Frederick2019ALINERs,Graham2019}. The discovery that $\sim1$\% of AGN gain or lose their BLR over human timescales has provided important context for understanding unobscured `True' Type 2 AGN, and the `AGN unification' classification scheme, where the visibility of the BLR is determined by viewing angle effects resulting from obscuration by a dusty torus. 

Coincident with the discovery of changing-look events in time-domain optical surveys, wide-field multi-epoch radio surveys such as the Very Large Array (VLA) Sky Survey \citep[VLASS;][]{Lacy2020TheDesign}, the Rapid ASKAP Continuum Survey \citep[RACS;][]{McConnell2020}, and the ASKAP Variable and Slow Transients Survey \citep[VAST;][]{Murphy2013VAST:Transients} have revealed that previously radio-quiet AGN can also drastically increase their radio flux on year to decade timescales \citep{Nyland2020QuasarsFIRST,Zhang2022TransientSurveys,Woowska2021Caltech-NRAOState}. For example, VLASS revealed radio jets that had been triggered in 18 previously radio-quiet galaxies in the last 5-20 years, 14 of which were classified as LINERs or normal galaxies in archival optical spectra \citep{Nyland2020QuasarsFIRST}. The onset of radio activity has been interpreted as the birth of a radio jet triggered by short, sporadic fueling of a supermassive black hole, which is predicted to last for a short period of $10^4-10^5$ years.

There are a range of models which may explain the fast increase in accretion rate observed in changing-look events. These include the excitation of spiral density waves induced by a companion MBH of lower mass entering the AGN accretion disk during its orbit \citep{Dodd2025}, nozzle shocks and disk tearing arising from precession of the SMBH \citep{Kaaz2025}, or thermal or magnetorotational instabilities in the disk \citep[see][for a review]{Ricci2022Changing-lookNuclei}. To test the various models for the accretion rate change in CLAGN and how they induce or suppress radio emission, we must understand the wider connection between the optically-selected CLAGN and the populations of peaked-spectrum radio AGN that show drastic changes in their radio emission over year to decade timescales. Radio follow-up of several notable CLAGN with optical, UV or X-ray variability has discovered radio dimming or brightening following the flares, somewhat analogous to the radio jets seen in some tidal disruption events. One such example is Mrk 590, which transitioned from Seyfert 1 to Seyfert 1.9/2 between the mid-1990s to 2013 \citep{Denney2014} before the broad lines re-emerged in 2016 \citep{Mathur2018TheAwakening}. VLA observations of the compact, parsec-scale radio emission from Mrk 590 showed a $\sim40$\% radio flux decrease over 20 years following the decrease in accretion rate observed in the optical, likely due to the expansion and fading of internal shocks in the radio-emitting outflow produced by the recent outburst \citep{Koay2016Parsec-scale590,Yang2021AMrk590}. 

1ES 1927+654 \citep{Trakhtenbrot20191ESMonths} showed a dimming in radio flux during the optical flare, interpreted as a temporary dissipation of the corona \citep{Saha2023MultiwavelengthSeyfert}, followed by the appearance of a compact, bipolar, sub-relativistic radio jet that increased in radio flux over the following 3 years \citep{Meyer2025}. NGC 2992 exhibited a radio flare following the loss and regaining of Balmer broad lines, with associated X-ray variability \citep{Marinucci2018TrackingStates,Guolo2021The2992,Fernandez2022FRAMEx.2992}. NGC 1566 changed its spectral state from type 1.9 to type 1 in June 2018 while simultaneously increasing its flux in the optical, UV, IR, and x-ray bands \citep{Oknyansky2019, Parker2019, Tripathi2022}. In 2020, NGC 1566 transitioned back to the type 2 state, and its fluxes in all wavebands subsequently decreased \citep{Xu2024}. \citet{Arghajit1566} observed NGC 1566 in the 230 GHz radio wavelengths from 2014-2023, finding that the 230 GHz flux increased during the type 1 to 2 state change, and decreased during subsequent type 2 to 1 state change. J154843.06+220812.6 was another radio transient associated with a mid-IR flare and the appearance of an H$\alpha$ broad line, and was attributed to the formation of a jet following either an obscured tidal disruption event or AGN flare \citep{Somalwar2022TheChange}.

Other CLAGN with radio core detections show no radio variations following a changing-look event. For example, NGC 2617, a turn-on CLAGN, had a flat-spectrum compact radio core that did not vary over 2 years following the outburst and a steep-spectrum jet component that extended 2 pc from the core and may have been associated with a previous outburst \citep{Yang2021A2617}.

In this paper, we undertake a population study of CLAGN initially identified by broad-line variations in optical spectroscopy and study their time-resolved radio properties using the VLASS and VAST radio surveys. We determine how frequently changing-look events have associated radio variability and how often they produce new radio jets, and we compare the radio properties of CLAGN to a broader AGN control sample. We discuss the relationship between the radio properties of our CLAGN sample and those of the AGN identified to transition from radio-quiet to radio-loud on decadal timescales in previous work. We discuss how population-level analysis of radio flares from CLAGN may provide future insights into the physical mechanism behind changing-state events. Through this population study, we aim to answer three main questions: 
\begin{enumerate}
    \item Do we find evidence for CLAGN having new radio jets appear after the changing-look event?
    \item How do the properties of optically and spectroscopically-selected CLAGN compare to those of a control AGN sample? 
    \item How do the properties of optically and spectroscopically selected CLAGN compare to those of AGN that have transitioned from radio quiet (RQ) to radio loud (RL)?
\end{enumerate}

\section{Selection of the CLAGN sample and control Samples} \label{ParentSampleConstruction} 

\subsection{A sample of spectroscopically confirmed CLAGN} \label{CLAGNSample}
We obtained a sample of 474 spectroscopically-confirmed CLAGN by first compiling samples from multi-epoch spectroscopic surveys: the Sloan Digital Sky Survey (SDSS)-IV/V \citep{Green2022TheSDSS-IV,Zeltyn2024ExploringResults}, SkyMapper \citep{Hon2022}, and the Dark Energy Spectroscopic Instrument \citep[DESI;][]{Guo2024Changing-lookData}. \citet{Green2022TheSDSS-IV} used SDSS-IV multi-epoch spectroscopy targeting broadline quasars to identify 61 newly-discovered CLAGN candidates. Because their parent sample consisted of quasars which previously showed the presence of broad lines, they were biased towards detecting turn-off CLAGN. \citet{Zeltyn2024ExploringResults} similarly used SDSS-V to find 116 CLAGN, where $\sim2/3$ of their objects were turn-off CLAGN, and $\sim1/3$ were turn-on CLAGN. \citet{Hon2022} searched for turn-on CLAGN by selecting Type 2 AGN from  the spectroscopic Six-degree Field Galaxy Survey (6dFGS), whose colors, observed about 15 years later by the SkyMapper Southern Survey, were suggestive of Type 1 AGN. They found 29 CLAGN candidates, 25 of which were turn-on CLAGN, and 4 of which were turn-off CLAGN. \citet{Guo2024Changing-lookData} compared new spectra from DESI to older SDSS spectra to find 56 CLAGN candidates, 8 of which show simultaneous appearance/disappearance of H$\alpha$, H$\beta$ or Mg II broad emission lines. 

We also sampled CLAGN identified from spectroscopic follow-up of AGN showing notable optical variability in photometric optical surveys such as the Catalina Real-time Transient Survey (CRTS) \citep{Graham2020UnderstandingQuasars} and the Zwicky Transient Facility (ZTF) \citep{Frederick2019ALINERs}. \citet{Frederick2019ALINERs} discovered six turn-on CLAGN upon spectroscopic follow-up of nuclear transients identified in the ZTF survey. These objects had host galaxies that were classified as low-ionization nuclear emission-line region galaxies (LINERs) using archival SDSS spectra. \citet{Graham2020UnderstandingQuasars} used CRTS to search for CLAGN exhibiting both strong photometric variability over a decade-long baseline and found 111 CLAGN that form a higher luminosity sample to complement existing sets of CLAGN in the literature. Of this 111-size sample, 63 were turn-on CLAGN, and 48 were turn-off CLAGN. The rest of the CLAGN in our 474-size sample were taken from a variety of different single-source papers, and their references can be seen in Table~\ref{tab:CLAGNSample}.

Table~\ref{tab:CLAGNSample} summarizes key properties of the 474-size CLAGN parent sample, including the CLAGN positions, redshifts, broad line state change dates, and whether the CLAGN was turn-on or turn-off, in addition to the ZTF optical variability properties. 

\subsection{Control sample of broad-line AGN that are not changing-look} \label{ControlSampleDescription}
We compiled a control sample of AGN based on the parent AGN sample used to search for the DESI CLAGN population \citep{Guo2024Changing-lookData}. The control AGN sample was compiled in order to investigate whether the radio properties of CLAGN are different to comparable populations of AGN which have not recently had a changing-look event. To do this, we regenerated the \citet{Guo2024Changing-lookData} parent sample by crossmatching all sources classified as a quasi-stellar object (QSO) or galaxy in the DESI early data release \citep{Adame2024TheInstrument} to objects classified as a QSO or galaxy in the SDSS DR16 catalog \citep{Ahumada2020TheSpectra} using a 1" radius. This reproduced the parent sample of 85,697 objects. As the CLAGN criteria required an AGN classification in at least one of the two surveys, we removed any objects with a galaxy classification in both catalogs and any objects identified as CLAGN. This left us with a control sample of 25,193 AGN for comparison to the DESI CLAGN sample.

\begin{figure*}
    \centering
    \gridline{\includegraphics[width=0.45\textwidth]{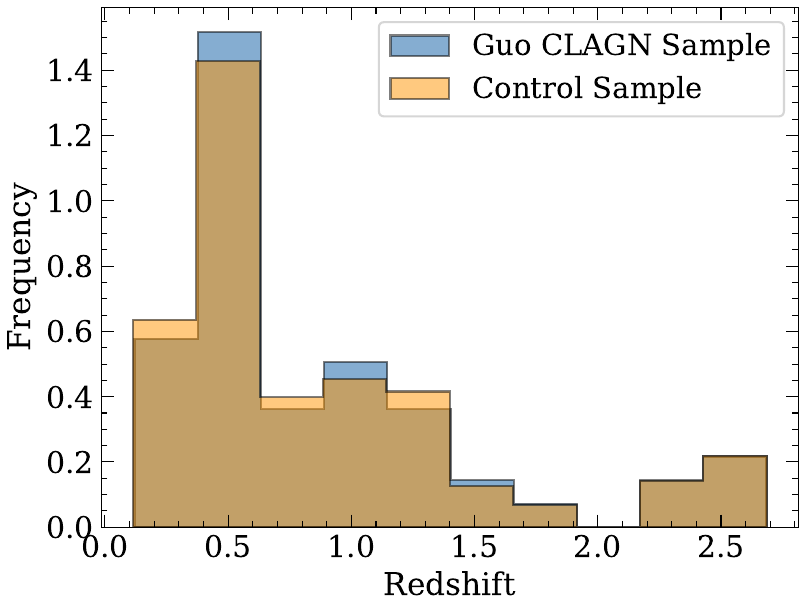}\includegraphics[width=0.51\textwidth]{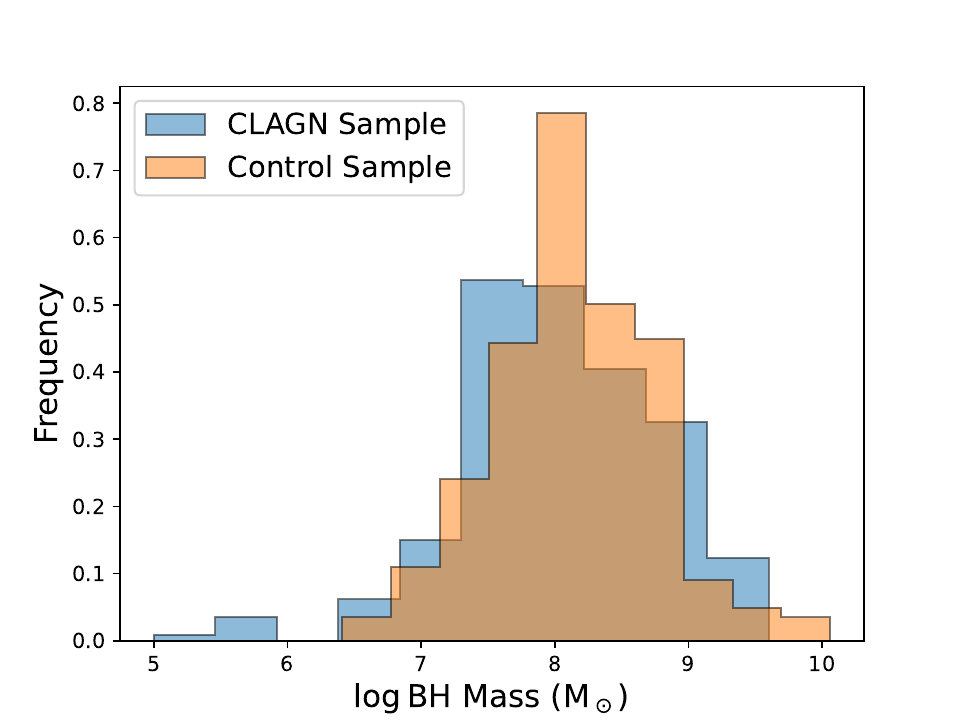}} 
    \caption{Normalized distribution of the \protect{\citet{Guo2024Changing-lookData}} CLAGN and control sample redshift distribution (left) and normalized distribution of the CLAGN Sample and control sample log black hole masses (right).}
    \label{fig:Control_Distribution}
\end{figure*}

This control sample was skewed to higher redshifts compared to the CLAGN sample. To correct this, we sampled the control AGN sample to generate a new sub-sample with a similar redshift distribution to the CLAGN sample from \citet{Guo2024Changing-lookData}. To create a 3000-size redshift-matched control AGN sample, we undertook the following procedure 3000 times: We randomly selected one of the CLAGN from the \citet{Guo2024Changing-lookData} sample, and found the 20 AGN from the control AGN sample with redshift closest to the selected CLAGN's redshift. We then randomly selected one of these 20 control sample AGN to include in the new sub-sample. Additionally, we obtained all available black hole masses reported in \citet{YangSMBH2019} and \citet{Kozlowski2017} for the broad-line control AGN sample. \citet{Kozlowski2017} reports black hole masses via the Mg II line, and \citet{YangSMBH2019} reports the fiducial virial BH mass obtained from the mean of the BH masses derived from H$\alpha$ and H$\beta$. Similarly, for the CLAGN sample, we obtained all available black hole masses from the discovery papers cited in Section~\ref{CLAGNSample}. In Figure~\ref{fig:Control_Distribution}, we display the \citet{Guo2024Changing-lookData} CLAGN and 3000-size control sample's redshifts along with the full CLAGN and control sample's black hole masses. We confirm that the redshifts of the control sample closely align with the redshifts of the \citet{Guo2024Changing-lookData} CLAGN sample, and we confirm that the peak and overall distributions of the black hole masses of our CLAGN sample and control sample are similar. We compare the properties of the optically and spectroscopically-selected CLAGN sample to this redshift-matched parent sample in Section~\ref{ControlComparison}. Table~\ref{tab:ControlSample} summarizes key properties of the control AGN sample. 

\subsection{Comparison sample of AGN transitioning from radio-quiet to radio-loud over decadal timescales} \label{Nyland Sample}
In order to compare our spectroscopically and optically-selected CLAGN sample to populations of AGN with changing radio loudness, we compiled a 52-size sample of AGN that were observed to change from radio-quiet to radio-loud in the VLA Sky Survey when compared to archival data from 1993–2009 in the Faint Images of the Radio Sky at Twenty cm (FIRST) survey \citep{Nyland2020QuasarsFIRST,Woowska2021Caltech-NRAOState,Zhang2022TransientSurveys}. \citet{Nyland2020QuasarsFIRST} used VLASS to identify 25 compact sources with peaked radio spectral shapes that were once radio-quiet but are now radio-loud. These objects are thought to be quasars hosting compact/young jets. \citet{Woowska2021Caltech-NRAOState} studied 12 gigahertz-peaked spectrum (GPS) sources that were revealed to be new radio sources after \textgreater5–20 yr of radio absence by the Caltech-NRAO Stripe 82 Survey. They also concluded that the observed spectral changes were a consequence of the evolution of newly born radio jets. \citet{Zhang2022TransientSurveys} detected 18 low-redshift (z\textless 0.3) galaxies that have brightened significantly in radio flux (by a factor of $\geq5$) in the epoch I (2017–2019) observations of VLASS compared to previous FIRST survey observations performed from 1993–2009. We note that none of these objects have been confirmed as CLAGN via optical spectroscopy. We compare the properties of the CLAGN sample to this sample of 52 AGN transitioning from radio-quiet to radio-loud in Section~\ref{RadioSelectedSample}. For the remainder of the paper, we refer to this comparison sample as the `RQ to RL AGN sample.' We obtained the black hole masses reported in the corresponding RQ to RL AGN sample's discovery papers, and found that the black hole mass distribution of the RQ to RL AGN objects is overall similar in nature to the CLAGN black hole masses. Table~\ref{tab:RQRLAGNSample} summarizes key properties of the RQ to RL AGN sample.


\section{Radio properties of the CLAGN sample} \label{Methods}

\subsection{VLASS detections} \label{VLASSDetections}
The VLASS survey covers the whole sky visible to the VLA (Declination \(> -40^\circ\)) for a total of 33,885 deg\(^2\). The survey provides us with two epochs of radio data in the 3 GHz band \citep{Lacy2020TheDesign}. VLASS has high angular resolution (\(\approx 2.5^{\prime\prime}\)), full linear Stokes polarimetry, time domain coverage, and wide bandwidth (2–4 GHz). 

We crossmatched our entire sample of 474 CLAGN with the VLASS VLASS Quick Look and Single Epoch Catalogues\footnote{https://cirada.ca/vlasscatalogueql0} using a 10.8 arcsecond radius. We filtered objects that had radio detections above 2 mJy and were inside the VLASS field of view. The VLASS detections of our CLAGN radio-detected sample are displayed in Table~\ref{tab:CLAGN_RQRL_RadioData}. We found that for our 474-size CLAGN sample, there were 61 valid VLASS epoch 1 and/or epoch 2 radio detections from a total of 478 sources within the VLASS footprint. Thus, the VLASS detection rate was $\sim13$\%.

\subsection{VAST detections} \label{VASTDetections}
The ASKAP VAST extragalactic survey provides bi-monthly observations of the Southern sky at 887.5 MHz, while the VAST pilot survey uses similar observing parameters (albeit with some observations at 1367.5 MHz) but on a more irregular cadence and covering a smaller fraction of sky \citep{Murphy2021TheSurvey}. The VAST survey gives us the capability to plot radio light curves across a $\sim6$ year baseline (for sources in the pilot survey footprint) for a large sample, allowing us to study how the radio emission of CLAGN changes over time. The typical VAST rms sensitivity is 0.25 mJy in each observation. 

We crossmatched our entire sample of 474 CLAGN with the VAST data to see if these objects had radio data detections above 2 mJy and were inside the field of view of the VAST survey. To do this, we accessed VAST radio data from the pilot survey and post-processed surveys, which cover a total of $74$ epochs. We used \textsc{vast-tools}\footnote{https://vast-survey.org/vast-tools} \citep{vasttools} to crossmatch each CLAGNs optical coordinates to the VAST data. In Table~\ref{tab:CLAGN_RQRL_RadioData}, we provide VAST radio data properties for our radio-detected CLAGN. We found that there were 24 of 147 sources that were within the VAST footprint with a radio detection, corresponding to a detection rate of $\sim16$\%. 



\begin{figure*}
  \centering
  \gridline{
    \fig{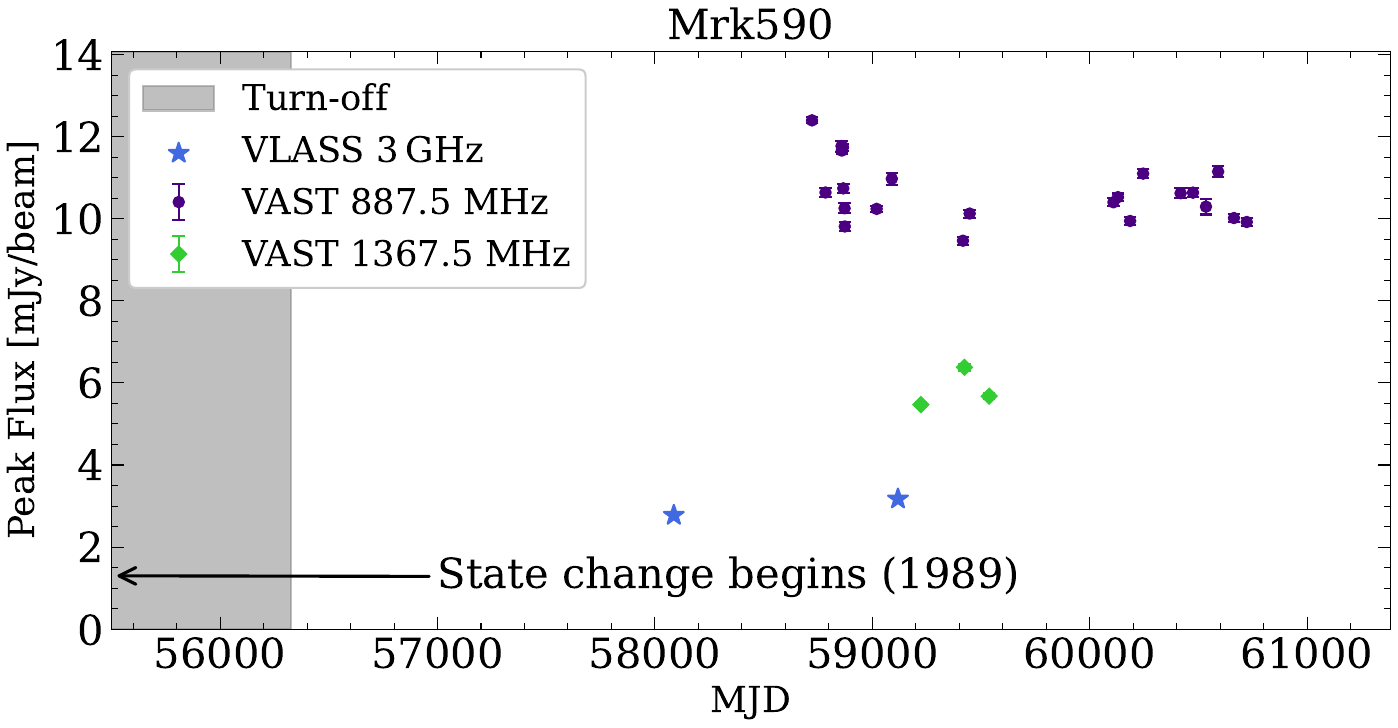}{0.50\textwidth}{Mrk\,590}
    \fig{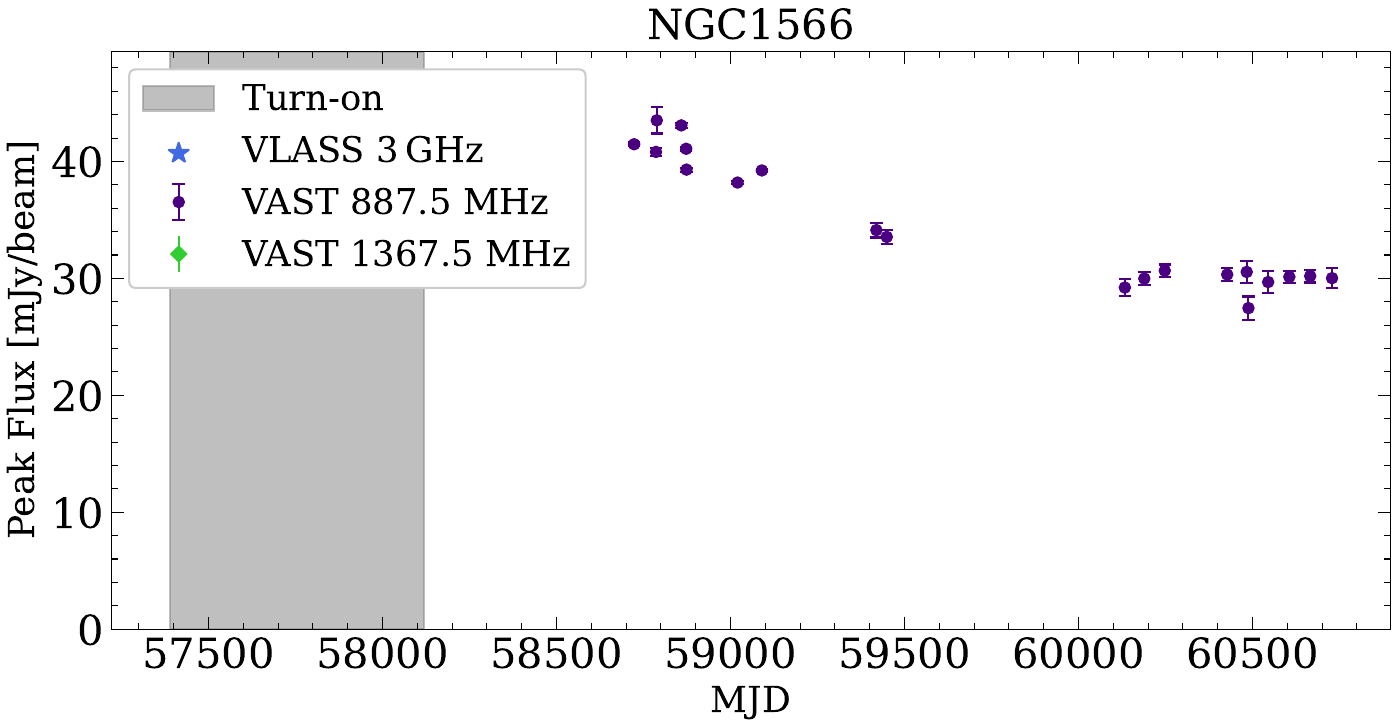}{0.50\textwidth}{NGC\,1566}}
  \caption{Radio light curves of Mrk\,590 and NGC\,1566 constructed using VAST and VLASS data. The gray region represents the dates of the change to the broad line fluxes identified in optical spectroscopy. For Mrk\,590, \protect{\citet{Denney2014TheRole}} reports that its broad lines disappeared between 2003/2006 and 2013, with prior fading observed since 1989. We indicate Mrk\,590's date range on the figure with the black arrow pointing towards 1989 labeling the beginning of the state change, and the gray region which ends at February 2013.}
  \label{fig:KnownCLAGNRadioLCs}
\end{figure*}

We obtained up to two epochs of VLASS radio fluxes in the 3 GHz wavelength, and VAST monthly cadence radio fluxes in up to two separate wavelengths (887.5 MHz and 1.367 GHz). We constructed monthly cadence VAST radio light curves from the years 2019 to 2025. In Figure~\ref{fig:KnownCLAGNRadioLCs}, we show the VAST light curves of known radio flaring CLAGN Mrk 590 and NGC 1566 respectively. For comparison, ~\citet{Koay2016Parsec-scale590} reported Mrk 590's VLA-detected flux density at 1.8 GHz in 2015 to be 2.75 $\pm$0.08 mJy. From 2019 to 2023, NGC 1566 was observed to steadily fade in the 230 GHz frequency \citep{Arghajit1566}. We confirm this fading behavior in the 887.5 MHz radio frequency, where we observe NGC 1566 to fade during 2019 to 2023 in VAST wavelengths.

\subsection{Radio source classification}

\subsubsection{Blazar catalog crossmatching} \label{Blazars}
We identified blazars by crossmatching all radio-detected CLAGN to the following catalogs with an 8" match radius: the Roma-BZCAT Multifrequency Catalogue of Blazars \citep{BZCAT2015}; a catalog of 
AGN detected in the $\gamma$-rays by the Fermi Large Area Telescope \citep{Fermi2022}; a catalog of radio-loud candidate blazars identified as having WISE mid-IR colors similar to the colors of confirmed $\gamma$-ray blazars \citep{DAbrusco2014ApJS}; and a sample of radio-bright AGN with jet angles measured to be $<$10\textdegree\ in very-long-baseline interferometry (VLBI) observations \citep{Plavin2022}. One object, J143041+364904, was identified in all three blazar catalogs, while J164829+410406 was identified in Roma-BZCat and the WISE-IR blazar candidate sample. Five other objects (J022931-000845, J134134+090356, J102818+211508, J111930+222650, J094232+233614) appeared in the WISE-IR blazar candidate sample only. J003849-011722 was identified as having a small jet angle in VLBI imaging and therefore a likely blazar. In summary, we identified that 8 of our radio-detected CLAGN are likely blazars using multi-wavelength blazar and blazar candidate catalogs. 

\subsubsection{Source compactness} \label{SourceCompactness}
In order to construct radio light curves of the CLAGN, we first checked if the radio sources are compact or extended. To calculate source compactness, we split our radio detections into two categories: sources with VLASS detections, and sources without VLASS detections (those with only VAST detections). This is because VLASS has a higher resolution than VAST, and we expect more compact emission regions at higher frequencies. For the sources with VLASS detections, we calculated the median integrated flux and the median peak flux across the two epochs in VLASS. For sources with only one epoch of data available, we simply use that singular epoch's detection to obtain the integrated and peak flux values. We calculated the ratio of the median integrated flux to the median peak flux. Sources in VLASS with a compactness ratio of $<2$ were classified as compact, and classified as extended otherwise. For our sources without VLASS detections, we calculated the median peak and the median integrated fluxes across all available epochs in VAST. Sources in VAST which had a compactness ratio (median integrated to median peak flux) of $<5$ were classified as compact, and classified as extended otherwise.


We used the Canadian Initiative for Radio Astronomy Data Analysis (CIRADA)'s Image Cutout Provider\footnote{http://cutouts.cirada.ca/}, which accesses VLASS and Rapid ASKAP Continuum Survey (RACS) radio imaging data, to generate radio cutout images of our radio-detected CLAGN. RACS is a radio survey in the declination range of -90 to 47 degrees in three frequency bands (RACS-low: 887.5 MHz, RACS-mid: 1367.5 MHz and RACS-high: 1655.5 MHz). The RACS tiles have a median $1\sigma$ sensitivity of $\sim$ 300 $\mu$Jy/beam. We obtained radio cutouts of radius 1-2 arcminutes to classify the morphologies of the CLAGN. In each case, we confirmed that the morphology observed in the cutout was consistent with the compactness classification based on the compactness ratio. 

\begin{figure*}
  \centering
  \gridline{
    \fig{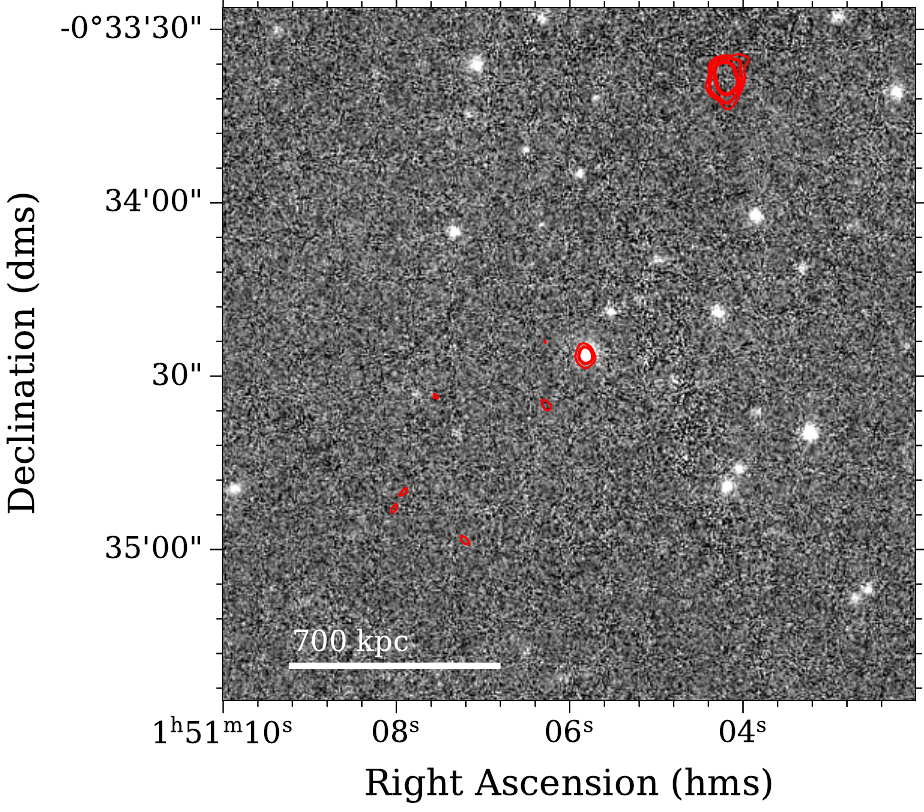}{0.28\textwidth}{J015106-003426}
    \fig{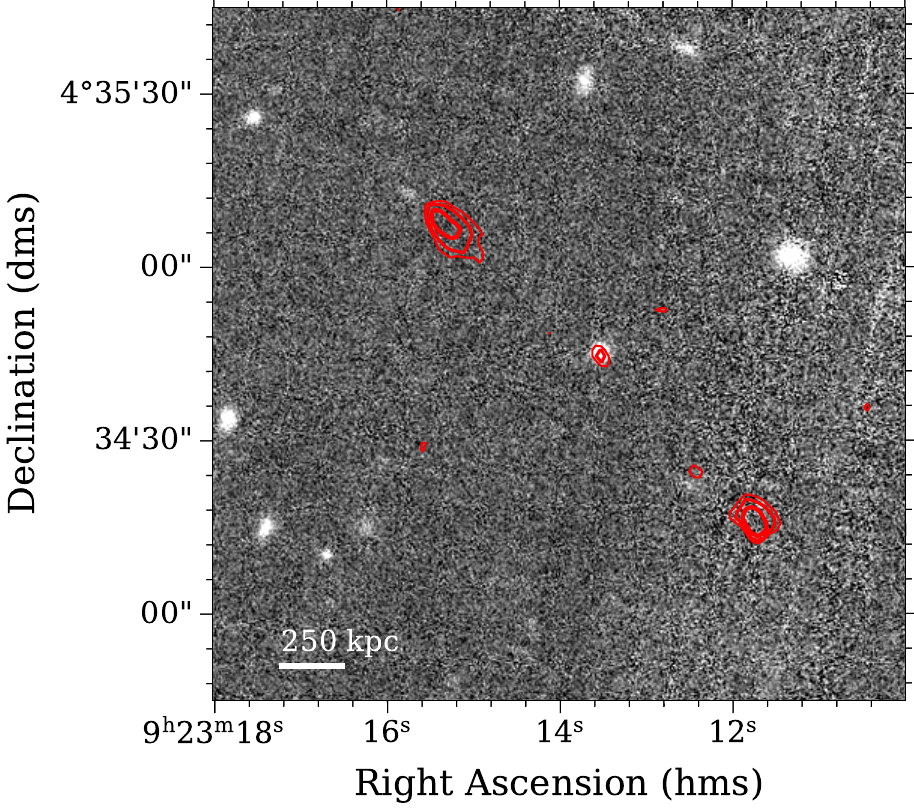}{0.28\textwidth}{J092314+043445}
    \fig{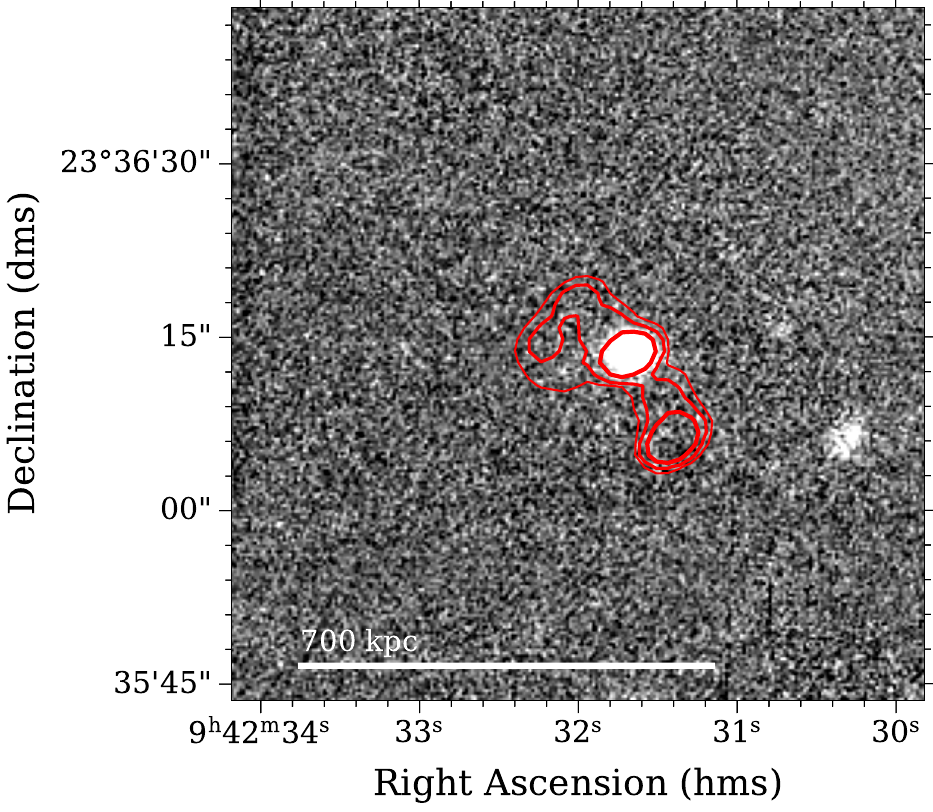}{0.28\textwidth}{J094232+233614}
  }
  \gridline{
    \fig{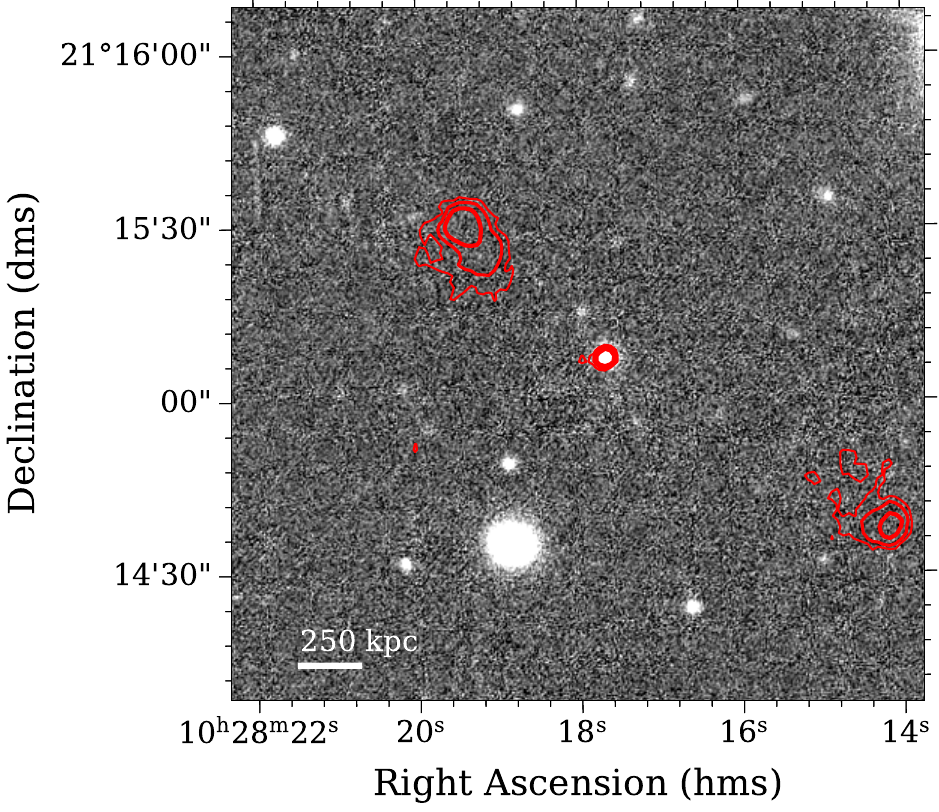}{0.28\textwidth}{J102818+211508}
    \fig{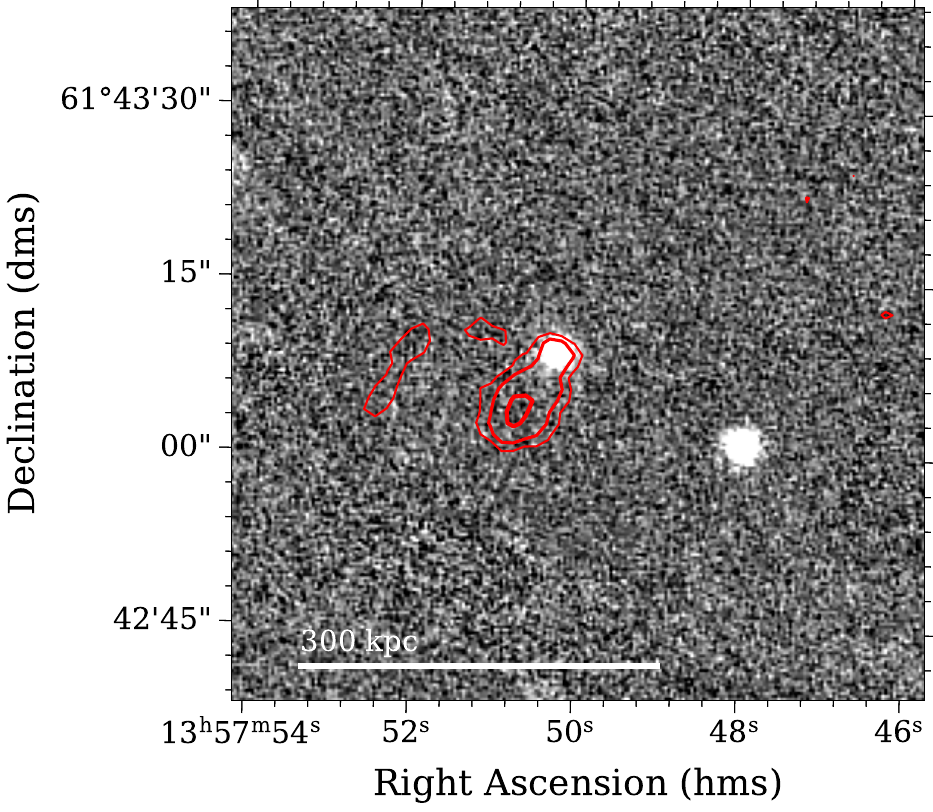}{0.28\textwidth}{J135750+614309}
    \fig{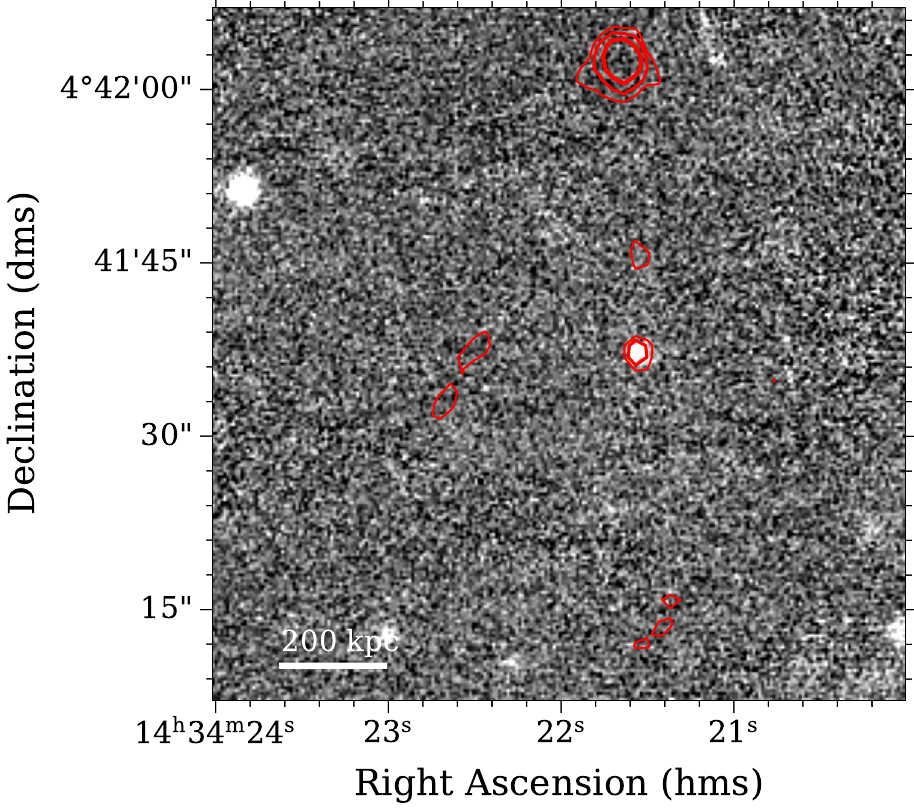}{0.28\textwidth}{J143422+044137}
  }
  \gridline{
    \fig{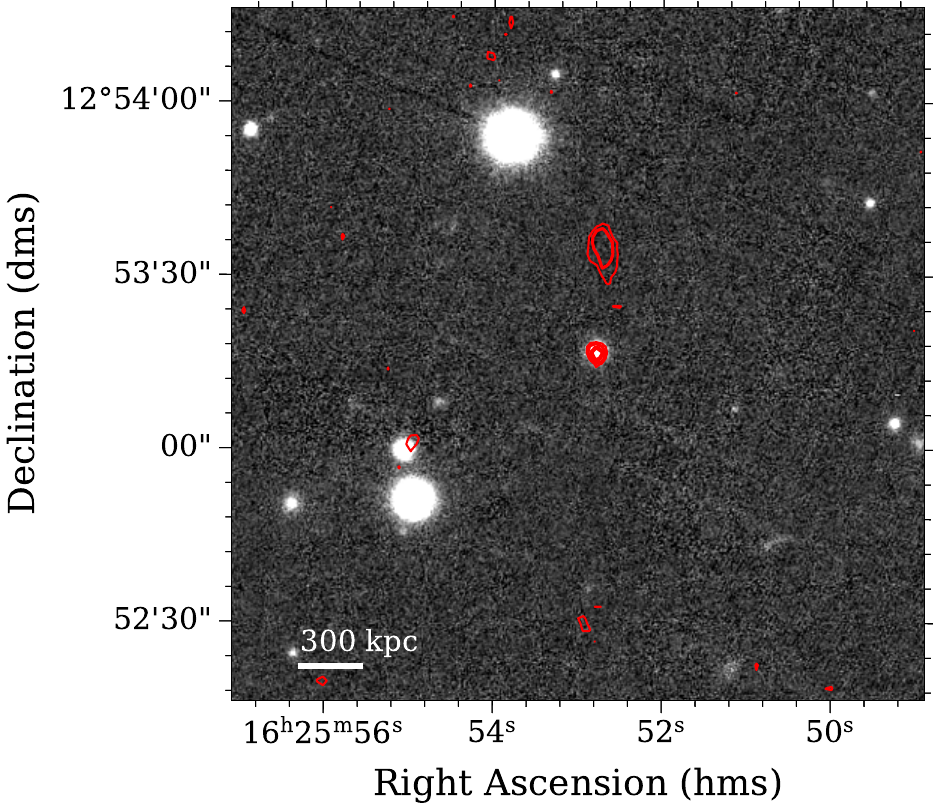}{0.28\textwidth}{J162553+125317}
    \fig{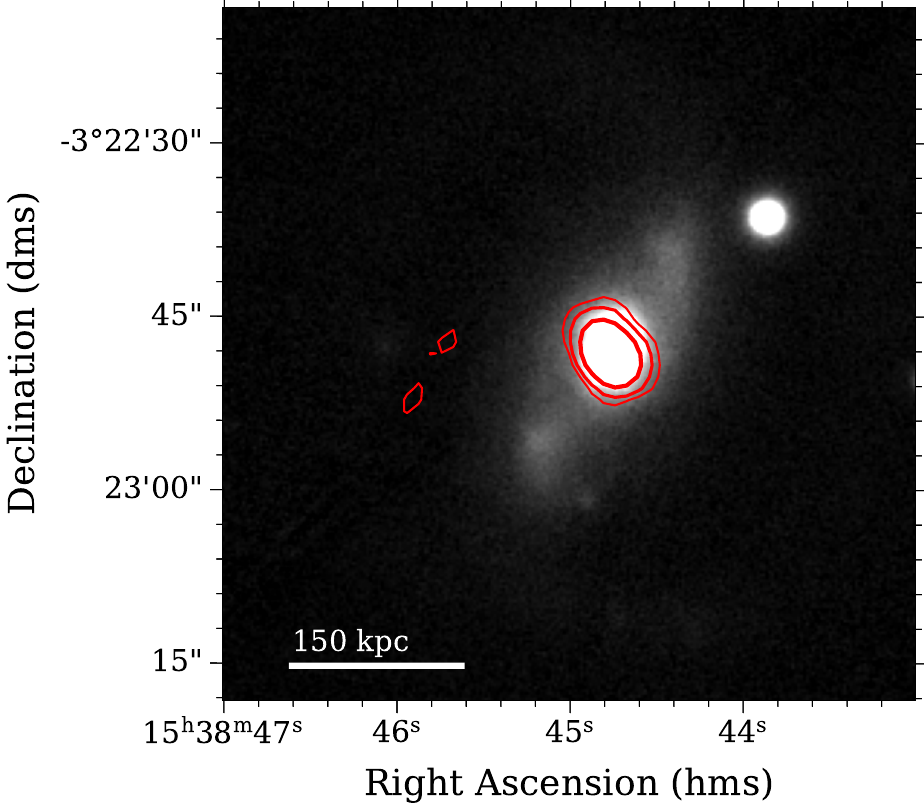}{0.28\textwidth}{J153845-032248}
    \fig{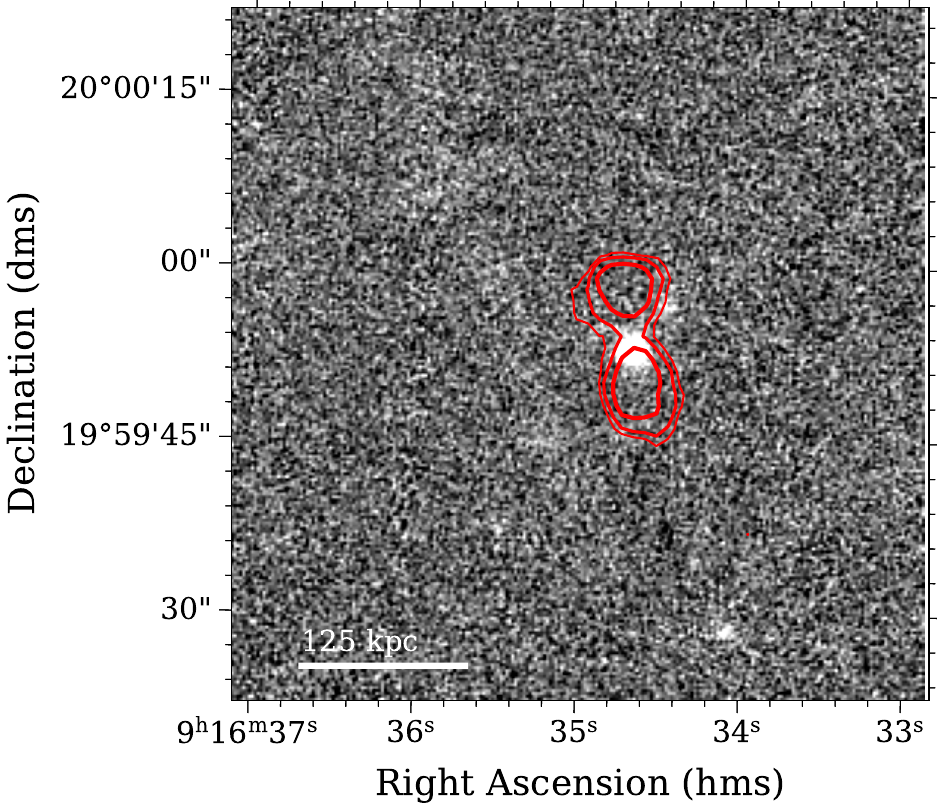}{0.28\textwidth}{J091635+195953}
  }
  \caption{Radio contours from VLASS images in red (at the 3$\sigma$, 5$\sigma$, and 10$\sigma$ RMS levels) overlaid onto optical background images for the 9 CLAGN objects which had extended jets or extended emission associated with the stellar population of the galaxy nucleus. The physical scale based on spectroscopic redshift is shown on the scalebar.}
  \label{fig:ExtendedJetFigure}
\end{figure*}

For the 68-size radio-detected CLAGN sample, we classified 59 CLAGN as compact sources, corresponding to a compactness fraction of $0.868^{+0.061}_{-0.100}$ (95\% Wilson score interval). Specifically, of our 24 VAST detections, 22 were classified as compact. We found that 9 CLAGN ($\sim13$\% of our 68-size sample) display kiloparsec-scale extended radio jets: J015106-003426, J091635+195953, J092314+043445, J094232+233614, J102818+211508, J135750+614309, J143422+044137, J162553+125317, and J153845-032248. Figure~\ref{fig:ExtendedJetFigure} displays the PanSTARRS-1\footnote{https://ps1images.stsci.edu/cgi-bin/ps1cutouts} optical background images and the CIRADA VLASS radio contours overlaid in red for the nine extended sources. We set radio contours at the 3$\sigma$, 5$\sigma$, and 10$\sigma$ significance levels, where $\sigma$ corresponds to the RMS noise. We overlay a scalebar to represent the physical size of the jets, calculated using the source redshifts in Table~\ref{tab:CLAGNSample}. The sizes of these jets (except CLAGN J153845-032248) range from $\sim100$ to $\sim700$ kiloparsecs, contrasting to point sources, which have smaller scales of tens of kiloparsecs. Using PanSTARRS-1 optical imaging, we confirmed that none of these nine sources are artifacts or multiple compact sources seen in projection. We note that for the low-redshift CLAGN J153845-032248, the radio emission traces the galaxy bulge. We note that two sources, CLAGN J084716+373219 and J090550+003948, have VLASS compactness ratios $> 2$, but we classified these as compact sources because visually, their radio cutouts were not convincingly extended. 

\subsubsection{Radio luminosities and radio-loudness} \label{RadioLuminosityLoudness}
To investigate whether radio emission is from an AGN jet, the X-ray emitting corona, or from star formation in the host galaxy, we calculated the radio loudness of CLAGN in our sample. We calculated \(R = S_{v,5GHz} / S_g\), defined as the ratio of the radio flux at 5 GHz to the optical luminosity at the g-band (4770 \AA) \citep{Kellermann1969}. Objects are classified as radio-loud if $R \geq 10$ and classified as radio-quiet otherwise. 

We obtained the $g$-band luminosities ($S_g$) for our radio-detected CLAGN using the following catalogs, in order of priority: the PanSTARRS DR2 catalog \citep{Flewelling2020TheProducts}, the SDSS DR16 catalog \citep{Ahumada2020TheSpectra}, and the Dark Energy Survey DR1 catalog \citep{Abbott2018The1}. For sources with no crossmatch within 5.0" in the three catalogs, we used the Simbad Search Catalogue to obtain a $g$-band magnitude from other catalogs. 

We obtained the radio flux at 5 GHz ($S_{v,5GHz}$) for the radio-detected CLAGN by extrapolating each CLAGN's median VLASS or VAST radio fluxes to the 5 GHz frequencies. First, we calculated each CLAGN's median peak flux in the VLASS 3GHz, VAST 1.3675 GHz, and VAST 0.8875 GHz frequencies across all available epochs. Then, to extrapolate each flux measurement to 5 GHz, we assumed a radio spectral shape of a power law with exponent $\alpha = -0.7$ where the flux density \(S_{\nu} \sim \nu^{\alpha}\) \citep{Kellermann1969,Amirkhanian1985}. To choose which flux measurement to extrapolate down to 5 GHz for each CLAGN, our first choice was the VLASS 3 GHz median flux (if available), VAST 1.3675 GHz median flux (if the first one was not available), and finally the median VAST 0.8875 GHz (if neither of the previous two fluxes were available). The radio loudness ratio for our radio-detected compact CLAGN sources is presented in Table~\ref{tab:Radio_Metrics_CLAGN_RQRLAGN}. We classify an object as radio-quiet if its radio loudness ratio $R < 10$ and as radio-loud otherwise.

We found that 31 of our 58 CLAGN were classified as radio-loud, corresponding to a fraction of $0.534^{+0.122}_{-0.126}$ (95\% Wilson score interval). There are 12 objects with very high radio loudness ratios ($R \geq 100$): J003849-011722, J022931-000845, J025515+003740, J094232+233614, J102914+271854, J111930+222650, J113111+373709, J123820+412420, J143422+044137, J164332+304836, J164829+410406, J214613+000931. Four of these objects, J111930+222650, J003849-011722, J022931-000845, and J164829+410406 are known and documented blazars.

\begin{figure}[hbt!]
    \centering
    \includegraphics[width=0.4\textwidth]{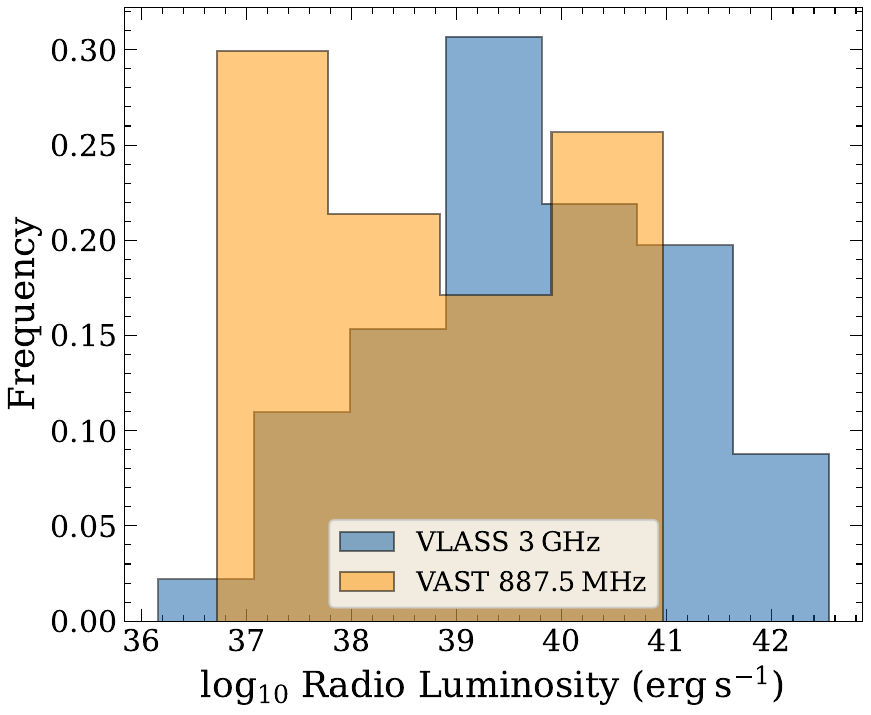}	
    \caption{Normalized distribution of the log of the radio luminosities for the 50 VLASS and 22 VAST-detected CLAGN samples.}
    \label{fig:RadioLuminosityHist}
\end{figure}

Furthermore, we calculated the radio luminosities of all objects with radio-detections. We used the spectroscopic redshifts reported in Table~\ref{tab:CLAGNSample} to calculate distances to the CLAGN. Using a 2 GHz bandwidth for VLASS and 288 MHz bandwidth for VAST, we calculated the VAST $L_{887.5\text{MHz}}$ radio luminosity for CLAGN with VAST detections, and calculated the VLASS $L_{3\text{GHz}}$ radio luminosity for CLAGN with VLASS detections. The radio luminosities for the 3 GHz VLASS detections and 887.5 MHz VAST detections are reported in Table~\ref{tab:Radio_Metrics_CLAGN_RQRLAGN}. The distribution of the log of our radio luminosities among the 3 GHz VLASS detections and 887.5 MHz VAST detections is shown in Figure~\ref{fig:RadioLuminosityHist}.

We found that the VLASS and VAST-calculated radio luminosities lie within the $10^{36-43}$ ergs s$^{-1}$ range. We observe that the two distributions of $\log(L_{\text{3GHz}})$ and $\log(L_{887.5\text{MHz}})$ have different peaks. Radio luminosities in the $\nu = 887.5$ MHz VAST frequency are skewed towards lower values (peaked at $10^{38.5}$ ergs$^{-1}$) compared to the luminosities in the $\nu = 3$~GHz VLASS frequency (peaked at $10^{39.7}$ ergs$^{-1}$). A two-sample Anderson–Darling test comparing the $\log(L_{\text{3GHz}})$ and $\log(L_{887.5\text{MHz}})$ distributions gives $A^2=4.87$ with $p = 3.8 \times 10^{-3}$. We therefore reject the null hypothesis that the two distributions are drawn from the same parent population at \textgreater 99.5\% confidence.

\subsubsection{Radio spectral index} \label{RadioSpectralIndex} 

We estimated the power-law spectral index of the radio SED for compact CLAGN which have both VAST 887.5 MHz and VLASS 3 GHz detections. We assumed a smooth radio spectral shape of a power law, as described in Section~\ref{RadioLuminosityLoudness}, where the radiative flux density depends on frequency as \(S_{\nu} \sim \nu^{\alpha}\). We used the formula:   

\begin{equation}
    \alpha = \frac{\log S_{3GHz} /S_{0.887GHz}}{\log 3 GHz/0.887 GHz}
\end{equation}

where $S_{3GHz}$ is the median 3GHz VLASS flux value, and $S_{0.887GHz}$ is the median 887.5 MHz VAST flux value. 

For the 15 compact CLAGN which have both VAST 887.5 MHz and VLASS 3 GHz detections, we calculate the power law spectral index $\alpha$. We report these values in Table~\ref{tab:Radio_Metrics_CLAGN_RQRLAGN}. We find that 14 out of our 15 compact CLAGN have negative power law spectral index indices, and that the CLAGN are distributed around $\alpha = -1.0$ to 0.

\subsubsection{Radio Variability} \label{RadioNEV} 
We calculated the normalized excess variance (NEV) of the compact VAST-detected CLAGN \citep{Vagnetti2011}:

\begin{equation}
    \sigma^2_{NXS} = \frac{S^2 - \sigma^2_{n}}{\overline{f}^2}
\end{equation}

where $\overline{f}$ is the mean flux computed over all available flux measurements of the source, $S$ is the total variance of the light curve, and $\sigma^2_{n}$ is the mean square photometric error associated with each measured flux. 

We calculated the NEV in the VAST 887.5 MHz band and the VAST 1.3675 GHz band for the CLAGN which had at least two detections in the corresponding frequency. We classified a CLAGN as variable in the 887.5 MHz and 1.3675 GHz frequencies if its NEV was $> 0.02$, and unvarying otherwise. We chose this variability cutoff value based on visually inspecting the radio light curves, and determining which cutoff appropriately reflects whether a CLAGN is displaying gradual flux evolution beyond what is expected for interstellar scintillation. These values are reported in Table~\ref{tab:Radio_Metrics_CLAGN_RQRLAGN}.

We also calculated the percentage change in VLASS flux across the two epochs for our radio-detected CLAGN objects. We classified a source as variable in VLASS if it had a greater than 10\% change in VLASS 3 GHz flux. We display the percentage change in flux across the two VLASS epochs, and the variability classification for the CLAGN in Table~\ref{tab:Radio_Metrics_CLAGN_RQRLAGN}. In Section~\ref{RadioSelectedSample}, we display the distribution of the NEV values calculated in the VAST 887.5 MHz and VAST 1.3675 GHz frequencies, for samples of size 22 and 10 respectively.

We found that 2 of the 22 VAST-detected CLAGN are variable in the 887.5 MHz frequency: NGC 1566 and J100849-095451. This corresponds to a radio variability rate of $0.091^{+0.187}_{-0.066}$ (95\% Wilson score interval). We observe NGC 1566 to fade from $\sim$ 45 to 30 mJy between 2019 and 2023, and from 2023 onward, its flux hovers around 30 mJy. We observe that J100849-095451 sporadically rises and falls from $\sim$ 6 to 3.5 mJy between 2019 and 2021, and from 2023 onwards, its flux hovers around 3 to 4 mJy. We found that none of the 10 VAST-detected CLAGN are variable in the 1.3675 GHz frequency. 
Overall, the majority of CLAGN in our sample do not exhibit significant flux variability in VAST wavelengths.

We calculated the percentage change in VLASS flux across the two epochs for 42 VLASS-detected CLAGN. We found that 33 of our 42 sources are variable in VLASS, corresponding to a fraction of $0.786^{+0.097}_{-0.145}$ (95\% Wilson score interval).

\subsubsection{Classification of source type} \label{SourceClassificationDesc}
We classified our radio-detected CLAGN using 6 categories: 
\begin{itemize}
\item Blazar (if it crossmatched to any catalogs described in Section \ref{Blazars}); 
\item Extended radio emission: Failed compactness criteria or had extended structure in VLASS or RACS imaging;
\item RL-V: Radio-loud, variable, compact source (if it was radio-loud, passed all compactness criteria, and had statistically significant variability);
\item RQ-V: Radio-quiet, variable, compact source;
\item RL-NV Radio-loud, non-variable compact source; 
\item RQ-NV: radio-quiet, non-variable, compact source. 
\end{itemize}

We present these results in Table~\ref{tab:Radio_Metrics_CLAGN_RQRLAGN}. If VLASS data across two epochs was available for a source, we classified it as variable if its VLASS percentage flux change was greater than 10\% and as non-variable otherwise. If VLASS data was not available and VAST data was available, then we classified a source as variable if its VAST 887.5 MHz NEV value was greater than 0.02 (see reasoning for cutoff choice in Section~\ref{RadioNEV}). 

Not all of our objects had variability metrics available, nor did they have radio loudness ratios available, since they were either not detected in VAST, nor did they have VLASS detections in both epochs, nor did they have $g$-band fluxes available. For these objects, we only specified if they were radio-quiet or radio-loud, or variable or unvarying. 

We found that for our 68-size radio-detected CLAGN sample, eight were classified as blazars (B), and that two of these blazars (J102818+211508 and J094232+233614) display visibly extended jets. 
We found that 7 non-blazar objects were classified as extended radio emission sources (E). Non-blazar extended radio emission sources represent $\sim10$\% of our sample. We found that 13 were classified as radio-loud, variable, compact sources (RL-V), comprising the largest portion, $\sim19$\%, of the radio-detected CLAGN sample. 12 CLAGN were classified as radio-quiet, variable, compact sources (RQ-V), representing $\sim18$\% of the radio-detected CLAGN sample. 7 CLAGN were classified as radio-loud, non-variable compact sources (RL-NV), representing $\sim10$\% of the radio-detected CLAGN sample. Finally, 9 CLAGN were classified as radio-quiet, non-variable, compact sources (RQ-NV), representing $\sim13$\% of the radio-detected CLAGN sample. For the remaining 10 compact radio-detected objects which did not have VLASS or VAST variability metrics available, or radio loudness ratios, we found that 6 ($\sim86$\%) of our 7 sources were classified as radio-quiet, and 2 of our 3 sources were classified as unvarying. Finally, we label the two sources, J084716+373219 and J090550+003948, that were classified as extended but displayed no extended radio jets simply as C. We do not calculate radio metrics for these two objects.



\section{Optical variability of the CLAGN sample} \label{ZTFLightCurves} 


We obtained optical data from the Zwicky Transient Facility (ZTF) optical time-domain survey   \citep{Bellm2019, Graham2019, Dekany2020TheSystem}. We constructed optical light curves from ZTF difference imaging using the ZTF photometry service \citep{Masci2019}. We extracted all available photometry from the ZTF public fields between 2018-01-01 and 2024-03-01. After removing poor quality images by requiring the \texttt{procstatus} flag be equal to 0, we measured the baseline flux from  the reference images, applied zero points, and combined the baseline flux measured from the reference images and the single epoch fluxes to produce g- and r-band light curves of the two samples.  We found ZTF detections associated with 268 of the 474-size spectroscopically and optically-selected CLAGN sample, and 38 of the 52-size RQ to RL VLASS AGN sample. ZTF data coverage was largely during or after the state change. 


In order to quantify the level of optical variability of the objects using ZTF photometry, we calculated the NEV \citep{Vagnetti2011} of the full set of ZTF flux measurements ($g$-, $r$-, and $i$-band photometry). We filtered out outlier ZTF data points which had error values $> 0.3$, and outlier data points which were defined as points with a z-score $> 3$. The ZTF-calculated NEVs are reported in Table~\ref{tab:CLAGNSample}.

\begin{figure}
    \centering
    \includegraphics[width=0.4\textwidth]{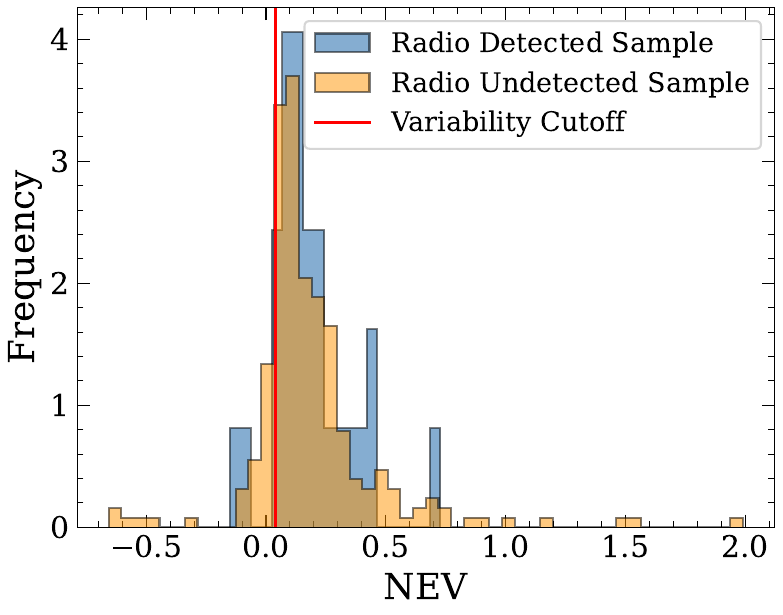}	
    \caption{Distributions of ZTF-calculated NEV values for the 28-size radio detected CLAGN sample and the 240-size radio undetected sample. The 0.04 variability cutoff value is overlaid.}
    \label{fig:ZTFNEVDist}
\end{figure}

We investigated whether the radio-detected CLAGN have different ZTF optical NEV values than the radio undetected CLAGN. To do this, we split the 268-size ZTF-detected CLAGN sample into two populations: 28 CLAGN that were detected in the radio wavelengths (in either VAST or VLASS), and 240 CLAGN that were not detected in the radio wavelengths. We compared the distribution of NEV values for radio detected and radio undetected CLAGN. We classified a CLAGN as optically variable if its ZTF NEV value was $> 0.04$, and unvarying otherwise. We visually chose this cutoff to appropriately reflect when the AGN light curves showed long-term AGN-like variability. In Figure~\ref{fig:ZTFNEVDist}, we display the two NEV distributions for the radio detected and radio undetected populations. Table~\ref{tab:CLAGNSample} contains the ZTF-calculated NEV values for our 268 ZTF-detected CLAGN sample. 


A two-sample Anderson–Darling test comparing the ZTF-NEV distributions of radio-detected and radio-undetected CLAGN yields $A^2=-0.94$ with $p > 0.25$. The Anderson–Darling test fails to reject the null hypothesis and thus we find no statistically significant evidence that the overall ZTF-NEV distributions differ between radio-detected and radio-undetected CLAGN. This is confirmed by the fact that we found that 26 of our 28-size radio detected CLAGN are variable in ZTF bands, corresponding to a ZTF variability fraction of $0.929^{+0.052}_{-0.155}$ (95\% Wilson score interval), and found that 200 of our 240-size radio undetected CLAGN are variable in ZTF bands, corresponding to a ZTF variability fraction of $0.833^{+0.042}_{-0.052}$ (95\% Wilson score interval). We do not find statistically significant evidence that radio-detected CLAGN display higher levels of ZTF variability compared to radio-undetected CLAGN. 



\section{Follow-up observations of selected CLAGN and RQ-to-RL AGN} 

\subsection{Spectroscopic follow-up with Magellan \label{Magellan}}
\begin{figure*}
\gridline{ 
 \fig{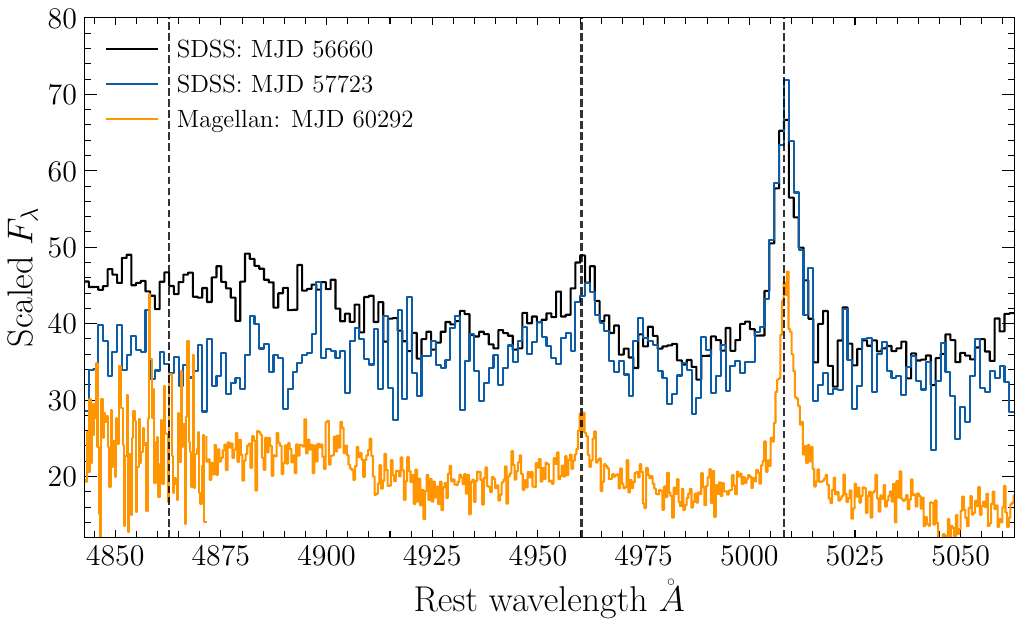}{0.48\textwidth}{a) J020515-045640} \fig{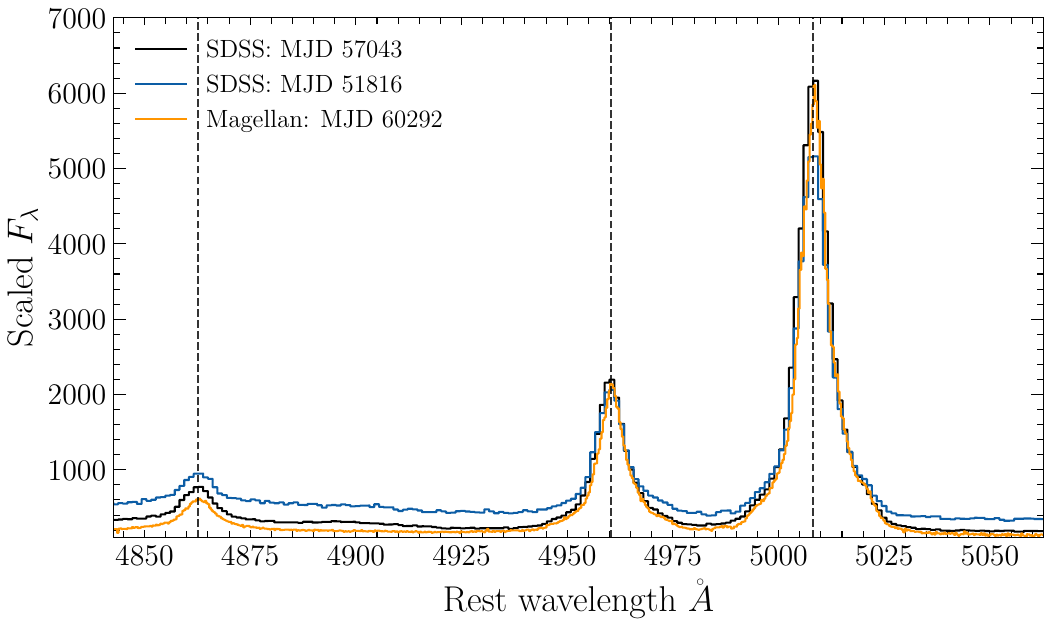}{0.48\textwidth}{b) J021400+004227} }
 \gridline{\fig{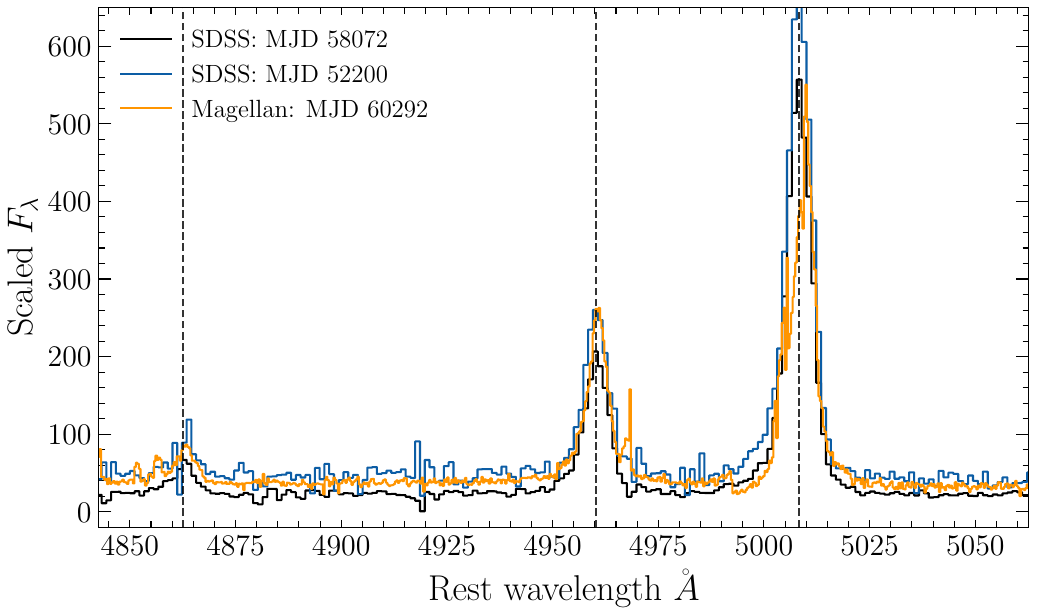}{0.48\textwidth}{c) J022931-000845}
\fig{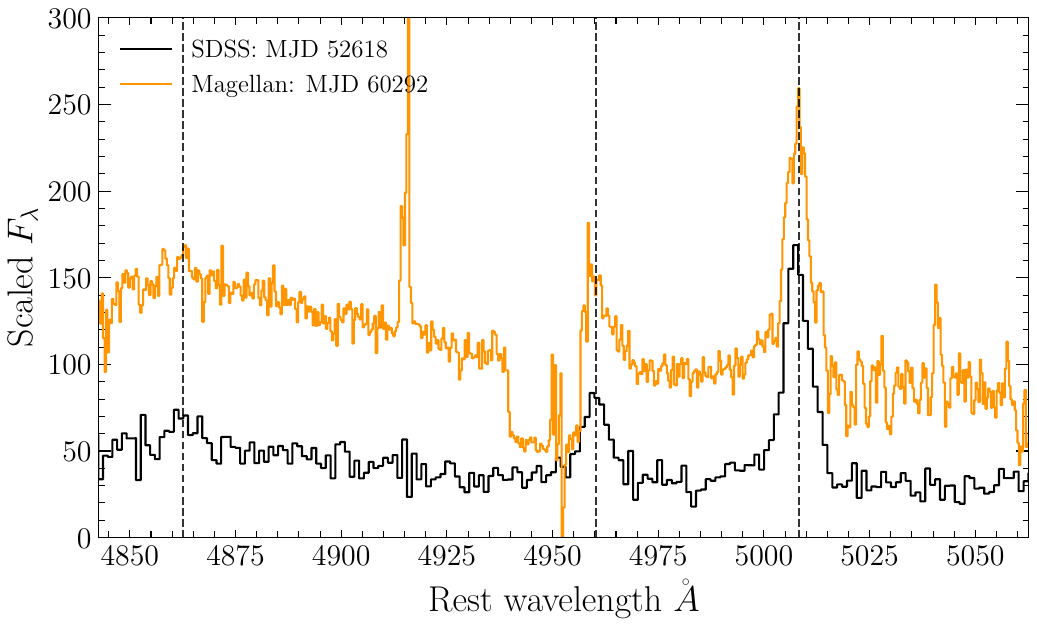}{0.48\textwidth}{d) J074248+270412} }
\gridline{ \fig{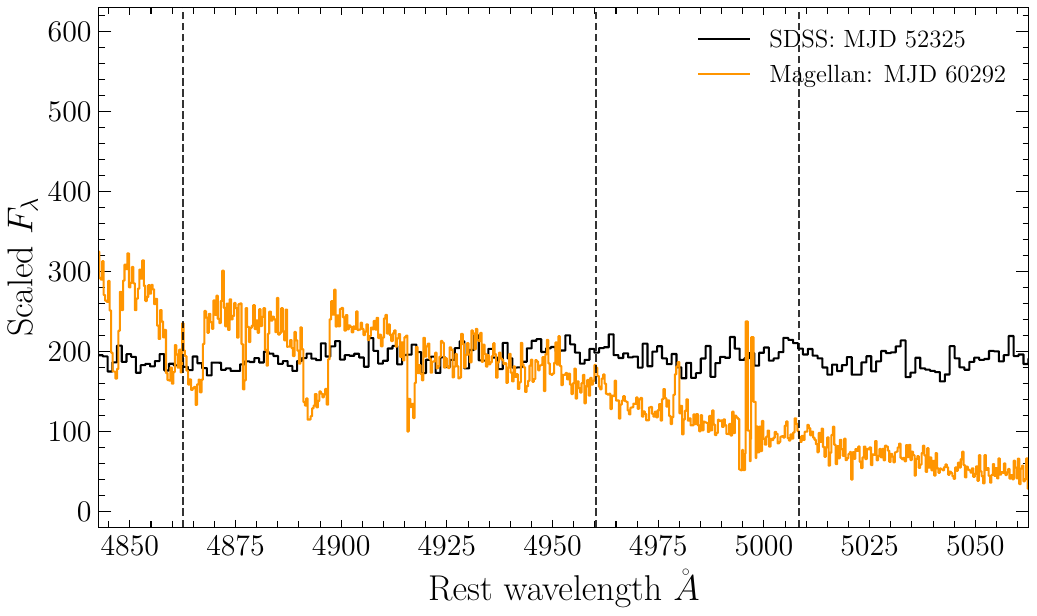}{0.48\textwidth}{e) J102951+043658} \fig{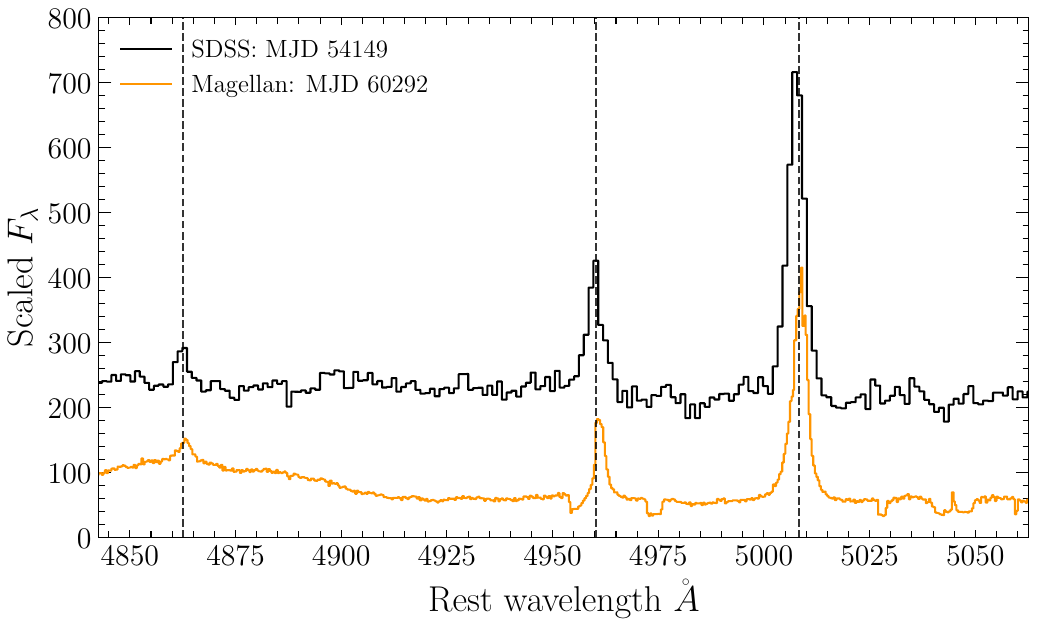}{0.48\textwidth}{f) ZTF18aaabltn}}
\caption{Zoom into the H$\beta$ regions for archival SDSS spectra and recent MAgE spectra of 3 CLAGN identified in SDSS-IV that have radio variability in VAST (a, b, and c), two AGN that switched from radio-quiet to radio-loud in VLASS (d-e), and one radio-undetected CLAGN identified via follow-up of a flare detected by ZTF (f).}
\label{fig:magellan}
\end{figure*}

In order to identify the timescales of broad line variability in different objects of interest: 3 radio-variable CLAGN from SDSS-IV (J020515-045640, J021400+004227, J022931-000845), one radio-undetected CLAGN from ZTF (ZTF18aaabltn), and 2 radio-quiet to radio-loud AGN from VLASS (J074248+270412 and J102951+043658), we obtained optical spectroscopy using the Magellan Echellette (MagE) spectrograph on the Magellan Baade telescope on December 14 2023. We observed over a 3200-10000\,\AA\ wavelength range with a 0.75" slit, with total exposure times ranging from 2400 to 5400 s. The recent spectra are shown compared to archival SDSS spectra in Figure \ref{fig:magellan}. The SDSS-IV CLAGN were found to have returned to their initial `bright' state, after dimming of the broad lines was observed in intermediate epochs of mjd 58000 - 59000 in SDSS-IV \citep{Green2022TheSDSS-IV}. ZTF18aaabltn was also found to be in the process of returning to its initial dim state after broad lines were observed to appear in 2018 \citep{Frederick2019ALINERs}. The broad lines have since faded relative to the narrow emission lines, but are still present. Both objects that were observed to switch from radio-quiet to radio-loud in VLASS \citep{Nyland2020QuasarsFIRST,Zhang2022TransientSurveys} did not show the appearance of broad lines compared to their archival spectra. The finding that the SDSS-IV CLAGN have returned to their original state, combined with the lack of changes to their radio properties, suggests that no long term changes to the AGN BLR geometry and radio jet behavior have occurred. Meanwhile, the finding that no broad lines have appeared for the RQ to RL AGN suggests that their long term radio transition was not associated with an optical state-change.

\subsection{VLA follow-up of radio-detected CLAGN} \label{VLA}
We performed high-resolution, multiband VLA observations for 4 radio-detected CLAGN and 8 RQ to RL AGN. These observations took place from March 3 – April 17, 2025, during the VLA D configuration. For each source, the observations at L (1–2 GHz), S (2–4 GHz), C (4–8 GHz), X (8–12 GHz) and \textit{Ku} (12–18 GHz) band were performed within a single scheduling block (SB) no longer than 1 hr, so the observations were ``quasi-simultaneous" in nature, following the observing approach of \citet{Nyland2020QuasarsFIRST}. Imaging and self-calibration were performed in CASA 5.6.0 following standard heuristics for widefield, broadband VLA data.

\begin{figure*}
\centering
\gridline{\fig{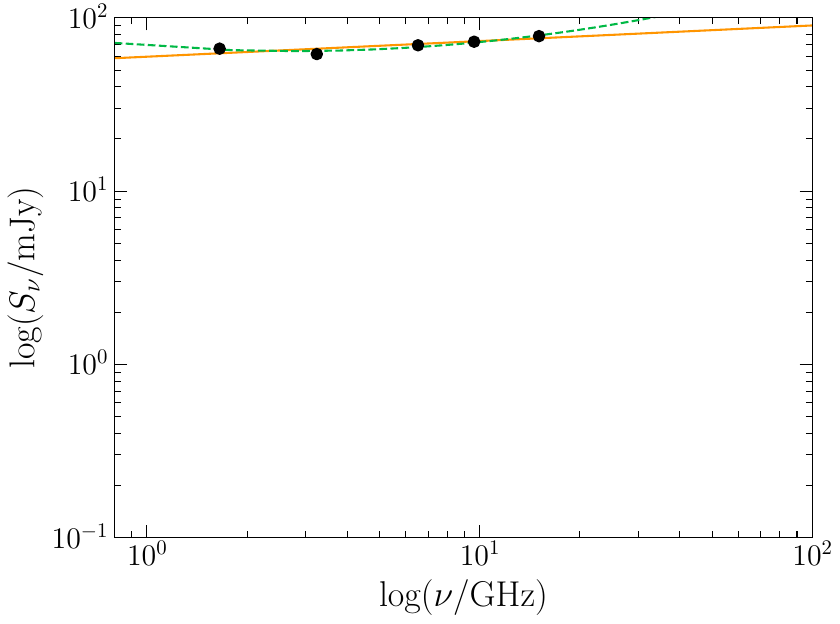}{0.4\textwidth}{a) J003849-011722}\fig{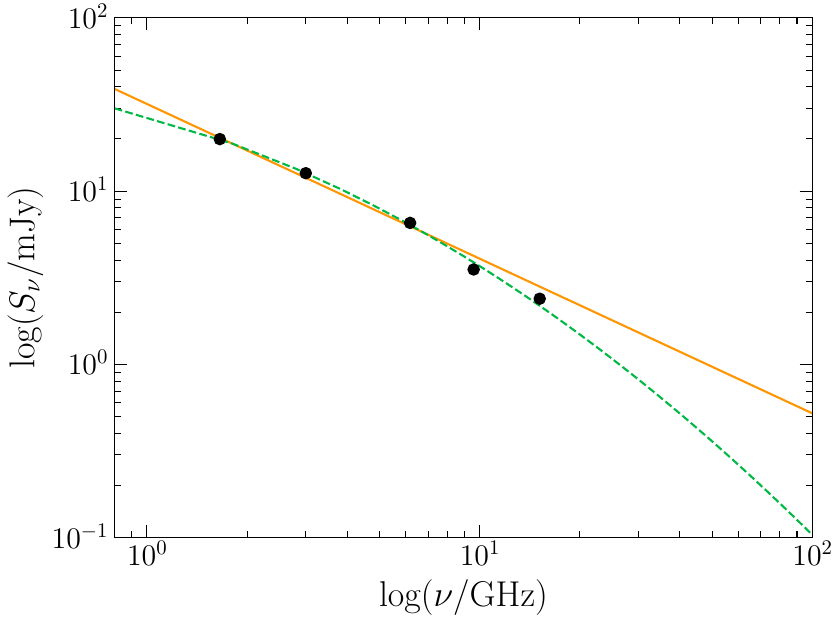}{0.4\textwidth}{b) J020515-045640}}
\gridline{\fig{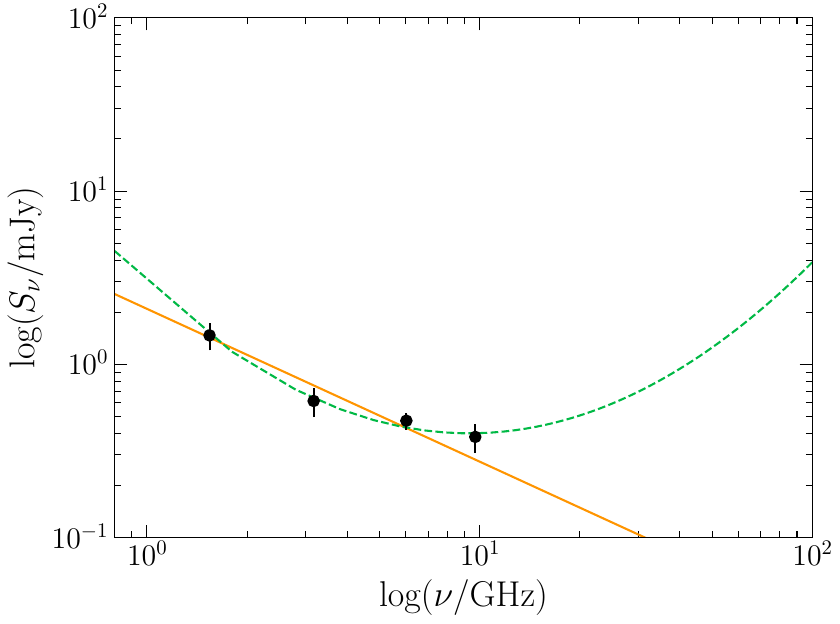}{0.4\textwidth}{c) J090536-004040}\fig{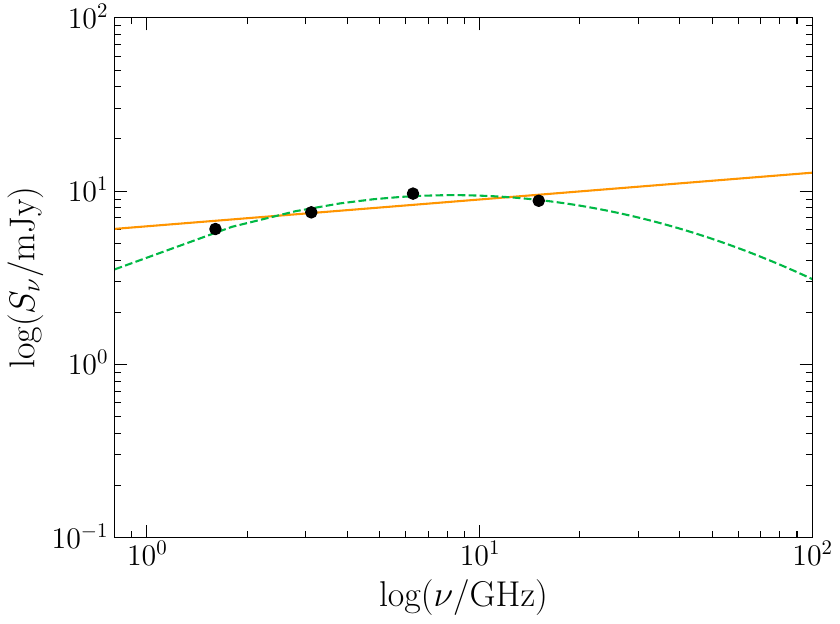}{0.4\textwidth}{d) J214613+000931}}
\caption{Broadband radio SEDs showing quasi-simultaneous multiband VLA imaging of optically-selected CLAGN. For each source, two models based
on the quasi-simultaneous VLA data are shown: a standard non-thermal power-model (green dotted line) and a curved power-law model (solid orange line).}
\label{fig:VLASEDs}
\end{figure*}

\begin{figure*}
\centering
\gridline{\fig{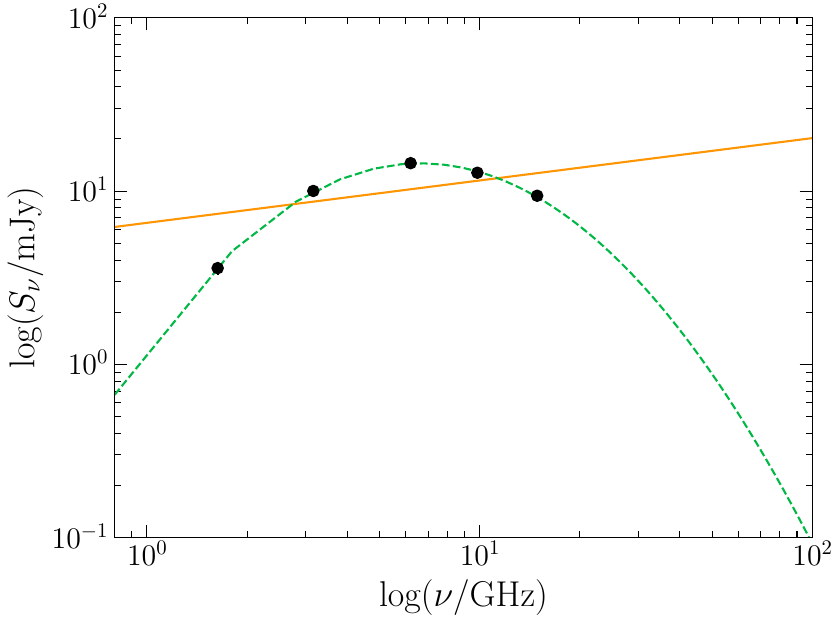}{0.33\textwidth}{a) J004044+082352 (Z22)}\fig{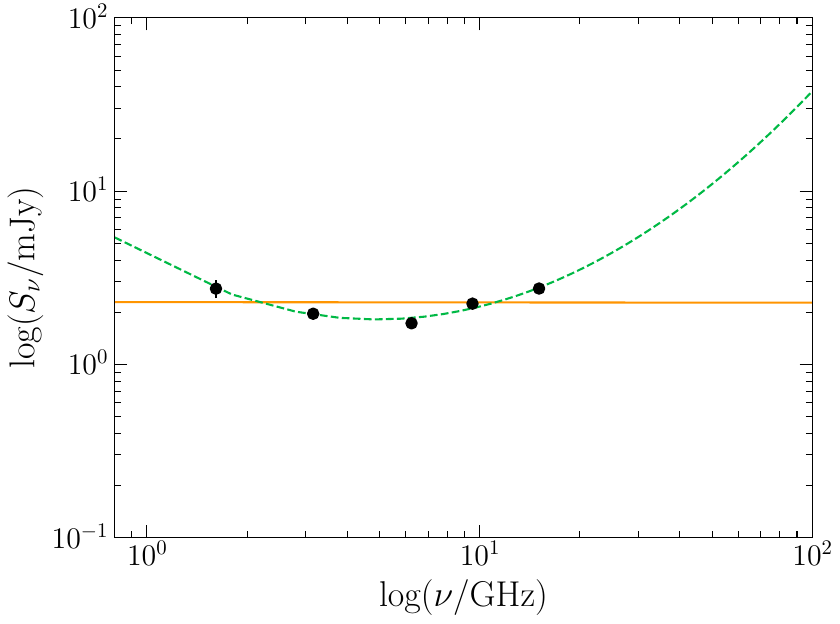}{0.33\textwidth}{b) J013815+002914 (W21)}\fig{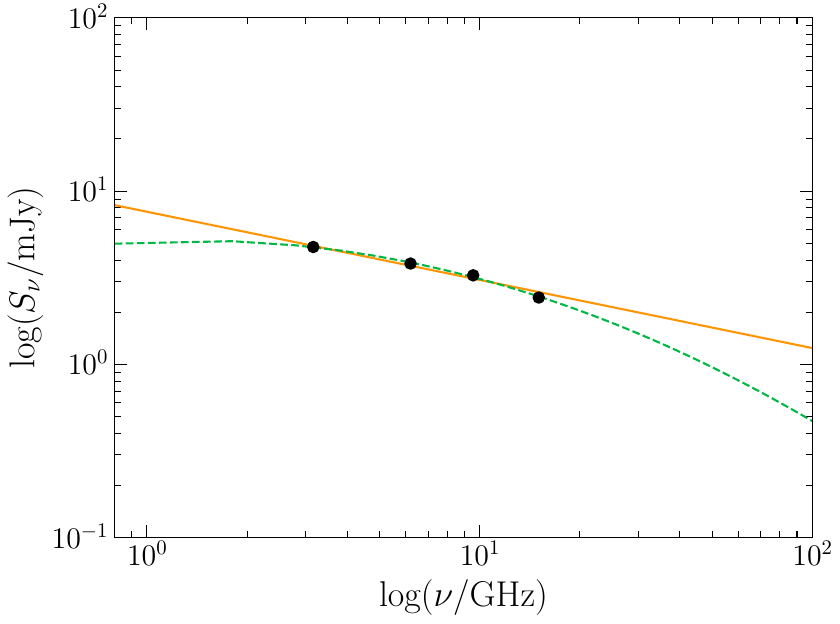}{0.33\textwidth}{c) J015412-011150 (W21 \& Z22)}}
\gridline{\fig{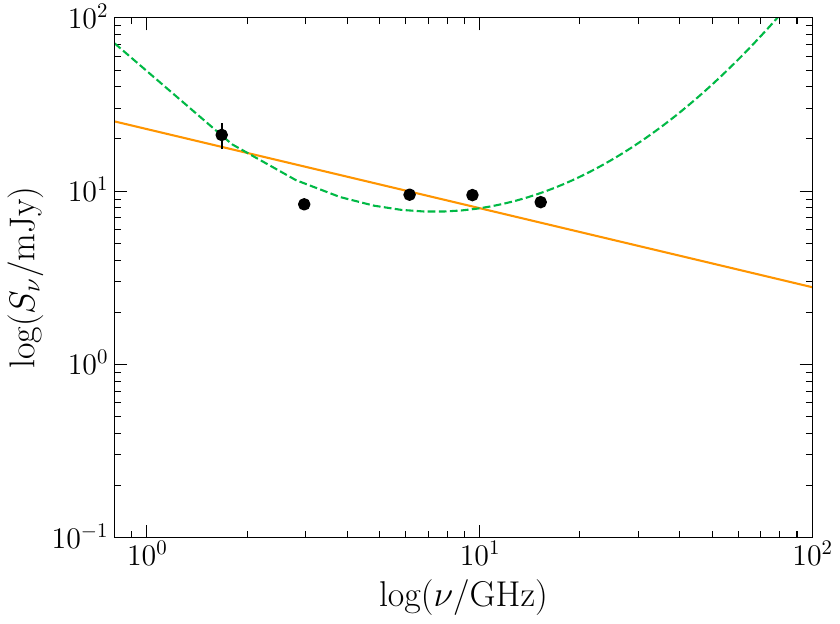}{0.33\textwidth}{d) J102951+043658 (Z21)}\fig{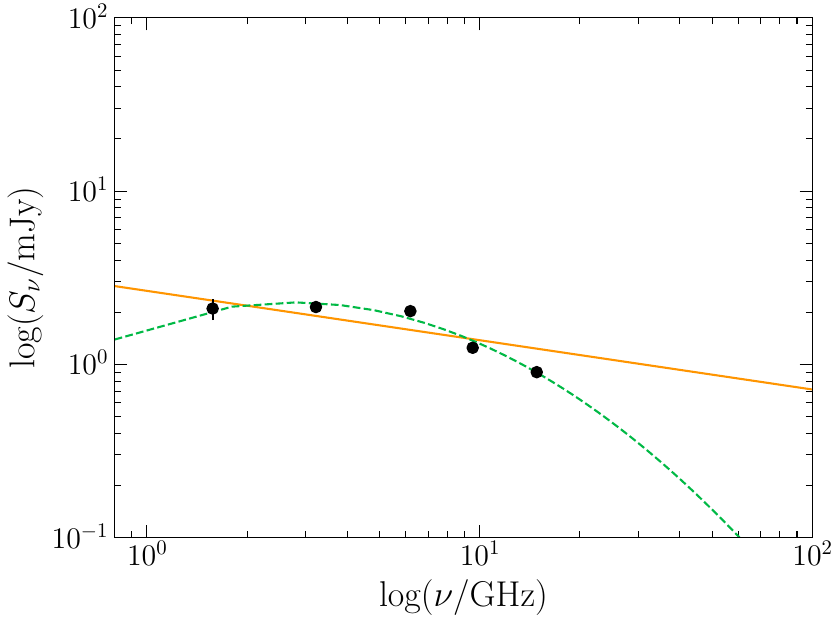}{0.33\textwidth}{e) J133320-034956 (N20)}\fig{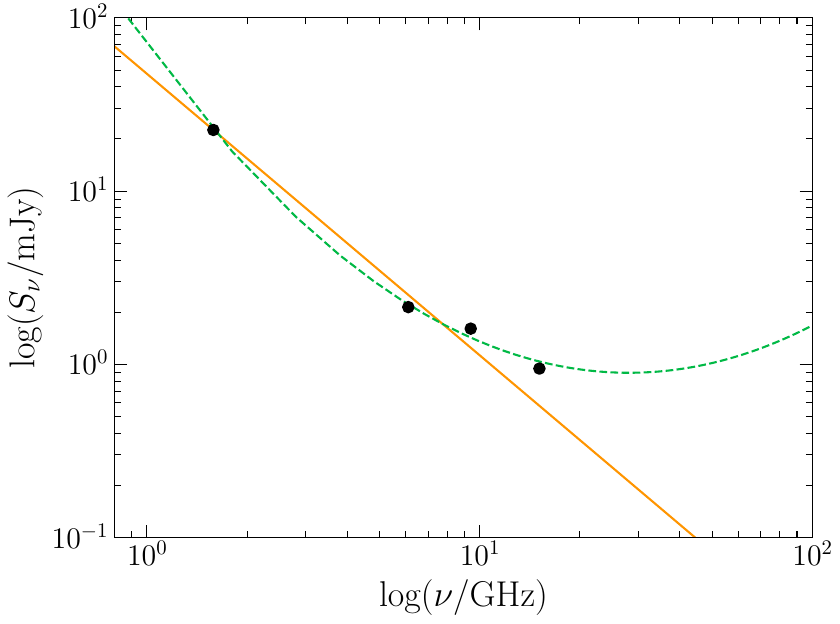}{0.33\textwidth}{f) J150751-054911 (N20)}}
\gridline{\fig{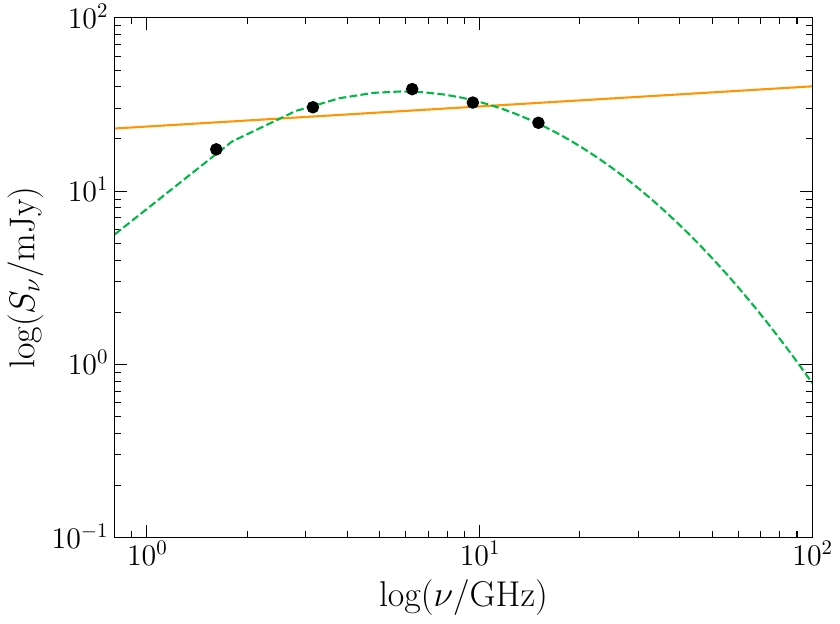}{0.33\textwidth}{g) J210917-064437 (N20)}\fig{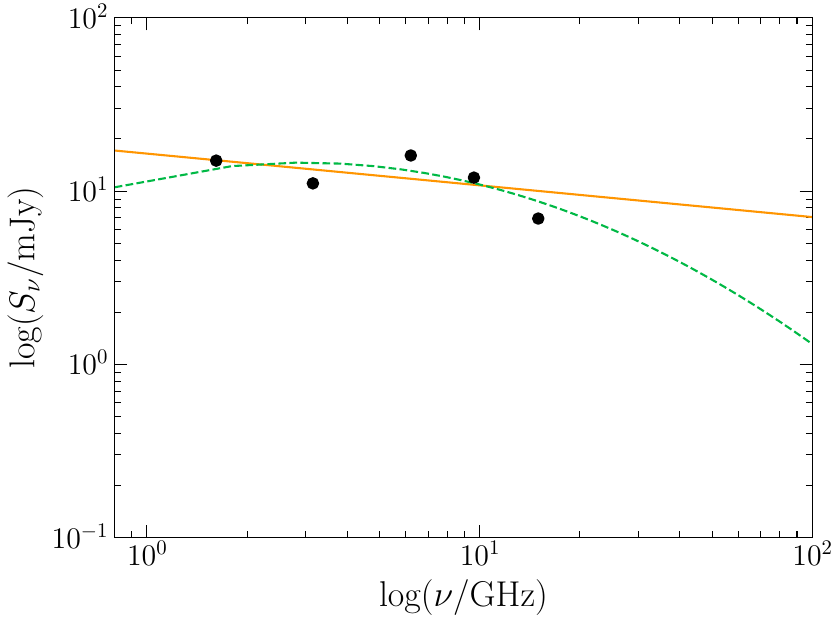}{0.33\textwidth}{h) J221813-010344 (N20)}}
\caption{Broadband radio SEDs from quasi-simultaneous multiband VLA imaging obtained within March 3 - April 27 2025 for a subset of the RQ to RL AGN. For each source, two models based on the quasi-simultaneous VLA data are shown: a standard non-thermal power-model (green dotted line) and a curved power-law model (solid orange line).}
\label{fig:VLASEDsradioAGN}
\end{figure*}

Following the methods presented in \citet{Nyland2020QuasarsFIRST} we fit the VLA SEDs with two models. We first fit a standard non-thermal power-law model defined by $S_{\nu} = a\nu ^{\alpha}$, where $S_{\nu}$ is the flux at frequency $\nu$, $a$ represents the amplitude, and $\alpha$ is the spectral index. Secondly, we fit a curved power-law model defined by $S_{\nu} = a\nu ^{\alpha} e^{q (\ln \nu)^{2}}$ where $q$ represents the degree of spectral curvature (the breadth of the peak of the radio SED). The $q$ parameter is defined by $\nu_{\text{peak}} = e^{-\alpha/2q}$, where $\nu_{\text{peak}}$ is the turnover frequency. The SEDs and best-fit models are shown in Figure \ref{fig:VLASEDs} for the 4 radio-detected CLAGN and in Figure \ref{fig:VLASEDsradioAGN} for the 8 RQ to RL AGN.

For the five of the eight RQ to RL AGN in Section~\ref{VLA}, we compared their radio SEDs reported in the discovery papers to the updated radio SEDs to determine if they showed SED evolution over time. We find that for RQ to RL AGN J0040+0823, our VLA SED has a different shape compared to the radio SED reported in \citet{Zhang2022TransientSurveys}, which was obtained in $\sim$2017-19. We find that the SED in \citet{Zhang2022TransientSurveys} is flatter than our SED, and more closely follows a power law compared to our SED which has a more concave shape. Our radio SED has a concave spectra. For RQ to RL AGN J015412-011150, our VLA SED has a very similar shape to the SED reported in \citet{Zhang2022TransientSurveys} which was obtained from 2015-19. Both SEDs are consistent with a power law, albeit our SED is slightly flatter than the \citet{Zhang2022TransientSurveys} SED. For RQ to RL AGN J102951+043658, our VLA SED also has a very similar shape to the SED reported in \citet{Zhang2022TransientSurveys} which was obtained from 2017-20. Both SEDs are roughly flat and consistent with a power law. For RQ to RL AGN J210917-064437, our VLA SED is also very similar in shape to the SED reported in \citet{Nyland2020QuasarsFIRST}, which is concave, and was obtained in 2019. Finally, for RQ to RL AGN J221813-010344, our VLA SED has a different shape compared to the radio SED reported in \citet{Woowska2021Caltech-NRAOState}, which was obtained from 2014-19. We find that although both SEDs are concave in shape, our recently observed SED is much more flat than the previously reported SED, and follows a power law more closely. For the three RQ to RL AGN which did not have previous SEDs available for comparison, we find that J013815+002914 has a relatively flat SED, J133320-034956 has a concave SED, and J150751-054911 has a relatively flat SED as well. 

Notably, we find that all four of our VLA-observed CLAGN SEDs are consistent with a power law rather than having concave shapes consistent with peaked-spectrum radio AGN. On the other hand, the majority of the RQ to RL AGN have concave SED shapes, which are indicative of young recently formed radio jets. Thus, we conclude that CLAGN differ from RQ to RL objects in that their radio SEDs are much flatter and are not concave. 

\section{Results} \label{Results}

\subsection{CLAGN Light Curves} \label{NotableObjects}
Previous literature has identified individual CLAGN that displayed notable radio variability; in particular, CLAGN Mrk 590 and NGC 1566 have well studied flare and fading events. The radio light curves of Mrk 590, and NGC1566 were displayed in Figure~\ref{fig:KnownCLAGNRadioLCs}. Mrk 590 is well known for its interesting radio behavior \citep{Denney2014TheRole, Mathur2018TheAwakening, Koay2016Parsec-scale590, Yang2021AMrk590}. We constructed 22 light curves of compact sources detected in VAST wavelengths. We found that Mrk 590 was classified as variable in VLASS 3 GHz wavelengths, but as unvarying in VAST 887.5 MHz wavelength. NGC 1566 was classified as variable and was observed to fade in the VAST 887.5 MHz wavelengths.

We do not find any additional sources with detected flare and fading events in our sample. We find that none of the 20 newly studied CLAGN have statistically significant fading detections after the changing-look event in VAST wavelengths. Thus, we find that the radio behavior of 1ES 1927+654, Mrk 590, and NGC 1566 is not reflected in the population of CLAGN.  

\begin{figure*}
\gridline{ 
    \fig{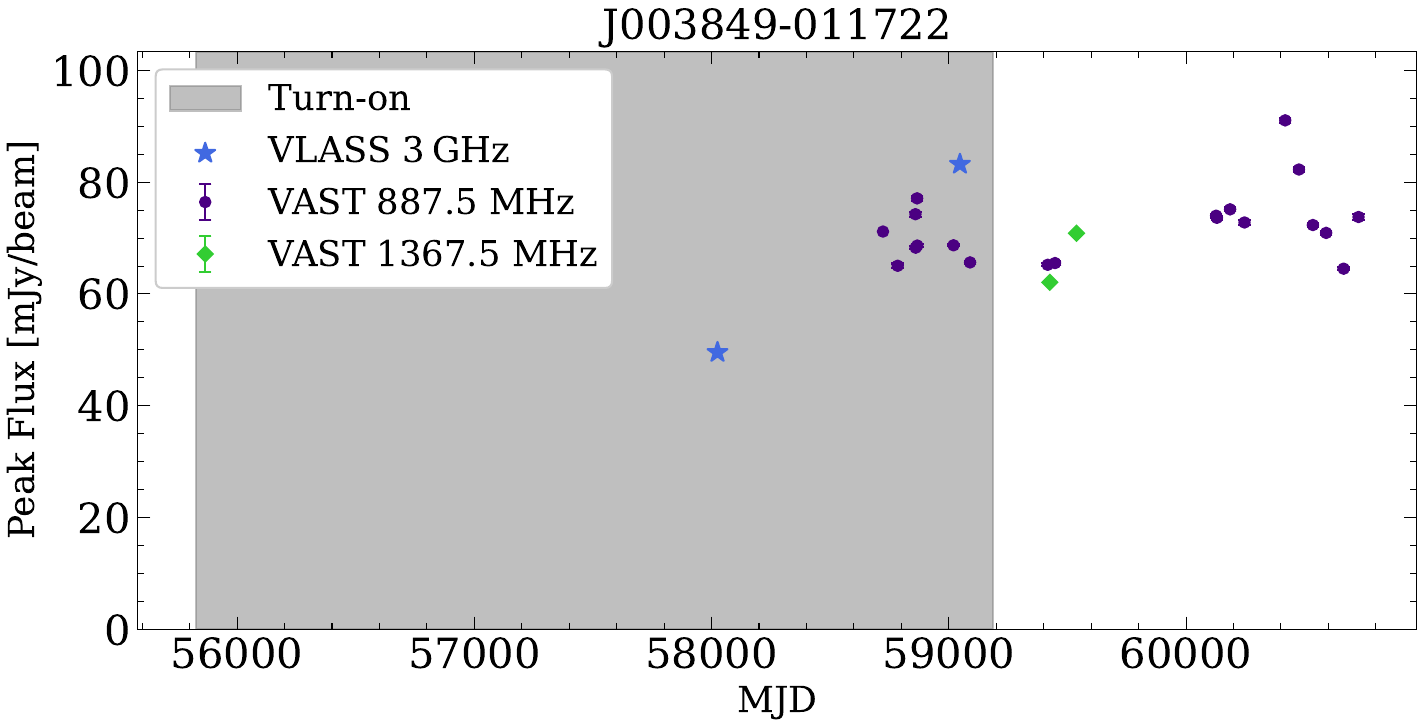}{0.49\textwidth}{}
    \fig{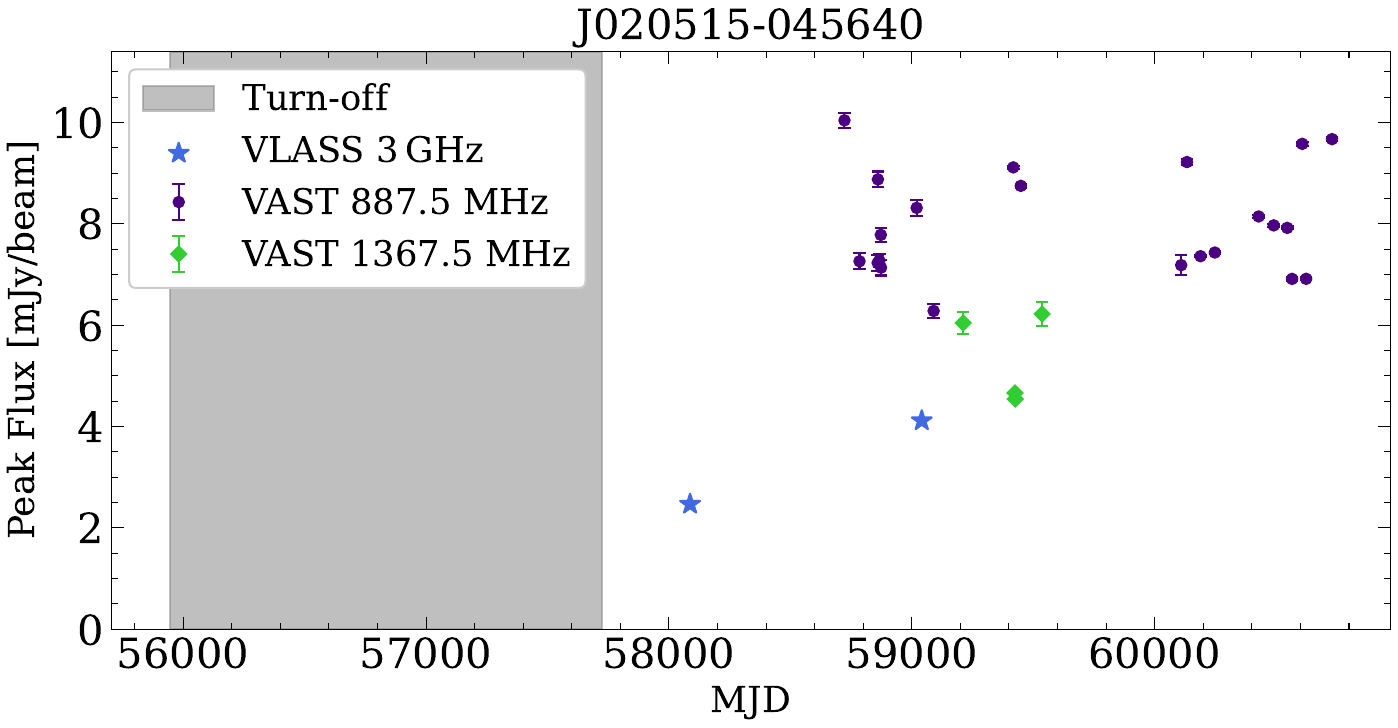}{0.49\textwidth}{}}
\gridline{
    \fig{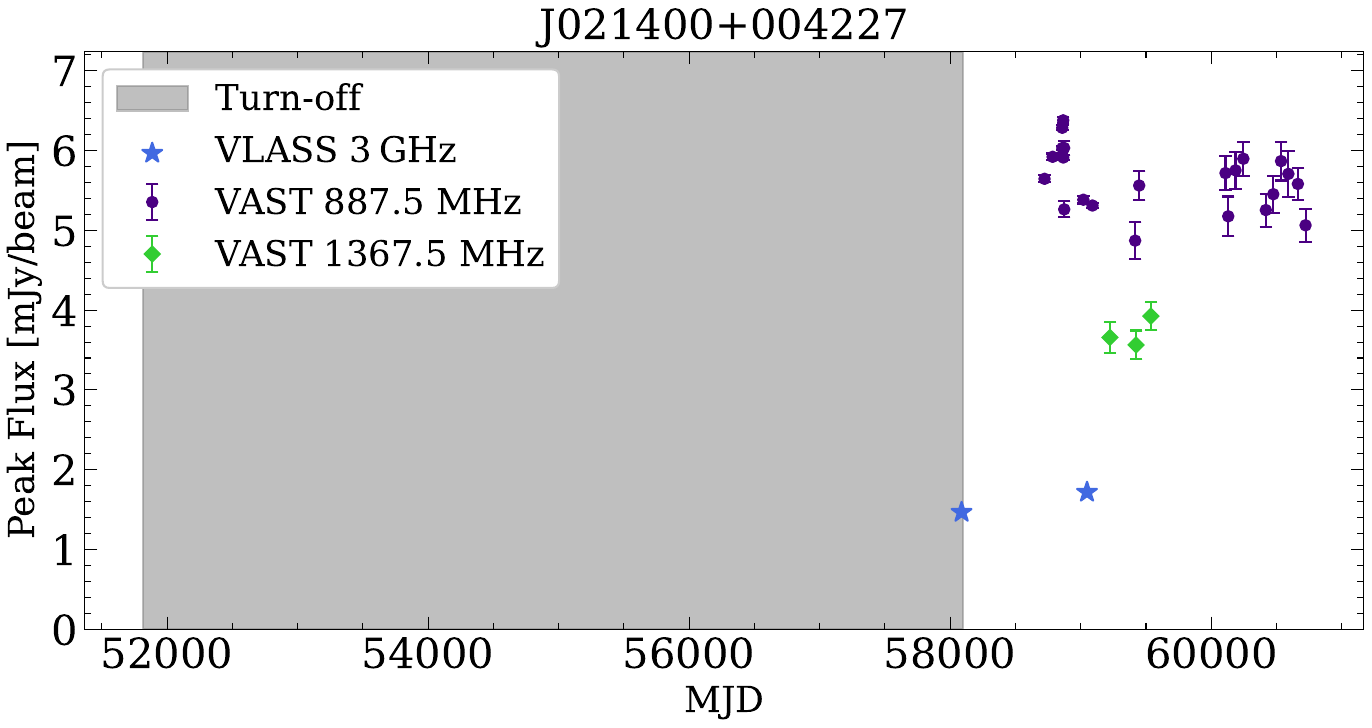}{0.49\textwidth}{}
    \fig{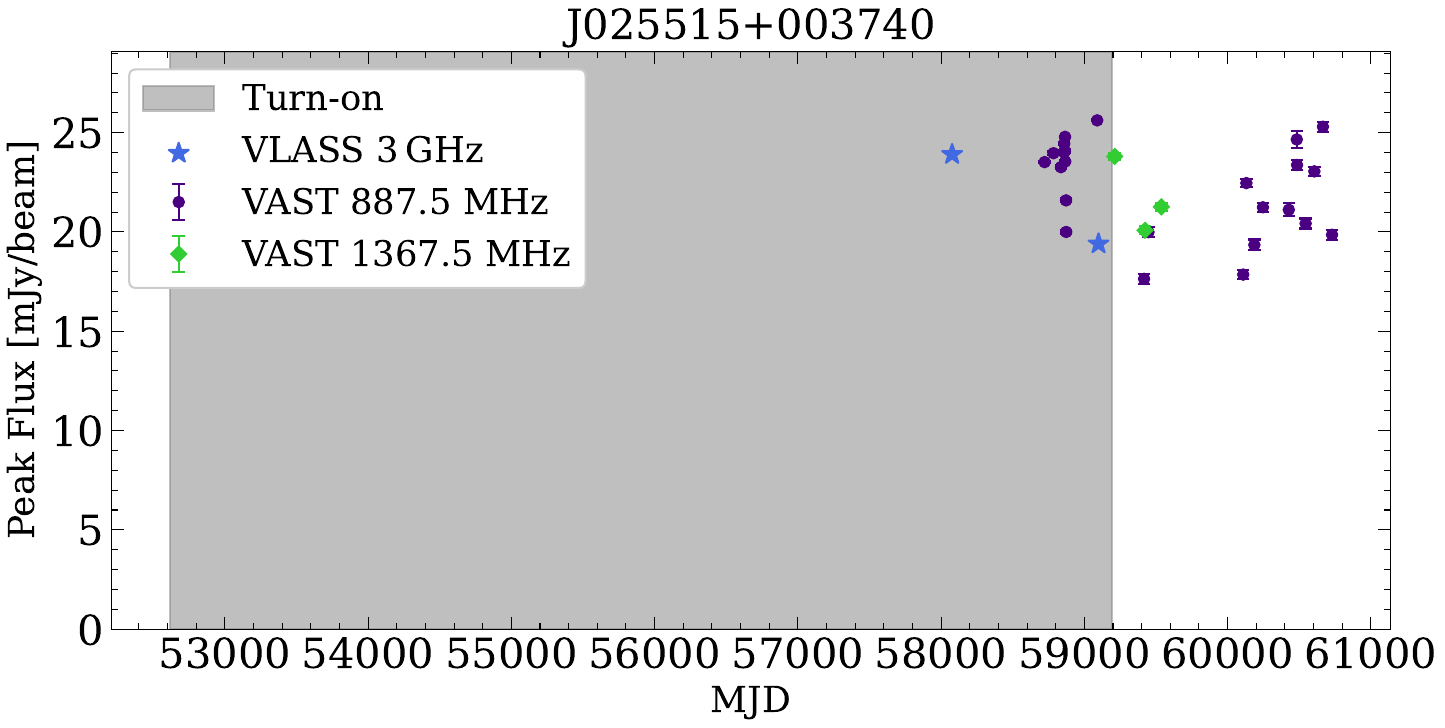}{0.49\textwidth}{}}
\gridline{
    \fig{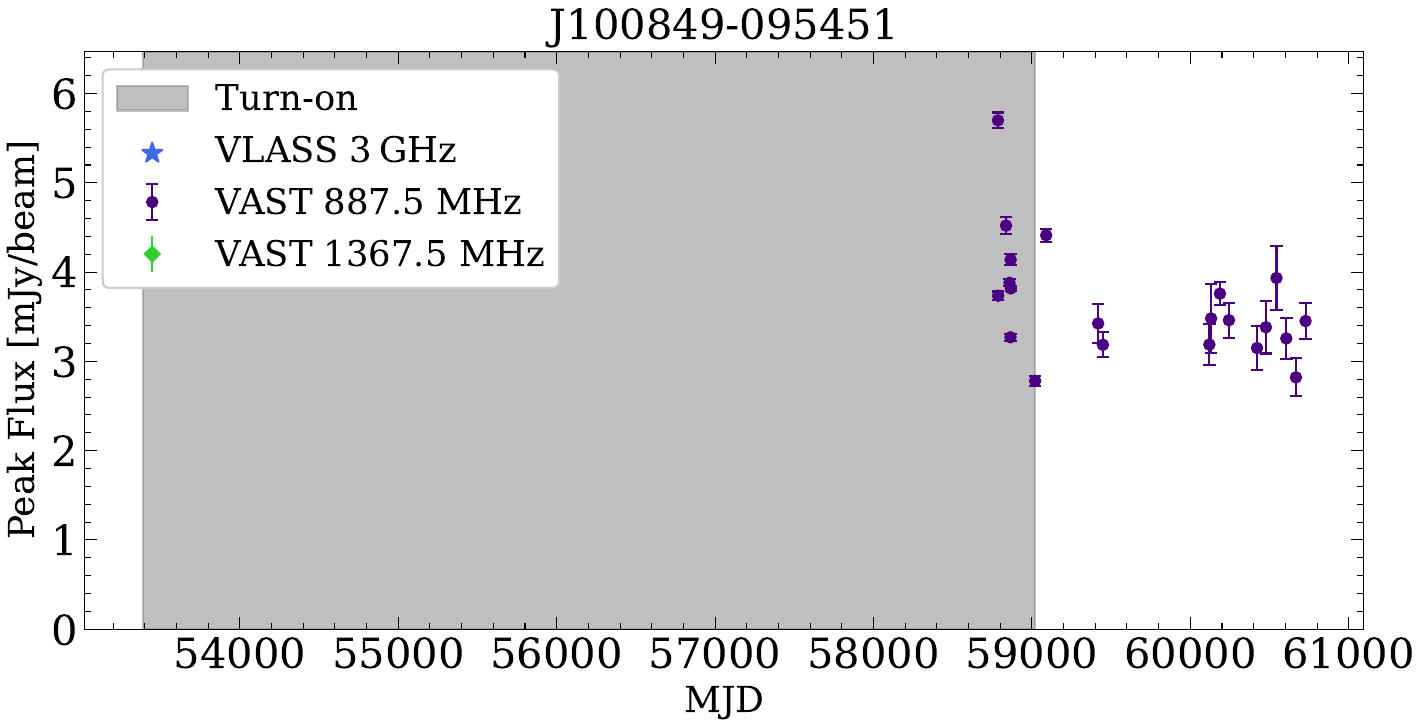}{0.49\textwidth}{}
    \fig{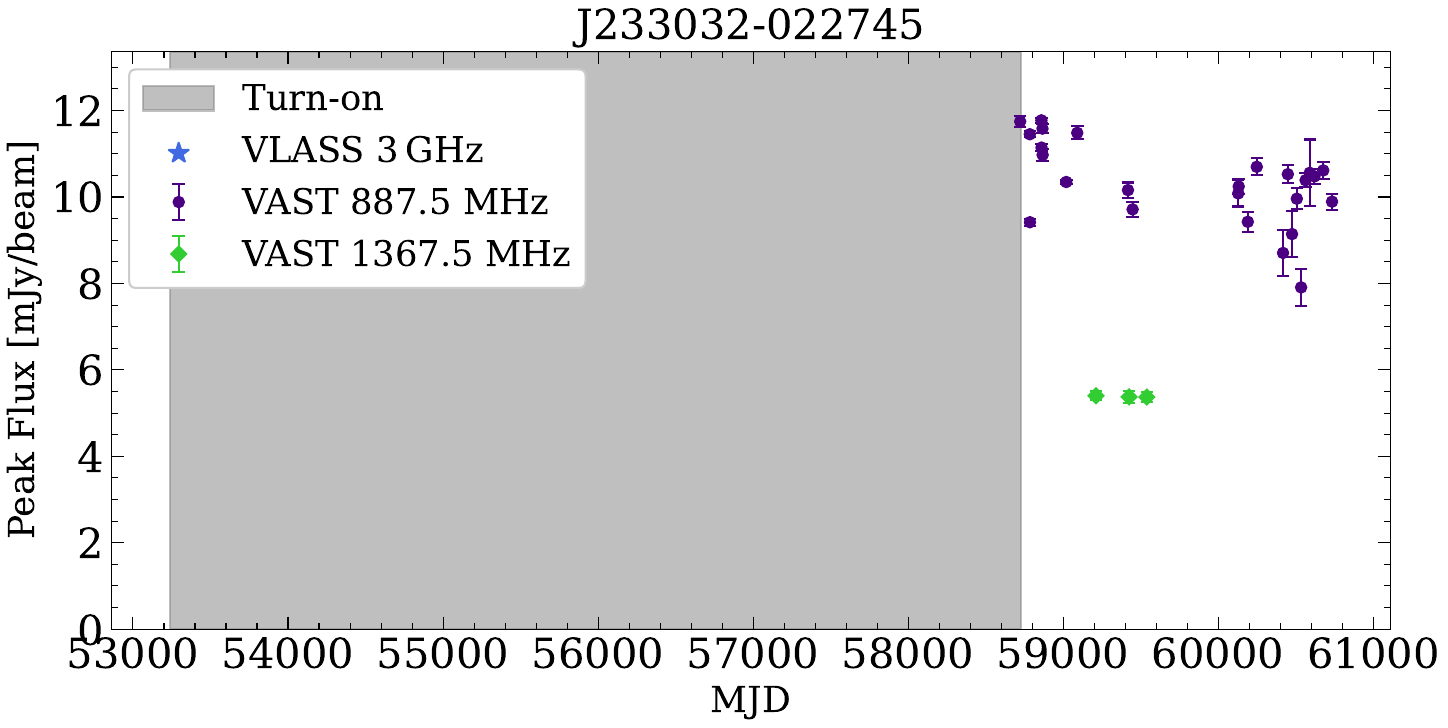}{0.49\textwidth}{}
}
\caption{Radio Light Curves of six CLAGN that display typical radio behavior of the CLAGN sample. We plot the peak flux in the VAST 887.5 MHz, 1367.5 MHz, and VLASS 3 GHz fluxes with the state change overlaid in gray. J1008-0954 is the only other CLAGN in our sample besides NGC 1566 that has a 887.5 MHz NEV value $>0.02$}
\label{fig:VAST_Radio_LCs}
\end{figure*}

We find that the CLAGN in our population do not display long term radio variability in VAST frequencies. In Figure~\ref{fig:VAST_Radio_LCs}, we display six VAST light curves which are representative of  the CLAGN population.

\subsection{Comparison to AGN that transitioned from RQ to RL in VLASS} \label{RadioSelectedSample}
We compared the radio-properties of the radio-detected CLAGN to the sample of 52 AGN transitioning from RQ to RL described in Section~\ref{Nyland Sample}. As described in Section~\ref{VLASSDetections} and ~\ref{VASTDetections}, we crossmatched the RQ to RL AGN objects to the VLASS and VAST catalogs. Of the 52 objects, there were 11 radio detections from a total of 19 sources that were within the VAST footprint. The 1 GHz detection rate for our 52 RQ to RL AGN is $\sim58$\%. 

We calculated the radio metrics described in Section~\ref{Methods} for 51 out of the 52 RQ to RL AGN sample to facilitate a comparison to the CLAGN sample. We calculated the source compactness and found that all 51 RQ to RL AGN were classified as compact radio sources. We calculated the radio loudness ratios for 48 compact RQ to RL AGN sample with $g$-band fluxes available. We confirmed that all 48 compact RQ to RL AGN were classified as radio-loud with our pipeline, consistent with the findings reported in the discovery papers. Of the 48, there were 16 objects with high radio loudness ratios ($R \geq 100$). None of these 16 are known or documented blazars. 

\begin{figure}[hbt!]
    \centering
    \includegraphics[width=0.4\textwidth]{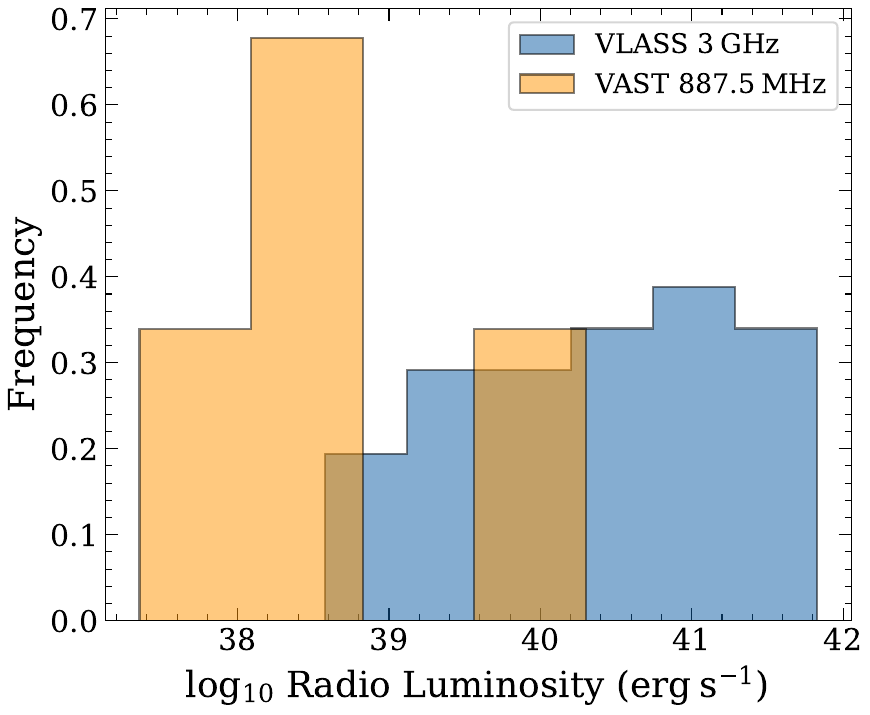}	
    \caption{Normalized distribution of the log of the radio luminosities for the 38 VLASS and 8 VAST-detected RQ to RL VLASS AGN.}
    \label{fig:RadioLuminosityHistRadioSelect}
\end{figure}

We calculated the VAST $L_{887.5\text{MHz}}$ radio luminosity for 8 RQ to RL AGN with VAST detections and spectroscopic redshifts available. We also calculated the VLASS $L_{3\text{GHz}}$ radio luminosity for the 38 RQ to RL AGN with VLASS detections and redshifts available. The distribution of the log of our radio luminosities among the 3 GHz VLASS detections and 887.5 MHz VAST detections is shown in Figure~\ref{fig:RadioLuminosityHistRadioSelect}. We found that the distribution of the log of the radio luminosities lies within the $10^{37-42}$ ergs s$^{-1}$ range. We observe that the RQ to RL AGN's $\log(L_{\text{3GHz}})$ and $\log(L_{887.5\text{MHz}})$ distributions have different peaks. The radio luminosity in the $\nu = 887.5$ MHz VAST frequency is skewed towards lower luminosities (with a median of $10^{38.6}$ ergs s$^{-1}$) compared to the radio luminosity values in the $\nu = 3$ GHz VLASS frequency (with a median of $10^{40.3}$ ergs s$^{-1}$). A two-sample Anderson-Darling test comparing the $\log(L_{\text{3GHz}})$ and $\log(L_{887.5\text{MHz}})$ distributions gives $A^2=8.76$ with $p < 10^{-3}$. We therefore reject the null hypothesis that the two distributions are drawn from the same parent population at \textgreater 99.98\% confidence. 

\begin{figure}[hbt!]
    \centering
    \includegraphics[width=0.4\textwidth]{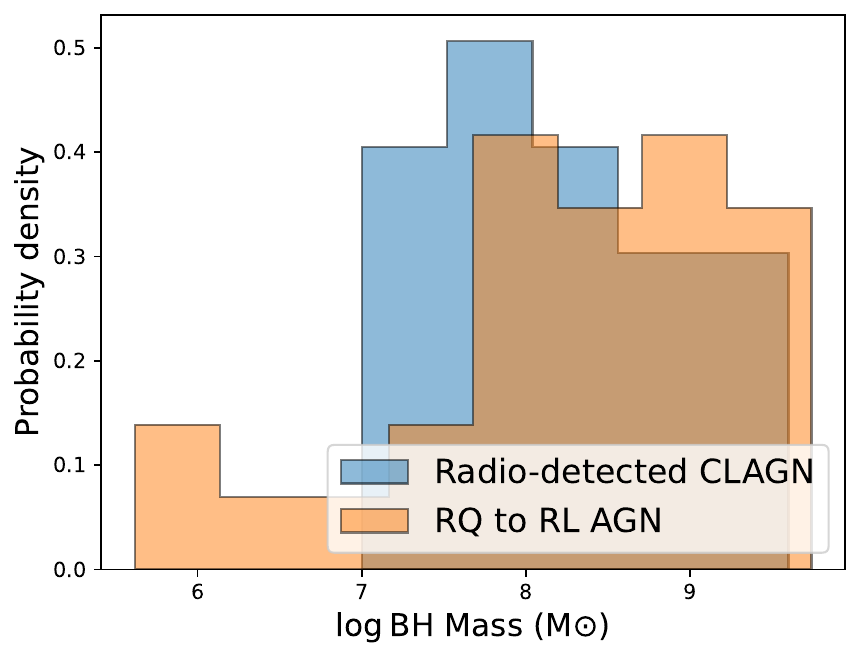}	
    \caption{Normalized distribution of the radio-detected CLAGN Sample and RQ to RL AGN Sample's black hole masses.}
    \label{fig:SMBH_RQRLAGN}
\end{figure}

We find that the 3 GHz radio luminosity distribution of the CLAGN sample (which had a median of $10^{39.7}$ ergs s$^{-1}$) is similar to that of the RQ to RL AGN Sample (which had a median of $10^{40.3}$ ergs s$^{-1}$). A two-sample Anderson-Darling test comparing the RQ to RL AGN $\log(L_{\text{3GHz}})$ distribution to the CLAGN $\log(L_{\text{3GHz}})$ distribution gives $A^2=2.60$ with $p=2.8$x$10^{-2}$. We therefore reject the null hypothesis that the two distributions are drawn from the same parent distribution at the 95\% confidence level. We also find that the median 887.5 MHz radio luminosity value of our CLAGN sample (which had a median of $10^{38.3}$ ergs s$^{-1}$) is also very similar to that of the RQ to RL AGN sample (which had a median of $10^{38.6}$ ergs s$^{-1}$). A two-sample Anderson-Darling test comparing the RQ to RL AGN $\log(L_{887.5\text{MHz}})$ distribution to the CLAGN $\log(L_{887.5\text{MHz}})$ distribution gives $A^2=0.68$ with $p > 0.25$. We therefore cannot reject the null hypothesis since the two distributions are statistically indistinguishable at more than the 75\% confidence level. Thus we find no statistically significant evidence that the overall $\log(L_{887.5\text{MHz}})$ distributions differ between CLAGN and RQ to RL AGN. We note that as shown in Figure~\ref{fig:SMBH_RQRLAGN}, the SMBH masses of the radio-detected CLAGN sample and the RQ to RL AGN sample do not noticeably differ, and thus any differences in luminosity between these two samples would not be due to a difference in SMBH masses. 

\begin{figure}[hbt!]
    \centering
    \includegraphics[width=0.4\textwidth]{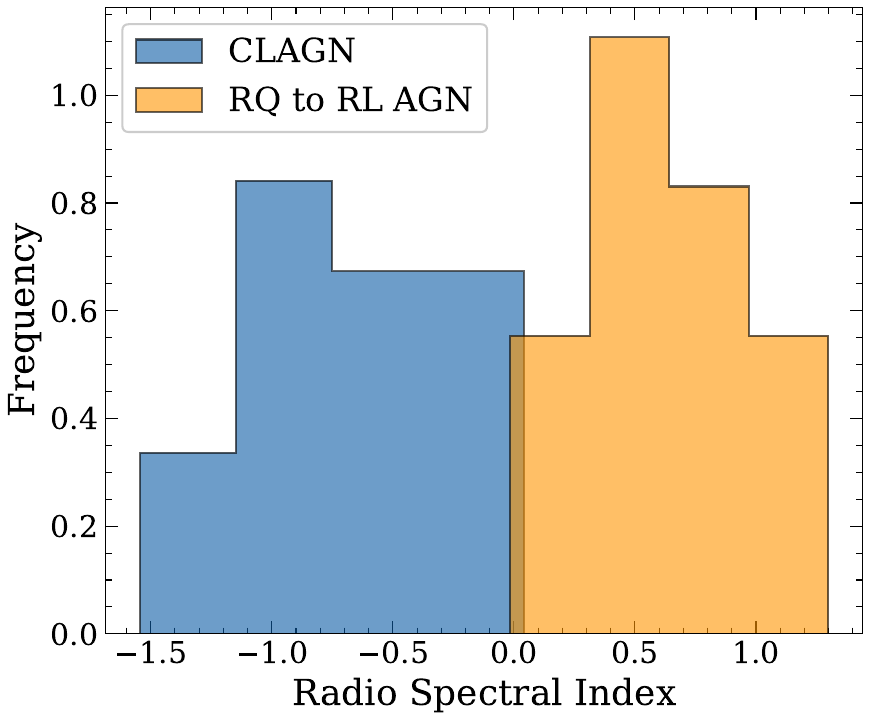}	
    \caption{Normalized distribution of the estimated radio SED power law spectral index $\alpha$ for 11 compact RQ to RL VLASS AGN and for 15 CLAGN with both compact VLASS 3 GHz and VAST 887.5 MHz detections.}
    \label{fig:Spectral_Index_Combined}
\end{figure}

The distribution of the power law spectral index for 11 compact RQ to RL AGN and 15 CLAGN with VLASS and VAST data is shown in Figure~\ref{fig:Spectral_Index_Combined}. We find that the RQ to RL AGN sample's spectral index distribution is largely positive with a median of 0.59. On the other hand, the CLAGN spectral index distribution has mostly negative values, with a median of -0.71. Thus, we find that the spectral indices of the RQ to RL AGN sample are distinctly positive while the indices of the CLAGN sample are distinctly negative. A two-sample Anderson-Darling test comparing the spectral index distribution of the CLAGN to the RQ to RL AGN sample gives $A^2=12.32$ with $p < 10^{-3}$. We therefore reject the null hypothesis that the two distributions are drawn from the same parent distribution at $>$ 99.9\% confidence. We note that these spectral index values are dependent on the redshift of the objects, and these differences could be an artifact of the higher-redshift RQ to RL AGN objects due to the difference in rest-frame frequencies for the two samples. 

\begin{figure*}
  \centering
  \gridline{
    \fig{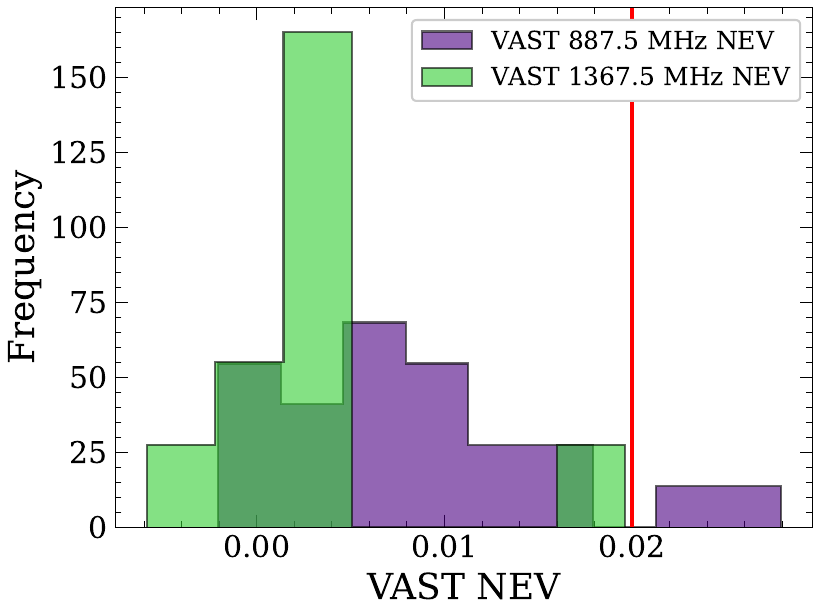}{0.42\textwidth}{Optical CLAGN Sample}
    \fig{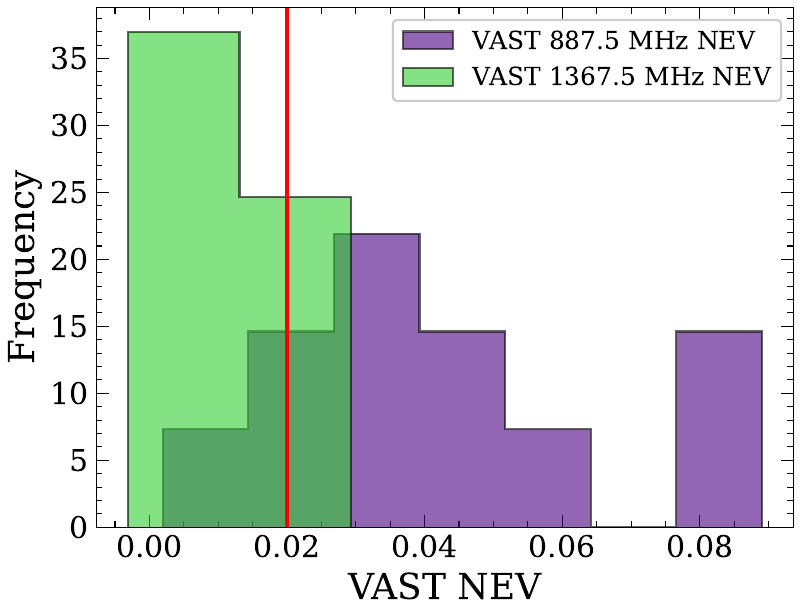}{0.41\textwidth}{RQ to RL VLASS AGN Sample}
  }
  \caption{Normalized distribution of radio NEV values of the spectroscopically and optically-selected CLAGN sample (left) and the RQ to RL VLASS AGN sample (right) calculated using VAST 887.5 MHz and 1.3675 GHz flux data. The 0.02 red line marks the boundary between variant and invariant CLAGN.}
  \label{fig:RadioNEVComparison}
\end{figure*}

We calculated the 887.5 MHz NEV values for 11 RQ to RL VLASS AGN detected in 887.5 MHz VAST, and the 1.3675 GHz NEV values for 5 RQ to RL VLASS AGN detected in 1.3675 GHz VAST. We also calculated the VLASS percentage flux change of 49 radio-selected RQ to RL VLASS AGN detections across two epochs. Figure~\ref{fig:RadioNEVComparison} displays the distribution of the NEV values calculated in the VAST 887.5 MHz and VAST 1.3675 GHz frequencies for the RQ to RL VLASS AGN and CLAGN samples. 

We found that 10 of our 11 VAST-detected RQ to RL VLASS AGN were classified as variable at the 887.5 MHz frequencies, corresponding to a variability fraction of $0.909^{+0.075}_{-0.286}$ (95\% Wilson score interval). We also found that 1 of our 5 VAST-detected CLAGN were classified as variable in the 1.3675 GHz frequency, corresponding to a variability fraction of $0.200^{+0.424}_{-0.164}$ (95\% Wilson score interval). Compared to the CLAGN population which was largely unvarying in VAST wavelengths, the vast majority of the RQ to RL VLASS AGN exhibit 1 GHz radio flux variability within error limits. Furthermore, we calculated the percentage change in VLASS flux across the two epochs for 49 VLASS-detected CLAGN. We found that 31 of our 49 sources were classified as variable across the two VLASS epochs, corresponding to a variability fraction of $0.633^{+0.121}_{-0.140}$ (95\% Wilson score interval).

We classified the radio source types of the RQ to RL AGN using the source type classification scheme described in Section~\ref{SourceClassificationDesc}. The results are shown in Table~\ref{tab:Radio_Metrics_CLAGN_RQRLAGN}. We found that 31 ($\sim61$\%) of the 51 RQ to RL AGN were classified as radio-loud, variable, compact sources (RL-V), comprising the largest portion of sources in our sample. We observe RL-V sources at a rate $\sim2$x higher in the RQ to RL AGN sample compared to the CLAGN sample. The remaining 17 ($\sim33$\%) of our 51-size RQ to RL VLASS AGN were classified as radio-loud, non-variable compact sources (RL-NV). Three RQ to RL AGN did not have radio loudness ratios available.  

We created 38 optical light curves for all ZTF-detected RQ to RL VLASS AGN and calculated their ZTF NEV values. These values are reported in Table~\ref{tab:RQRLAGNSample}. We used the same ZTF variability classification cutoff as in Section~\ref{ZTFLightCurves}. We found that 36 of our 38 CLAGN are variable in ZTF bands, corresponding to a variability fraction of $0.947^{+0.038}_{-0.120}$ (95\% Wilson score interval). We observed that the ZTF NEV values of the radio-detected CLAGN show similar levels of optical variability to the RQ to RL VLASS AGN. 

We recognize that it is difficult to rigorously compare the RQ to RL AGN and CLAGN samples due to small sample sizes and the differences in selection techniques between RQ to RL AGN (radio-selected) and CLAGN (optically/spectroscopically-selected).

\subsection{Control Sample Comparison} \label{ControlComparison}
In this section, we present a comparison between the 3000-size broad-line AGN control sample described in Section~\ref{ControlSampleDescription}, the full 474-size CLAGN sample, and the 56-size \cite{Guo2024Changing-lookData} subset of the CLAGN sample (whose parent sample was used to produce the broad-line AGN control sample). As described in Section~\ref{VLASSDetections} and ~\ref{VASTDetections}, we crossmatched the control sample AGN objects to the VLASS and VAST catalogs. The properties of the control sample are provided in Table~\ref{tab:ControlSample}.

We found that for our broad-line AGN control sample, there were 52 VAST detections from a total of 458 sources that were within the VAST footprint (detection rate of $\sim11$\%). We also found that there were 167 VLASS detections from a total of 3000 sources within the VLASS footprint (detection rate of $\sim6$\%). We found that both the VLASS and VAST-detection rates were slightly higher for the CLAGN sample compared to the control AGN sample. Additionally, we found that although none of the \cite{Guo2024Changing-lookData} CLAGN were detected in VAST, there were 2 out of 56 ($\sim4$\%) VLASS detection rate) \citet{Guo2024Changing-lookData} CLAGN detected in VLASS. These were CLAGN J153714+454348 and J164332+304836. The VLASS and VAST radio data for the control sample is displayed in Table~\ref{tab:ControlDetections}. 

As before, we calculated the radio metrics described in Section~\ref{Methods} for the 177 radio-detected broad-line AGN to facilitate a comparison to both the CLAGN and \citet{Guo2024Changing-lookData} samples. We calculated the source compactness and found that 144 broad-line AGN were classified as compact sources out of a total of 172 broad-line AGN with source compactness metrics available. The two \citet{Guo2024Changing-lookData} objects were both classified as compact sources. 

\begin{figure}[hbt!]
    \centering
    \includegraphics[width=0.4\textwidth]{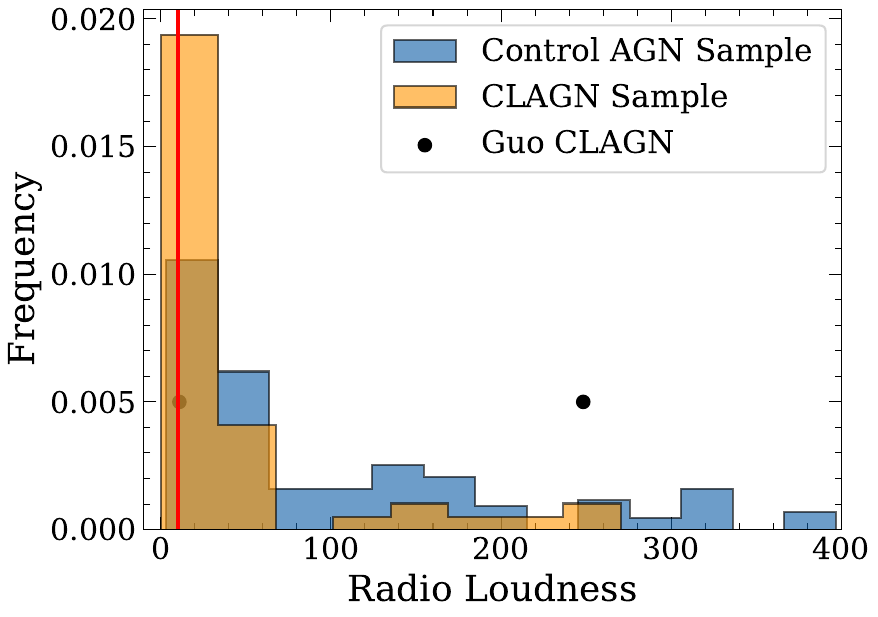}	
    \caption{Normalized distribution of the radio loudness ratio R for the CLAGN, the control AGN objects, and the two \protect{\citet{Guo2024Changing-lookData}} CLAGN zoomed into radio loudness ratios $R < 400$. The R = 10 cutoff value for radio loudness classification is overlaid.}
    \label{fig:RadioLoudnessControl}
\end{figure}

We calculated the radio loudness ratio for 144 broad-line control AGN objects that had g-band fluxes available, and we report these values in Table~\ref{tab:Radio_Metrics_Control}. Figure~\ref{fig:RadioLoudnessControl}, displays the distribution of the radio loudness ratios for the full CLAGN sample, the broad-line control AGN objects, and two \citet{Guo2024Changing-lookData} CLAGN objects. 

We found that the overall shapes of the radio loudness distributions of broad-line control AGN and CLAGN are similar. Additionally, we found that both of the \citet{Guo2024Changing-lookData} CLAGN objects were classified as radio-loud. We found that 129 out of 144 broad-line control AGN were classified as radio-loud, corresponding to a fraction of $0.896^{+0.040}_{-0.061}$ (95\% Wilson score interval). Thus, we found that at the population level, the radio loudness rate for the broad-line control AGN sample is roughly $1.7$ times higher than that of the CLAGN sample.

\begin{figure}[hbt!]
    \centering
    \includegraphics[width=0.4\textwidth]{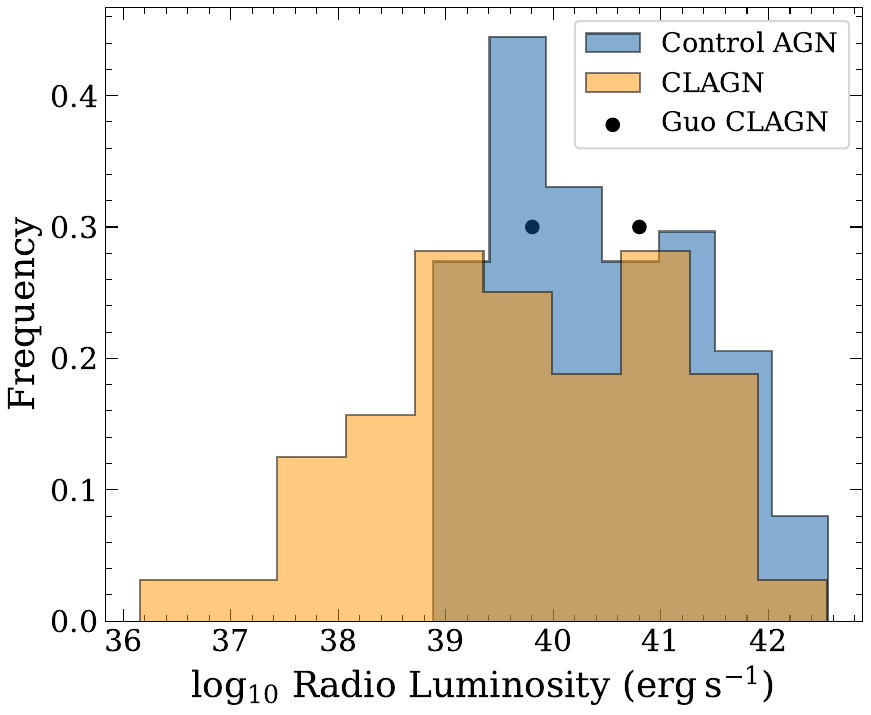}	
    \caption{Normalized distribution of the VLASS 3 GHz-calculated radio luminosities $L_{3GHz}$ for the broad-line control AGN, the full CLAGN sample, and the two \protect{\citet{Guo2024Changing-lookData}} CLAGN.}
    \label{fig:RadioLuminosityControl}
\end{figure}

We calculated the VLASS 3 GHz radio luminosity $L_{3GHz}$ for 167 control AGN sample objects which had redshifts and VLASS 3 GHz flux detections available. These values are reported in Table~\ref{tab:Radio_Metrics_Control}. In Figure~\ref{fig:RadioLuminosityControl}, we display the distribution of $L_{3GHz}$ for the broad-line control AGN along with the full CLAGN sample, and the two \citet{Guo2024Changing-lookData} CLAGN.

We found that the broad-line control AGN sample has $L_{3GHz}$ luminosities in the range of $10^{39-42}$ ergs s$^{-1}$, whereas the full CLAGN sample has a wider spread of luminosity values lying between $10^{36-42}$ ergs s$^{-1}$. The two \citet{Guo2024Changing-lookData} objects have $L_{3GHz}$ luminosities that lie within the range of the control sample luminosities. The distribution of the broad-line control AGN sample luminosities is slightly skewed towards higher luminosity values (median of $10^{40.2}$ ergs s$^{-1}$) compared to the CLAGN sample (median of $10^{39.7}$ ergs s$^{-1}$). A two-sample Anderson-Darling test comparing the distribution of log $L_{3GHz}$ luminosities of the CLAGN sample to the control AGN sample gives $A^2=10.41$ with $p < 10^{-3}$. We therefore reject the null hypothesis that the two distributions are drawn from the same parent distribution at $>$ 99.9\% confidence. We note that any differences in the luminosities are not associated with a difference in the distribution of SMBH masses between the CLAGN and control sample, as described in Section~\ref{ControlSampleDescription}.

\begin{figure}[hbt!]
    \centering
    \includegraphics[width=0.4\textwidth]{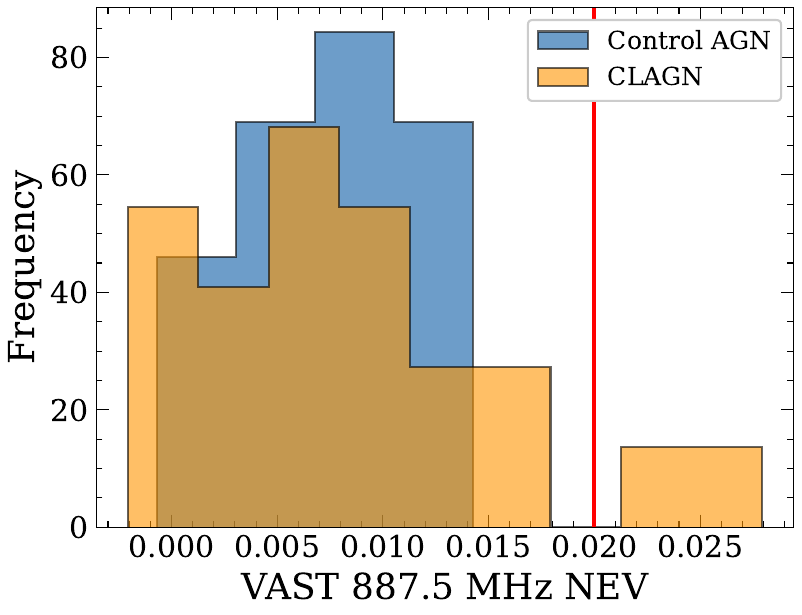}	
    \caption{Normalized distribution of the VAST 887.5 MHz NEV values for the broad-line control AGN and the CLAGN sample with a 0.20 radio variability cutoff.}
    \label{fig:VAST_v_variance}
\end{figure}

We calculated the 887.5 MHZ NEV values for 35 broad-line control AGN detected in 887.5 MHz VAST, and the 1.3675 GHz NEV values for 11 broad-line control AGN detected in 1.3675 GHz VAST. These values can be found in Table~\ref{tab:Radio_Metrics_Control}.  Figure~\ref{fig:VAST_v_variance} displays the distribution of the VAST 887.5 MHz NEV values for the broad-line control AGN and the CLAGN sample. The \citet{Guo2024Changing-lookData} were not detected with VAST. 

We found that none of the 35 VAST-detected broad-line control AGN were classified as variable at the 887.5 MHz frequencies. Similarly, we found that none of the 11 VAST-detected broad-line control AGN were classified as variable in the 1.3675 GHz frequency. We find statistically significant evidence indicating that the CLAGN population has a higher fraction of variable objects in VAST 887.5 MHz frequencies compared to the broad-line control AGN sample. Furthermore, we also calculated the VLASS percentage flux change of 142 broad-line control AGN across two epochs. We found that 97 of the 142 sources were classified as variable across the two VLASS epochs, corresponding to a fraction of $0.683^{+0.071}_{-0.080}$ (95\% Wilson score interval). We conclude that the VLASS variability rates of CLAGN and broad-line control AGN are within the same error range of each other, and thus do not significantly differ. Finally, we found that for the 56-size \citet{Guo2024Changing-lookData} CLAGN sample, there was only one CLAGN candidate that had VLASS detections in both epochs, and it was classified as variable.



\section{Discussion} \label{Discussion}
We have presented a population study of CLAGN initially identified by spectroscopic and optical methods, and have studied their time-resolved radio properties with VLASS and VAST. The majority of the radio-detected CLAGN were compact radio-loud point sources. We classified the population of CLAGN based on their radio loudness and radio variability (RL-V, RL-NV, RQ-V, RQ-NV), and found that the proportion of CLAGN that belong to each class is roughly equal, and that the marginally largest classification were radio-loud, variable, compact sources (RL-V). 

The literature includes both studies of individual CLAGN which display radio activity following a changing look event \citep{Yang2021AMrk590,Saha2023MultiwavelengthSeyfert,Meyer2025}, and CLAGN which do not display radio variations following a changing look event \citep{Yang2021A2617}. We constructed 22 monthly cadence VAST radio light curves in the 887.5 MHz frequency for compact CLAGN. For the 20 newly studied objects, we do not see statistically significant evidence that radio fading has occurred after the state change event. Thus, CLAGN such as 1ES 1927+654, Mrk 590, and NGC 1566 display atypical radio behavior compared to the broader population of CLAGN. 

\citet{Arghajit1566} observed NGC 1566 in the 230 GHz radio wavelengths from 2014-2023, finding that the 230 GHz flux increased during the type 1 to 2 state change, and decreased during subsequent type 2 to 1 state change. We observe NGC 1566 in the 887.5 MHz frequency, and find that it fades from $\sim$ 45 to 30 mJy between 2019 and 2023. NGC 1566 displays a $\sim$33\% decrease in its 887.5 MHz flux after the turn-on state change event. VAST's monthly cadence observations extend the previous knowledge of NGC 1566's behavior in the radio wavelengths by providing greater time resolution. Highly studied object Mrk 590, displays a $\sim40$\% radio flux decrease in the 20 years following a turn-off state change event. Thus, NGC 1566 and Mrk 590 are similar in that they both display a long term radio flux decrease following a state change event. The type of state change differs (turn-on for NGC 1566 and turn-off for Mrk 590), but the radio behavior after the state change is similar.  

Our sample suggests that changing-look events may launch a temporary jet for some CLAGN (Mrk 590, NGC 1566), but overall the majority of these events do not appear to initiate an episode of long term radio activity. This finding implies that the physical mechanism behind such CLAGN activity is temporary disk instabilities that cause an increase in short term accretion. Our sample provides support for the variable accretion scenario for CLAGN as a population. This is the preferred physical model compared to other explanations such as a global change in gas supply. Other studies of CLAGN are consistent with this interpretation, and have observed radio variability that is explained by disk instability models. \citet{Saha2023MultiwavelengthSeyfert} determines that J0428-00's radio variability is not likely to be caused by long-term or systematic accretion changes. Instead, their observations indicate that a temporary change to J0428-00's existing accretion flow, potentially caused by a disk instability, explains the observed radio variability. CLAGN 1ES 1927+654's radio behavior is evidence that a young short-lived jet was launched after the changing-look event \citep{Meyer2025}, similar to our objects Mrk 590 and NGC 1566. Ultimately, since we do not find evidence for sustained jets that have been launched after the changing-look event, we conclude that temporary disk instabilities drive CLAGN activity.

To interpret the radio variability that we observe at the 887.5 MHz frequency, we refer to the methods of \citet{Walker1998} to determine if the observed variability is consistent with refractive interstellar scintillation. At the positions of our sources, the critical frequency is in the range of $\nu_0\sim8-50$ GHz, meaning our compact VAST-detected CLAGN are in the strong scattering regime. The expected flux modulation for refractive scintillation at 887.5 MHz is $m=(\nu/\nu_0)^{17/30} \sim 0.1-0.3$. The modulation timescale is $t_r\sim2(\nu_0/\nu)^{11/5}$ hours $\sim$ 10 days to a couple of years. The observed modulation index for our 22 CLAGN is at most 0.18, and the observed timescales are about months to years. Comparing the $m$ and $t_r$ values calculated from \citet{Walker1998} to our observed values, this implies that interstellar scintillation cannot be ruled out as the origin of 887.5 MHz variability for our 22 sources. In contrast, \citet{Nyland2020QuasarsFIRST} ruled out scintillation as the origin of variability for their RQ to RL AGN objects.

In order to determine how well we can rule out radio flares associated with changing-look events for the objects in our sample, we consider the radio variability properties of 1ES 1927+654, Mrk 590, and NGC 1566 and whether similar events would be detectable for the other sources given their redshifts and the VAST single-epoch sensitivity of 1-2 mJy. 1ES 1927+654, at redshift $z=0.017$ and with BH mass of $\sim10^6 M_{\odot}$, exhibited an increase in 5 GHz radio luminosity from $\log L_{5\text{GHz}} = 37.7$ erg s$^{-1}$ to $\log L_{5\text{GHz}} = 39.0$ erg s$^{-1}$ over 1 year, with this rebrightening taking place 5 years after an optical outburst associated with the temporary appearance of broad lines \citep{Meyer2025}. Mrk 590, at redshift $z=0.026$ and with BH mass of $\sim10^8M_{\odot}$, varied from a 5 GHz radio luminosity from $\log L_{5\text{GHz}} = 38.25$ erg s$^{-1}$ in 1995 to $\log L_{5\text{GHz}}=37.65$ erg s$^{-1}$ in 2014 while its broad lines disappeared between 2003/2006 and 2013, with prior fading observed since 1989 \citep{Denney2014TheRole}. NGC 1566, at redshift $z=0.005$ and with BH mass of $\sim10^7M_{\odot}$, varied from a 887.5 MHz radio luminosity from $\log L_{887\text{MHz}}=$ 36.83 to 36.70 erg s$^{-1}$ while transitioning from a type 1 to 2 state. With our VAST light curves, we observe the tail-end of this trend of radio fading for Mrk 590 and NGC 1566 in 887.5 MHz. We conclude that for the 14 CLAGN with z $>$ 0.05, we would not actually detect Mrk 590-like luminosities, since they are too faint to see with the VAST 2 mJy limit. The $\log_{10}$ mass of Mrk 590 is 8.29 and the $\log_{10}$ masses of the 14 other z $>$ 0.05 VAST-detected CLAGN lie in the range of 7.40 to 9.32.  For the 6 other z $<$ 0.05 VAST-detected CLAGN, we can rule out a Mrk 590-like flare but not an NGC 1566-like flare. Additionally, 11 of our CLAGN objects are much more radio luminous and radio loud than Mrk 590 and NGC 1566.

We have compared the radio properties of this CLAGN sample to both those of a redshift-matched AGN control sample, and those of AGN identified to transition from RQ to RL in previous work. We have found that radio properties (radio detection rates, radio loudness, radio luminosity, radio variability, and spectral index) of CLAGN distinctly differ from those of the control AGN sample and the RQ to RL AGN. \citet{Yang2021A2617} compared the FIRST survey radio detection rates between two larger AGN samples and their own CLAGN sample, assuming that CLAGN and general AGN have similar contamination fractions. They found that their CLAGN sample had lower radio detection rates compared to the two larger AGN samples. However, we found that both the VLASS and VAST detection rates were higher for our CLAGN sample compared to the redshift-matched AGN sample. 

Based on this study, there does not seem to be an obvious link between the RQ to RL AGN sample and the CLAGN sample. Upon performing Magellan spectroscopic follow up of two RQ to RL AGN from this sample (J074248+270412 and J102951+043658), we did not find broad lines to appear compared to these objects' archival spectra. This suggests that the RQ to RL AGN sample's long term radio transition was not associated with an optical state-change. The VLA follow up shows that the SEDs of the RQ to RL AGN sample are different than the CLAGN sample. Thus, we find no evidence of a clear relation between two samples. This prompts the idea that changing look events may not produce long term changes in jet activity, which was the previous explanation of the RQ to RL AGN sources. 

As mentioned previously, we have obtained radio SED shapes and energetics and have compared a sample of CLAGN SEDs to SEDs of AGN that have been known to transition from radio-quiet to radio-loud using the VLA. We find that all four of our CLAGN SEDs are consistent with a power law as opposed to having concave shapes that are typical of young recently formed jets. On the other hand, the majority of the RQ to RL AGN have concave SED shapes. 

We acknowledge the limitations to the archival data used for our population study. Because we identify CLAGN events through archival SDSS and new spectra, we do not know the exact time frame in which state changes occur for a portion of these CLAGN. Additionally, our relatively small sample size and survey overlap poses challenges to our population study. This demonstrates the importance of multi-wavelength follow-up contemporaneous with the start of the original flare.

\section{Conclusions} \label{ResultsStatistics}
Our population study of CLAGN was motivated by individual CLAGN that were well studied in multiple wavelengths. The interesting behavior seen in these CLAGN prompted us to pose three main questions in Section~\ref{Introduction} about the properties of CLAGN at a population level. We found that the vast majority of our CLAGN sample were radio point sources and more than half of our sample were radio-loud, with a fraction of roughly $21\%$ of the sample having high radio loudness ratios ($R>100$). This implies that CLAGN radio emission is coming from the base of the compact jet close to the SMBH (rather than the X-ray emitting corona) for these objects. Through our multi-wavelength population study of 474 optically and spectroscopically-selected CLAGN, and comparing this sample to a RQ to RL AGN and to a control AGN sample, we present our answers to the three questions presented in the Introduction.

We do not find statistically significant evidence that the 20 newly studied compact-VAST detected CLAGN show the appearance of radio jets after the changing-look event. Besides confirming that the previously known CLAGN NGC 1566 is fading after the state change, we have no new reports of fading events in the CLAGN population. We cannot confidently say that any of the 20 newly studied CLAGN (those besides well-studied Mrk 590 and NGC 1566) have faded in VAST wavelengths after the changing-look event. Thus, we conclude that the radio behavior of 1ES 1927+654, Mrk 590, and NGC 1566 is not typical of the CLAGN population. The radio-detected CLAGN population generally displays low radio variability consistent with interstellar scintillation, with no long term, high amplitude variability in response to the changing-look event. The caveat of this conclusion is that some changing-look events happened longer ago, before the timeframe of the VAST survey, and that our observations are limited to 2 mJy sensitivity. This motivates more comprehensive radio-followup of CLAGN shortly after the broad-line changes are detected.
    
On the population level, CLAGN have slightly higher VAST and VLASS radio-detection rates than a broad-line control AGN sample. We find statistically significant evidence that CLAGN are notably less radio loud compared to the control AGN. The fraction of radio loud objects in the CLAGN sample was roughly 1.7 times less than that of the control AGN sample. We find statistically significant evidence that the $\log(L_{\text{3GHz}})$ distributions of the control AGN sample and CLAGN sample were not drawn from the same parent distribution. We also find statistically significant evidence that CLAGN display a higher fraction of variability in the 887.5 MHz frequency compared to the control AGN. We do not find significant evidence that the VLASS variability rates of CLAGN and broad-line control AGN differ. This motivates further time domain analysis to probe the properties of CLAGN compared to their regular AGN counterparts.

Finally, we compared the radio properties of CLAGN to the radio properties of a sample of AGN that were observed to transition from radio quiet to radio loud on decadal timescales, see Section~\ref{Nyland Sample}. While there are biases due to the different selection methods for the two samples, we provide the results of our population study. We find statistically significant evidence that the $\log(L_{\text{3GHz}})$ distributions of CLAGN and RQ to RL AGN are not drawn from the same parent distributions. We do not find statistically significant evidence that the $\log(L_{887.5\text{MHz}})$ distributions differ between CLAGN and RQ to RL AGN. We find that the power law spectral index for RQ to RL AGN are much more positive compared to the negative CLAGN power law spectral indices, and find statistically significant evidence that the spectral index distributions of CLAGN and RQ to RL AGN were not drawn from the same parent distribution. However, we note that there is a difference in redshift values of the two samples, where the RQ to RL AGN have higher redshifts than the CLAGN. We find statistical evidence that RQ to RL AGN have a significantly higher fraction of variability in the VAST 887.5 MHz frequency compared to CLAGN. We found that the optical ZTF variability fraction of the RQ to RL AGN sample is comparable to that of the CLAGN sample. With optical spectroscopic follow-up and comparison to archival SDSS spectra, we found that a subset of the RQ to RL did not show the appearance of new broad lines. VLA SEDs of 4 CLAGN found either power law or only slightly concave SEDs, in contrast to the steeply concave SEDs of some RQ to RL AGN. Overall, we do not find evidence of a clear link between CLAGN and the RQ to RL AGN, implying that changing-look events may not be the drivers of the radio-loudness transitions identified for the AGN reported in \citet{Nyland2020QuasarsFIRST, Zhang2022TransientSurveys, Woowska2021Caltech-NRAOState}.

The difference we observe in CLAGN behavior at a population level could hint at a heterogeneous population of CLAGN. We detect radio fading in two known CLAGN objects (Mrk 590 and NGC 1566), and rule out Mrk 590-like flares for 6 objects. Our sample suggests that changing-look events may launch a temporary jet for some CLAGN (Mrk 590, NGC 1566), but overall these events do not appear to initiate an episode of long term radio activity. This finding implies that the physical mechanism behind such CLAGN activity is temporary disk instabilities that cause an increase in short term accretion. Future time domain surveys such as the Legacy Survey of Space and Time at Rubin Observatory \citep{Ivezic2019LSST:Products} and DSA-2000 \citep{Hallinan2019} will provide optical and radio light curves of many AGN that overlap with VAST, enabling the identification of radio variability following changing-look events in a larger sample.  

\section{Acknowledgments}
The authors would like to thank Benny Trakhtenbrot and Grisha Zeltyn for helpful discussions on the SDSS-V CLAGN parent sample. Basic research in radio astronomy at the U.S. Naval Research Laboratory is supported by 6.1 Base Funding. Based on observations obtained with the Samuel Oschin Telescope 48-inch and the 60-inch Telescope at the Palomar Observatory as part of the Zwicky Transient Facility project. ZTF is supported by the National Science Foundation under Grant No. AST-2034437 and a collaboration including Caltech, IPAC, the Weizmann Institute for Science, the Oskar Klein Center at Stockholm University, the University of Maryland, Deutsches Elektronen-Synchrotron and Humboldt University, the TANGO Consortium of Taiwan, the University of Wisconsin at Milwaukee, Trinity College Dublin, Lawrence Livermore National Laboratories, and IN2P3, France. Operations are conducted by COO, IPAC, and UW. The ZTF photometry service was funded under the Heising-Simons Foundation grant 12540303 (PI: Graham). 

This research has made use of the CIRADA cutout service at URL cutouts.cirada.ca, operated by the Canadian Initiative for Radio Astronomy Data Analysis (CIRADA). CIRADA is funded by a grant from the Canada Foundation for Innovation 2017 Innovation Fund (Project 35999), as well as by the Provinces of Ontario, British Columbia, Alberta, Manitoba and Quebec, in collaboration with the National Research Council of Canada, the US National Radio Astronomy Observatory and Australia’s Commonwealth Scientific and Industrial Research Organisation.

This scientific work uses data obtained from Inyarrimanha Ilgari Bundara, the CSIRO Murchison Radio-astronomy Observatory. We acknowledge the Wajarri Yamaji People as the Traditional Owners and native title holders of the Observatory site. CSIRO’s ASKAP radio telescope is part of the Australia Telescope National Facility (https://ror.org/05qajvd42). Operation of ASKAP is funded by the Australian Government with support from the National Collaborative Research Infrastructure Strategy. ASKAP uses the resources of the Pawsey Supercomputing Research Centre. Establishment of ASKAP, Inyarrimanha Ilgari Bundara, the CSIRO Murchison Radio-astronomy Observatory and the Pawsey Supercomputing Research Centre are initiatives of the Australian Government, with support from the Government of Western Australia and the Science and Industry Endowment Fund.

\software{
Astropy \citep{TheAstropyCollaboration2022}, 
}

\section{Appendix} \label{Appendix}

\setlength{\tabcolsep}{3pt}
\setlength{\extrarowheight}{5pt}
\startlongtable
\begin{deluxetable*}{ccccccccccc}
\tabletypesize{\scriptsize}
\tablecolumns{11}
\tablewidth{0pt}
\tablecaption{CLAGN Sample + ZTF Data \label{table:ztfcands}}
\tablehead{
  \colhead{Name} 
& \colhead{RA} 
& \colhead{Dec} 
& \colhead{z} 
& \colhead{Epoch 1} 
& \colhead{Epoch 2} 
& \colhead{Class} 
& \colhead{ZTF NEV} 
& \colhead{ZTF Variable?} 
& \colhead{Log Black Hole Mass} 
& \colhead{Reference} 
\\[-0.3cm]
  \colhead{} 
& \colhead{(hms)} 
& \colhead{(dms)} 
& \colhead{} 
& \colhead{(MJD)} 
& \colhead{(MJD)} 
& \colhead{} 
& \colhead{} 
& \colhead{(Y / N)} 
& \colhead{($M_\odot$)} 
& \colhead{}
}
\startdata
J002331+055355 & 00:23:31.3   & +05:53:54.8   & 0.637    & 55829  & 58662  & Turn-off      & \dots       & \dots                     & 8.5                          & G20       \\
J002354-025200 & 00:23:53.5   & -02:51:59.1   & 0.246    & 57362  & 58054  & Turn-off      & \dots       & \dots                     & 7.5                          & G20       \\
J005346+223222 & 00:53:46.1   & +22:32:21.9   & 0.149    & 56904  & 58348  & Turn-off      & \dots       & \dots                     & 8.4                          & G20       \\
J011919-093722 & 01:19:19.3   & -09:37:21.6   & 0.383    & 52163  & 58013  & Turn-off      & \dots       & \dots                     & 7.4                          & G20       \\
J012327+241844 & 01:23:26.8   & +24:18:44.1   & 0.914    & 55885  & 58662  & Turn-on       & \dots       & \dots                     & 9                            & G20       \\
J022015-072900 & 02:20:14.6   & -07:28:59.3   & 0.213    & 52162  & 57360  & Turn-off      & \dots       & \dots                     & 7.8                          & G20       \\
J024534-000745 & 02:45:33.6   & -00:07:45.2   & 0.655    & 52177  & 52946  & Turn-off      & \dots       & \dots                     & 8.2                          & G20       \\
J025003+010931 & 02:50:03.0   & +01:09:30.7   & 0.194    & 52177  & 52295  & Turn-off      & \dots       & \dots                     & 7.2                          & G20       \\
J025410+034912 & 02:54:10.1   & +03:49:12.5   & 0.774    & 55477  & 57399  & Turn-off      & \dots       & \dots                     & 8.1                          & G20       \\
J025506+002524 & 02:55:05.7   & +00:25:23.5   & 0.353    & 51816  & 51877  & Turn-off      & \dots       & \dots                     & 7.6                          & G20       \\
\enddata
\vspace{0.1cm}
\tablecomments{Identifying information of the CLAGN sample and ZTF optical properties and variability metrics. Column (1): source name. Column (2) and (3) right ascension and declination. Column (4): Redshift. Column (5) and (6): Epoch 1 and 2 are the date range for when the spectroscopic state change occured. Column (7): Class of the state change representing whether the broad lines appeared or disappeared. Column (8): ZTF-calculated NEV value. Column (9): Variability of the source in ZTF. Column (10): $log_{10}$ of the SMBH mass. Column (11): Reference. References are formatted as follows: G20 = Graham 2020, Z24 = Zeltyn 2024, Gu24 = Guo 2024, Mi19 = Miniutti 2019, Li18 = Lin 2018, Su18 = Sun 2018, La22 = Laha 2022, Ok21 = Oknyansky 2021, Go21 = Guolo 2021, Ma21 = Mandal 2021, LM15 = LaMassa 2015, Wa18 = Wang 2018, Ge16 = Gezari 2016, ML16 = MacLeod 2016, Fr19 = Frederick 2019, Zl18 = Zetzl 2018, Gr22 = Green 2022, H22 = Hon 2022. Table 1 is published in its entirety in the machine-readable format. A portion is shown here for guidance regarding its form and content.}  
\label{tab:CLAGNSample}
\end{deluxetable*}

\setlength{\tabcolsep}{3pt}
\setlength{\extrarowheight}{5pt}
\startlongtable
\begin{deluxetable*}{ccccccc}
\tabletypesize{\scriptsize}
\tablecolumns{7}
\tablewidth{0pt}
\tablecaption{Broad-line Control AGN Sample \label{table:fiber_bh}}
\tablehead{
  \colhead{RA} 
& \colhead{Dec} 
& \colhead{Mean Fiber RA} 
& \colhead{Mean Fiber Dec} 
& \colhead{Redshift} 
& \colhead{Separation} 
& \colhead{Log Black Hole Mass} 
\\[-0.3cm]
  \colhead{(hms)} 
& \colhead{(dms)} 
& \colhead{(degrees)} 
& \colhead{(degrees)} 
& \colhead{} 
& \colhead{(arcsec)} 
& \colhead{($M_\odot$)} 
}
\startdata
16:57:40 & 34:11:29 & 254.4147      & 34.1914        & 0.706    & 0.05741    & \dots                            \\
12:31:53 & 55:26:26 & 187.9704      & 55.4406        & 0.416    & 0.03333    & 8.62                         \\
11:37:39 & 56:22:52 & 174.4114      & 56.3809        & 0.604    & 0.47667    & 8.96                         \\
16:25:01 & 42:15:35 & 246.2555      & 42.2598        & 0.730    & 0.05748    & 8.23                         \\
12:10:43 & 1:04:54  & 182.6805      & 1.0818         & 1.389    & 0.04242    & 9.44                         \\
14:27:17 & -1:02:06 & 216.8190      & -1.0351        & 2.328    & 0.09387    & \dots                            \\
16:44:29 & 35:34:26 & 251.1209      & 35.5738        & 0.511    & 0.01800    & \dots                            \\
16:26:59 & 42:44:50 & 246.7476      & 42.7473        & 1.201    & 0.03289    & \dots                            \\
16:42:51 & 24:11:27 & 250.7128      & 24.1908        & 1.201    & 0.10990    & \dots                            \\
14:31:58 & 34:16:50 & 217.9915      & 34.2806        & 0.716    & 0.03532    & 8.89                         \\
12:37:12 & 56:44:15 & 189.2988      & 56.7374        & 1.156    & 0.13136    & \dots                            \\
12:09:03 & 58:24:38 & 182.2641      & 58.4106        & 1.155    & 0.12608    & \dots                            \\
11:33:46 & 50:38:58 & 173.4417      & 50.6496        & 1.718    & 0.03770    & \dots                            \\
\enddata
\vspace{0.1cm}
\tablecomments{Identifying information of the broad-line AGN control sample. Column (1) and (2): right ascension and declination of sources. Column (3) and (4) mean fiber right ascension and declination of sources. Column (5) redshifts. Column (6): source separation. Column (7): $log_{10}$ of the SMBH mass. Table 2 is published in its entirety in the machine-readable format. A portion is shown here for guidance regarding its form and content.}
\label{tab:ControlSample}
\end{deluxetable*}

\setlength{\tabcolsep}{3pt}
\setlength{\extrarowheight}{5pt}
\startlongtable
\begin{deluxetable*}{cccccccc}
\tabletypesize{\scriptsize}
\tablecolumns{8}
\tablewidth{0pt}
\tablecaption{RQ to RL AGN Sample  \label{table:RQRLAGN}}
\tablehead{
  \colhead{Name} 
& \colhead{RA} 
& \colhead{Dec} 
& \colhead{z} 
& \colhead{ZTF NEV} 
& \colhead{ZTF Variable?} 
& \colhead{Log Black Hole Mass} 
& \colhead{Reference} 
\\[-0.3cm]
  \colhead{} 
& \colhead{(hms)} 
& \colhead{(dms)} 
& \colhead{} 
& \colhead{} 
& \colhead{(Y / N)} 
& \colhead{($M_\odot$)} 
& \colhead{}
}
\startdata
J221650+005429 & 22:16:50.4  & +00:54:29.1  & 0.55   & 0.0172  & N                     & \dots                            & W21       \\
J221813-010344 & 22:18:12.9  & -01:03:44.2  & \dots      & \dots       & \dots                     & \dots                            & W21       \\
J223041-001644 & 22:30:41.4  & -00:16:44.3  & 0.84   & \dots       & \dots                     & \dots                            & W21       \\
J233002-002737 & 23:30:01.8  & -00:27:36.5  & 1.65   & 0.0466  & Y                     & 9.16                         & W21       \\
J010734+011223 & 01:07:33.5  & +01:12:22.8  & 0.12   & 0.1331  & Y                     & \dots                            & W21       \\
J013815+002914 & 01:38:15.1  & +00:29:14.1  & 0.94   & 0.0370  & N                     & 9.26                         & W21       \\
J015412-011150 & 01:54:11.7  & -01:11:49.7  & 0.05   & 0.4818  & Y                     & 8.9                          & W21       \\
J020827-005208 & 02:08:27.1  & -00:52:08.0  & 1.34   & 0.3504  & Y                     & 8.42                         & W21       \\
J030533+004610 & 03:05:33.1  & +00:46:09.9  & 0.42   & \dots       & \dots                     & \dots                            & W21       \\
\enddata
\vspace{0.1cm}
\tablecomments{Identifying information of the RQ to RL AGN sample and their ZTF optical properties and variability metrics. Column (1): source name. Column (2) and (3) right ascension and declination. Column (4): Redshift. Column (5): ZTF-calculated NEV value. Column (6): Variability of the source in ZTF. Column (7): $log_{10}$ of the SMBH mass. Column (8): Reference. References are formatted as W21 = Wolowska 2021, Ny20 = Nyland 2020, Zh22 = Zhang 2022. Table 3 is published in its entirety in the machine-readable format. A portion is shown here for guidance regarding its form and content.}  
\label{tab:RQRLAGNSample}
\end{deluxetable*}

\setlength{\tabcolsep}{-1pt}
\renewcommand{\arraystretch}{0.7}
\begin{longrotatetable}
\startlongtable
\begin{deluxetable*}{ccccccccccccc}
\tabletypesize{\tiny}
\tablecolumns{13}
\tablewidth{\textwidth}
\tablecaption{VLASS and VAST radio-detections of CLAGN + RQ to RL AGN Sample \label{table:CLAGNMetrics}}

\tablehead{
  \colhead{\begin{tabular}{@{}c@{}}Name\end{tabular}} 
& \colhead{\begin{tabular}{@{}c@{}}VLASS\\Epoch 1\\Total Flux\\(mJy)\end{tabular}}
& \colhead{\begin{tabular}{@{}c@{}}VLASS\\Epoch 1\\Peak\\Flux\\(mJy/\\beam)\end{tabular}}
& \colhead{\begin{tabular}{@{}c@{}}Epoch 1 \\(MJD)\end{tabular}}
& \colhead{\begin{tabular}{@{}c@{}}VLASS\\Epoch 2\\Total Flux\\(mJy)\end{tabular}}
& \colhead{\begin{tabular}{@{}c@{}}VLASS\\Epoch 2\\Peak\\Flux\\(mJy/\\beam)\end{tabular}}
& \colhead{\begin{tabular}{@{}c@{}}Epoch 2\\(MJD)\end{tabular}}
& \colhead{\begin{tabular}{@{}c@{}}g-band\\Flux\\(Jy)\end{tabular}}
& \colhead{\begin{tabular}{@{}c@{}}3 GHz \\Median\\Peak\\Flux\\(mJy/\\beam)\end{tabular}}
& \colhead{\begin{tabular}{@{}c@{}} 0.8875 GHz\\Median\\Peak\\Flux\\\\(mJy/\\beam)\end{tabular}}
& \colhead{\begin{tabular}{@{}c@{}}1.3675 GHz\\Median\\Peak\\Flux\\(mJy/\\beam)\end{tabular}}
& \colhead{\begin{tabular}{@{}c@{}}0.8875 GHz\\Median\\Int.\\Flux\\\\(mJy)\end{tabular}}
& \colhead{\begin{tabular}{@{}c@{}}1.3675 GHz\\Median\\Int.\\Flux\\\\(mJy)\end{tabular}}
}
\startdata
1ES1927+654    & $10.698\pm0.234$               & $9.345\pm0.122$                    & 58004        & $10.364\pm0.238$               & $10.13\pm0.131$                    & 59084        & 4.55E-04                & 9.738                                    & \dots                                             & \dots                                             & \dots                                              & \dots                                              \\
HE1136-2304    & $4.098\pm0.357$                & $2.528\pm0.147$                    & 58164        & $5.071\pm0.434$                & $2.984\pm0.173$                    & 59157        & 9.62E-04                & 2.756                                    & \dots                                             & \dots                                             & \dots                                              &\dots                                              \\
J001010-044238 & \dots                              & \dots                                  & \dots            & \dots                              & \dots                                  & \dots            & 5.62E-04                & \dots                                        & 2.4355                                        & \dots                                             & 11.2305                                        & 2.044                                          \\
J001418-051905 & \dots                              & \dots                                  & \dots            & \dots                              & \dots                                  & \dots            & 2.70E-04                & \dots                                        & 2.313                                         & \dots                                             & 3.207                                          & 2.606                                          \\
J003849-011722 & $52.386\pm0.324$               & $49.544\pm0.177$                   & 58025        & $84.803\pm0.584$               & $83.227\pm0.247$                   & 59047        & 2.30E-05                & 66.386                                   & 71.755                                        & 66.4745                                       & 79.604                                         & 73.909                                         \\
J004044+082352 & $5.343\pm0.265$                & $5.27\pm0.151$                     & 58084        & $6.588\pm0.239$                & $6.595\pm0.135$                    & 59111        & 5.62E-05                & 5.933                                    & 2.2305                                        & \dots                                             & 2.454                                          & \dots                                              \\
J010734+011223 & $2.755\pm0.253$                & $2.943\pm0.15$                     & 58042        & \dots                              & \dots                                  & \dots            & 2.30E-05                & 2.943                                    & 2.996                                         & 2.397                                         & 3.4345                                         & 3.619                                          \\
J013815+002914 & $3.433\pm0.187$                & $3.208\pm0.102$                    & 58023        & $2.884\pm0.276$                & $2.644\pm0.15$                     & 59096        & 5.71E-05                & 2.926                                    & \dots                                             & \dots                                             & \dots                                              & \dots                                              \\
J015106-003426 & $2.942\pm0.555$                & $0.991\pm0.144$                    & 58103        & $3.313\pm0.502$                & $1.806\pm0.186$                    & 59076        & 2.37E-05                & 1.399                                    & 3.892                                         & 2.243                                         & 9.945                                          & 4.8                                            \\
J015412-011150 & $5.952\pm0.234$                & $5.628\pm0.128$                    & 58023        & $6.472\pm0.299$                & $6.117\pm0.163$                    & 59076        & 3.39E-04                & 5.873                                    & 3.0535                                        & 4.013                                         & 3.3575                                         & 4.767                                          \\
J020515-045640 & $2.556\pm0.236$                & $2.467\pm0.121$                    & 58087        & $5.069\pm0.347$                & $4.115\pm0.167$                    & 59041        & 4.72E-05                & 3.291                                    & 7.852                                         & 5.3535                                        & 36.4945                                        & 13.9815                                        \\
J020827-005208 & $4.524\pm0.404$                & $4.774\pm0.238$                    & 58087        & $3.663\pm0.423$                & $3.608\pm0.235$                    & 59118        & 3.78E-05                & 4.191                                    & \dots                                             & \dots                                             & \dots                                              & \dots                                              \\
J021400+004227 & $3.315\pm0.493$                & $1.466\pm0.158$                    & 58087        & $1.907\pm0.29$                 & $1.718\pm0.149$                    & 59048        & 1.78E-04                & 1.592                                    & 5.678                                         & 3.658                                         & 6.146                                          & 4.121                                          \\
J022015-072900 & $1.439\pm0.261$                & $1.341\pm0.142$                    & 58087        & $1.285\pm0.313$                & $1.327\pm0.183$                    & 59041        & 3.34E-04                & 1.334                                    & \dots                                             & \dots                                             & \dots                                              & \dots                                              \\
J022931-000845 & $20.209\pm0.283$               & $18.854\pm0.154$                   & 58087        & $25.032\pm0.327$               & $23.792\pm0.179$                   & 59118        & 3.62E-05                & 21.323                                   & 24.036                                        & 22.752                                        & 26.401                                         & 24.38                                          \\
\enddata
\vspace{0.1cm}
\tablecomments{VLASS and VAST detections of the CLAGN and RQ to RL AGN objects, along with the $g$-band fluxes. Column (1): source name. Column (2): VLASS Epoch 1 Total Flux (mJy). Column (3): VLASS Epoch 1 Peak Flux (mJy/beam). Column (4): Epoch 1 Date of VLASS observation. Column (5): VLASS Epoch 2 Total Flux (mJy). Column (6): VLASS Epoch 2 Peak Flux (mJy/beam). Column (7): Epoch 2 Date of VLASS. Column (8): $g$-band fluxes (janskys). Column (9): VLASS Median Peak Flux (3GHz) (mJy/beam). Column (10): VAST Median Peak Flux (0.8875 GHz) (mJy/beam). Column (11): VAST Median Peak Flux (1.3675 GHz) (mJy/beam). Column (12): VAST Median Integrated Flux (0.8875 GHz) (mJy). Column (13): VAST Median Integrated Flux (1.3675 GHz) (mJy). Table 4 is published in its entirety in the machine-readable format. A portion is shown here for guidance regarding its form and content.}
\label{tab:CLAGN_RQRL_RadioData}
\end{deluxetable*}
\end{longrotatetable}

\setlength{\tabcolsep}{-1pt}  
\renewcommand{\arraystretch}{0.5}  
\begin{longrotatetable}
\startlongtable
\begin{deluxetable*}{@{}ccccccccccccccc@{}}  
  \tabletypesize{\tiny}
  \tablecolumns{15}
  \tablewidth{\textwidth}
  \tablecaption{CLAGN + RQ to RL AGN Radio Metrics \label{table:CLAGNProps}}

  \tablehead{
    \colhead{\begin{tabular}{@{}c@{}}Name\end{tabular}}
  & \colhead{\begin{tabular}{@{}c@{}}Comp.\\Ratio\end{tabular}}
  & \colhead{\begin{tabular}{@{}c@{}}Comp.\\Class\end{tabular}}
  & \colhead{\begin{tabular}{@{}c@{}}RL\\Ratio\\(R)\end{tabular}}
  & \colhead{\begin{tabular}{@{}c@{}}RL\\Class\end{tabular}}
  & \colhead{\begin{tabular}{@{}c@{}}$L_{3\rm GHz}$\\(erg\,s$^{-1}$)\end{tabular}}
  & \colhead{\begin{tabular}{@{}c@{}}$L_{887.5\rm MHz}$\\(erg\,s$^{-1}$)\end{tabular}}
  & \colhead{\begin{tabular}{@{}c@{}}Spec.\\Index\end{tabular}}
  & \colhead{\begin{tabular}{@{}c@{}}887.5\\MHz\\NEV\end{tabular}}
  & \colhead{\begin{tabular}{@{}c@{}}887.5\\MHz\\Var?\\(Y/N)\end{tabular}}
  & \colhead{\begin{tabular}{@{}c@{}}1367.5\\MHz\\NEV\end{tabular}}
  & \colhead{\begin{tabular}{@{}c@{}}1367.5\\MHz\\Var?\\(Y/N)\end{tabular}}
  & \colhead{\begin{tabular}{@{}c@{}}VLASS\\\%Flux\\Chg.\end{tabular}}
  & \colhead{\begin{tabular}{@{}c@{}}VLASS\\Var?\\(Y/N)\end{tabular}}
  & \colhead{\begin{tabular}{@{}c@{}}Source\\Type\end{tabular}}
  }
  \startdata
1ES1927+654    & 1.081             & C                 & 14.958                   & RL                   & 38.096                   & \dots                            & \dots              & \dots             & \dots                   & \dots              & \dots                    & 8.400                  & N               & RL-NV             \\
HE1136-2304    & 1.663             & C                 & 2.004                    & RQ                   & 37.948                   & \dots                            & \dots              & \dots             & \dots                   & \dots              & \dots                    & 18.038               & Y               & RQ-V              \\
J001010-044238 & 4.611          & C                 & 1.292                 & RQ                   & \dots                        & 37.126                    & \dots              & 0.01658      & N                   & \dots              & \dots                    & \dots                    & \dots               & RQ-NV             \\
J001418-051905 & 1.3865          & C                 & 2.554                 & RQ                   & \dots                        & 38.005                    & \dots              & 0.009814      & N                   & -0.005851      & N                    & \dots                    & \dots               & RQ-NV             \\
J003849-011722 & 1.033             & C                 & 2018.60              & RL                   & 41.765                   & 40.958                    & -0.06386       & 0.007604      & N                   & 0.004351       & N                    & 67.986               & Y               & B                 \\
J004044+082352 & 1.006             & C                 & 73.826                & RL                   & 40.041                   & 38.776                    & 0.8032       & 0.04213      & Y                   & \dots              & \dots                    & 25.142               & Y               & RL-V              \\
J010734+011223 & 0.936             & C                 & 89.489                & RL                   & 39.254                   & 38.420                    & -0.01466      & 0.076738      & Y                   & -0.000075      & N                    & \dots                    & \dots               & RL-V              \\
J013815+002914 & 1.079             & C                 & 35.838                   & RL                   & 40.857                   & \dots                            & \dots              & \dots             & \dots                   & \dots              & \dots                    & -17.581              & Y               & RL-V              \\
J015106-003426 & 2.236             & E                 & \dots                        & \dots                    & \dots                        & \dots                            & \dots              & \dots             & \dots                   & \dots              & \dots                    & \dots                    & \dots               & E                 \\
J015412-011150 & 1.058             & C                 & 12.115                & RL                   & 38.807                   & 37.683                    & 0.5370       & 0.02314      & Y                   & 0.01920       & N                    & 8.689                & N               & RL-NV             \\
J020515-045640 & 1.158             & C                 & 48.763                & RL                   & 40.212                   & 39.750                    & -0.7140      & 0.01595       & N                   & 0.01961       & N                    & 66.802               & Y               & RL-V              \\
J020827-005208 & 0.977             & C                 & 77.541                   & RL                   & 41.233                   & \dots                            & \dots              & \dots             & \dots                   & \dots              & \dots                    & -24.424              & Y               & RL-V              \\
J021400+004227 & 1.64              & C                 & 6.255                 & RQ                   & 39.336                   & 39.048                    & -1.0440      & 0.00367       & N                   & -0.000726      & N                    & 17.190                & Y               & RQ-V              \\
J022015-072900 & 1.021             & C                 & 2.795                    & RQ                   & 39.389                   & \dots                            & \dots              & \dots             & \dots                   & \dots              & \dots                    & -1.044               & N               & RQ-NV             \\
J022931-000845 & 1.061             & C                 & 411.951               & RL                   & 41.418                   & 40.630                    & -0.09833      & 0.009564      & N                   & 0.004284       & N                    & 26.191               & Y               & B                 \\
J025515+003740 & 1.027             & C                 & 1547.04              & RL                   & 41.775                   & 40.964                    & -0.05495      & 0.01033       & N                   & 0.005021       & N                    & -18.831              & Y               & RL-V              \\
J030533+004610 & 1.051             & C                 & 7587.04                  & RL                   & 40.288                   & \dots                            & \dots              & \dots             & \dots                   & \dots              & \dots                    & -9.554               & N               & RL-NV             \\
  \enddata
  \vspace{0.1cm}
  \tablecomments{Radio metrics of the radio-detected CLAGN and RQ to RL AGN objects. Column (1): source name. Column (2): Source compactness ratio. Column (3): Source compactness classification. Column (4): Radio Loudness Ratio. Column (5): Radio loudness classification. Column (6): VLASS 3 GHz radio luminosity. Column (7): VAST 887.5 MHz radio luminosity. Column (8): Radio spectral index. Column (9): 887.5 MHz NEV. Column (10): Variable in 887.5 MHz. Column (11): 1367.5 MHz NEV. Column (12): Variable in 1367.5 MHz. Column (13): VLASS percent Flux Change between two epochs. Column (14): Variable in VLASS. Column (15): Radio source type. Table 5 is published in its entirety in the machine-readable format. A portion is shown here for guidance regarding its form and content.}
  \label{tab:Radio_Metrics_CLAGN_RQRLAGN}
\end{deluxetable*}
\end{longrotatetable}

\setlength{\tabcolsep}{-1pt}
\startlongtable
\begin{deluxetable*}{ccccccccccc}
\tabletypesize{\tiny}
\tablecolumns{11}
\tablewidth{0pt}
\tablecaption{VLASS and VAST Radio-Detected Control AGN Sample\label{table:ControlDetections}}
\tablehead{
  \colhead{\begin{tabular}{c}RA\\(hms)\end{tabular}} & 
  \colhead{\begin{tabular}{c}Dec\\(dms)\end{tabular}} & 
  \colhead{\begin{tabular}{c}VLASS \\Epoch 1\\Peak Flux\\(Jy)\end{tabular}} & 
  \colhead{\begin{tabular}{c}VLASS \\Epoch 2\\Peak Flux\\(Jy)\end{tabular}} & 
  \colhead{\begin{tabular}{c}VLASS Epoch 1\\Int. Flux\\(Jy)\end{tabular}} & 
  \colhead{\begin{tabular}{c}VLASS Epoch 2\\Int. Flux\\(Jy)\end{tabular}} & 
  \colhead{\begin{tabular}{c}g-band\\Flux\\(Jy)\end{tabular}} & 
  \colhead{\begin{tabular}{c}0.8875 GHz\\ Median\\Peak Flux\\(mJy/beam)\end{tabular}} & 
  \colhead{\begin{tabular}{c}1.3675 GHz \\ Median\\Peak Flux\\(mJy/beam)\end{tabular}} & 
  \colhead{\begin{tabular}{c}0.8875 GHz\\ Median\\Int. Flux\\(mJy)\end{tabular}} & 
  \colhead{\begin{tabular}{c}1.3675 GHz\\ Median\\Int. Flux\\(mJy)\end{tabular}}
}
\startdata
12:31:53 & 55:26:26 & 0.002179                         & 0.002666                          & 0.006766                           & 0.01002                              & 3.560E-05             & \dots                                             & \dots                                             & \dots                                              & \dots                                              \\
2:29:05  & -4:20:38 & 0.002693                         & 0.002331                          & 0.0030103                           & \dots                                       & 5.480E-06             & 4.095                                         & 3.752                                         & 4.299                                          & 3.852                                          \\
14:20:23 & 51:51:59 & 0.001978                         & 0.002144                          & 0.002322                           & 0.00263                              & 8.770E-06             & \dots                                             & \dots                                             & \dots                                              & \dots                                              \\
16:07:25 & 44:48:10 & 0.006545                         & 0.007785                          & 0.006911                           & 0.008006                              & 8.250E-05             & \dots                                             & \dots                                             & \dots                                              & \dots                                              \\
11:47:50 & 53:00:06 & 0.001090                         & \dots                                 & \dots                                       & \dots                                       & 5.010E-05             & \dots                                             & \dots                                             & \dots                                              & \dots                                              \\
14:32:37 & 1:20:51  & 0.002130                         & 0.001349                          & 0.002129                           & 0.002770                              & 1.170E-04             & 4.749                                         & \dots                                             & 4.913                                          & \dots                                              \\
12:53:15 & 30:36:08 & 0.0061830                         & 0.006684                          & 0.007116                           & 0.006851                              & 1.630E-05             & \dots                                             & \dots                                             & \dots                                              & \dots                                              \\
14:32:37 & 1:20:51  & 0.0021304                         & 0.001349                          & 0.002129                           & 0.002770                              & 1.170E-04             & 4.749                                         & \dots                                             & 4.913                                          & \dots                                              \\
10:58:09 & 33:53:03 & 0.03001                         & 0.04926                          & 0.03736                           & 0.05001                              & 3.230E-05             & \dots                                             & \dots                                             & \dots                                              & \dots                                              \\
14:15:01 & 51:09:46 & 0.0017040                         & 0.001919                          & 0.002085                           & 0.003887                              & 3.430E-05             & \dots                                             & \dots                                             & \dots                                              & \dots                                              \\
12:50:30 & 62:29:01 & 0.003126                         & 0.003889                          & 0.004352                           & 0.005250                              & 1.440E-05             & \dots                                             & \dots                                             & \dots                                              & \dots                                              \\
8:59:04  & 2:05:04  & \dots                                 & \dots                                 & \dots                                       & \dots                                       & 8.620E-05             & 2.024                                         & \dots                                             & 2.769                                          & \dots                                              \\
\enddata
\vspace{0.1cm}
\tablecomments{VLASS and VAST detections for the broad-line control AGN sample. Column (1) and (2): right ascension and declination. Column (3): VLASS Epoch 1 Peak Flux (janskys). Column (4): VLASS Epoch 2 Peak Flux (janskys). Column (5): VLASS Epoch 1 Integrated Flux (janskys). Column (6): VLASS Epoch 2 Integrated Flux (janskys). Column (7): $g$-band Flux (janskys). Column (8): VAST Median Peak Flux (0.8875 GHz) (mJy/beam). Column (9): VAST Median Peak Flux (1.3675 GHz) (mJy/beam). Column (10): VAST Median Integrated Flux (0.8875 GHz) (mJy). Column (11): VAST Median Integrated Flux (1.3675 GHz) (mJy). Table 6 is published in its entirety in the machine-readable format. A portion is shown here for guidance regarding its form and content.}
\label{tab:ControlDetections}
\end{deluxetable*}

\setlength{\tabcolsep}{-1pt}  
\renewcommand{\arraystretch}{0.5}  
\begin{longrotatetable}
\startlongtable
\begin{deluxetable*}{@{}ccccccccccccc@{}}  
  \tabletypesize{\tiny}
  \tablecolumns{13}
  \tablewidth{\textwidth}
  \tablecaption{Control AGN Sample Radio Metrics \label{table:Radio_Metrics_CLAGN_RQRLAGN}}
  \tablehead{
    \colhead{\begin{tabular}{@{}c@{}}RA\\(hms)\end{tabular}}
  & \colhead{\begin{tabular}{@{}c@{}}Dec\\(dms)\end{tabular}}
  & \colhead{\begin{tabular}{@{}c@{}}VLASS \%\\Flux \\Change\end{tabular}}
  & \colhead{\begin{tabular}{@{}c@{}}\\VLASS\\Variable?\\(Y/N)\end{tabular}}
  & \colhead{\begin{tabular}{@{}c@{}}Comp.\\Ratio\end{tabular}}
  & \colhead{\begin{tabular}{@{}c@{}}Comp.\\Class\end{tabular}}
  & \colhead{\begin{tabular}{@{}c@{}}RL\\Ratio\\(R)\end{tabular}}
  & \colhead{\begin{tabular}{@{}c@{}}RL\\Class\end{tabular}}
  & \colhead{\begin{tabular}{@{}c@{}}$L_{3\rm GHz}$\\(erg\,s$^{-1}$)\end{tabular}}
  & \colhead{\begin{tabular}{@{}c@{}}VAST\\887.5 MHz\\NEV\end{tabular}}
  & \colhead{\begin{tabular}{@{}c@{}}887.5 MHz\\Variable?\\(Y/N)\end{tabular}}
  & \colhead{\begin{tabular}{@{}c@{}}VAST\\1367.5 MHz\\NEV\end{tabular}}
  & \colhead{\begin{tabular}{@{}c@{}}1367.5 MHz\\Variable?\\(Y/N)\end{tabular}}
  }
  \startdata
12:31:53 & 55:26:26 & 22.365                  & Y                                     & 3.4630       & E                          & \dots                        & \dots                             & 40.187                                                & \dots                  & \dots                          & \dots                   & \dots                           \\
2:29:05  & -4:20:38 & -13.435                 & Y                                     & 1.1995       & C                          & 320.5863              & RL                            & 39.827                                                & 0.01008      & N                          & -0.0015091     & N                           \\
14:20:23 & 51:51:59 & 8.380                   & N                                     & 1.20102       & C                          & 164.3555               & RL                            & 40.996                                                & \dots                  & \dots                          & \dots                   & \dots                           \\
16:07:25 & 44:48:10 & 18.950                  & Y                                     & 1.0420       & C                          & 60.7391              & RL                            & 40.785                                                & \dots                  & \dots                          & \dots                   & \dots                           \\
11:47:50 & 53:00:06 & \dots                       & \dots                                     & \dots                 & \dots                          & \dots                        & \dots                             & 39.923                                                & \dots                  & \dots                          & \dots                   & \dots                           \\
14:32:37 & 1:20:51  & -36.655                 & Y                                     & 1.4078       & C                          & 10.4009              & RL                            & 39.142                                                & 0.002983     & N                          & \dots                   & \dots                           \\
12:53:15 & 30:36:08 & 8.107                   & N                                     & 1.0854       & C                          & 276.0574              & RL                            & 41.767                                                & \dots                  & \dots                          & \dots                   & \dots                           \\
14:32:37 & 1:20:51  & -36.655                 & Y                                     & 1.4078       & C                          & 10.4009              & RL                            & 39.142                                                & 0.002983     & N                          & \dots                   & \dots                           \\
10:58:09 & 33:53:03 & 64.153                  & Y                                     & 1.1021       & C                          & 858.2092              & RL                            & 41.041                                                & \dots                  & \dots                          & \dots                   & \dots                           \\
14:15:01 & 51:09:46 & 12.630                  & Y                                     & 1.6479       & C                          & 36.9462              & RL                            & 39.409                                                & \dots                  & \dots                          & \dots                   & \dots                           \\
12:50:30 & 62:29:01 & 24.423                  & Y                                     & 1.3690       & C                          & 170.3253              & RL                            & 40.426                                                & \dots                  & \dots                          & \dots                   & \dots                           \\
8:59:04  & 2:05:04  & \dots                       & \dots                                     & 1.3680       & C                          & 5.6587              & RQ                            & \dots                                                     & -0.0006720   & N                          & \dots                   & \dots                           \\
  \enddata
  \vspace{0.1cm}
  \tablecomments{Radio metrics of the broad-line control AGN sample. Column (1) and (2): right ascension and declination. Column (3): VLASS Percent Change in Flux. Column (4): VLASS variable. Column (5): Source compactness ratio. Column (6): Source compactness classification. Column (7): Radio loudness ratio. Column (8): Radio loudness classification. Column (9): $log_{10} L_{3GHz}$ VLASS radio luminosity. Column (10): VAST 887.5 MHz NEV. Column (11): 887.5 MHz Variable. Column (12): VAST 1367.5 MHz NEV. Column (13): 1367.5 MHz Variable. Table 7 is published in its entirety in the machine-readable format. A portion is shown here for guidance regarding its form and content.}
  \label{tab:Radio_Metrics_Control}
\end{deluxetable*}
\end{longrotatetable}

\bibliography{main,Mendeleyreferences,references}

\begin{thebibliography}{}
\expandafter\ifx\csname natexlab\endcsname\relax\def\natexlab#1{#1}\fi
\providecommand{\url}[1]{\href{#1}{#1}}
\providecommand{\dodoi}[1]{doi:~\href{http://doi.org/#1}{\nolinkurl{#1}}}
\providecommand{\doeprint}[1]{\href{http://ascl.net/#1}{\nolinkurl{http://ascl.net/#1}}}
\providecommand{\doarXiv}[1]{\href{https://arxiv.org/abs/#1}{\nolinkurl{https://arxiv.org/abs/#1}}}

\bibitem[{Abbott {et~al.}(2018)Abbott, Abdalla, Allam, Amara, Annis, Asorey, Avila, Ballester, Banerji, Barkhouse, Baruah, Baumer, Bechtol, Becker, Benoit-L{\'{e}}vy, Bernstein, Bertin, Blazek, Bocquet, Brooks, Brout, Buckley-Geer, Burke, Busti, Campisano, Cardiel-Sas, Rosell, Kind, Carretero, Castander, Cawthon, Chang, Chen, Conselice, Costa, Crocce, Cunha, D’Andrea, Costa, Das, Daues, Davis, Davis, Vicente, DePoy, DeRose, Desai, Diehl, Dietrich, Dodelson, Doel, Drlica-Wagner, Eifler, Elliott, Evrard, Farahi, Neto, Fernandez, Finley, Flaugher, Foley, Fosalba, Friedel, Frieman, Garc{\'{i}}a-Bellido, Gaztanaga, Gerdes, Giannantonio, Gill, Glazebrook, Goldstein, Gower, Gruen, Gruendl, Gschwend, Gupta, Gutierrez, Hamilton, Hartley, Hinton, Hislop, Hollowood, Honscheid, Hoyle, Huterer, Jain, James, Jeltema, Johnson, Johnson, Kacprzak, Kent, Khullar, Klein, Kovacs, Koziol, Krause, Kremin, Kron, Kuehn, Kuhlmann, Kuropatkin, Lahav, Lasker, Li, Li, Liddle, Lima, Lin, L{\'{o}}pez-Reyes, MacCrann, Maia, Maloney,
  Manera, March, Marriner, Marshall, Martini, McClintock, McKay, McMahon, Melchior, Menanteau, Miller, Miquel, Mohr, Morganson, Mould, Neilsen, Nichol, Nogueira, Nord, Nugent, Nunes, Ogando, Old, Pace, Palmese, Paz-Chinch{\'{o}}n, Peiris, Percival, Petravick, Plazas, Poh, Pond, Porredon, Pujol, Refregier, Reil, Ricker, Rollins, Romer, Roodman, Rooney, Ross, Rykoff, Sako, Sanchez, Sanchez, Santiago, Saro, Scarpine, Scolnic, Serrano, Sevilla-Noarbe, Sheldon, Shipp, Silveira, Smith, Smith, Smith, Soares-Santos, Sobreira, Song, Stebbins, Suchyta, Sullivan, Swanson, Tarle, Thaler, Thomas, Thomas, Troxel, Tucker, Vikram, Vivas, Walker, Wechsler, Weller, Wester, Wolf, Wu, Yanny, Zenteno, Zhang, Zuntz, Juneau, Fitzpatrick, Nikutta, Nidever, Olsen, \& Scott}]{Abbott2018The1}
Abbott, T. M.~C., Abdalla, F.~B., Allam, S., {et~al.} 2018, The Astrophysical Journal Supplement Series, 239, 18, \dodoi{10.3847/1538-4365/aae9f0}

\bibitem[{Adame {et~al.}(2024)Adame, Aguilar, Ahlen, Alam, Aldering, Alexander, Alfarsy, Allende~Prieto, Alvarez, Alves, Anand, Andrade-Oliveira, Armengaud, Asorey, Avila, Aviles, Bailey, Balaguera-Antol{\'{i}}nez, Ballester, Baltay, Bault, Bautista, Behera, Beltran, BenZvi, Beraldo~e Silva, Bermejo-Climent, Berti, Besuner, Beutler, Bianchi, Blake, Blum, Bolton, Brieden, Brodzeller, Brooks, Brown, Buckley-Geer, Burtin, Cabayol-Garcia, Cai, Canning, Cardiel-Sas, Carnero~Rosell, Castander, Cervantes-Cota, Chabanier, Chaussidon, Chaves-Montero, Chen, Chen, Chuang, Claybaugh, Cole, Cooper, Cuceu, Davis, Dawson, de~Belsunce, de~la Cruz, de~la Macorra, Della~Costa, de~Mattia, Demina, Demirbozan, DeRose, Dey, Dey, Dhungana, Ding, Ding, Doel, Doshi, Douglass, Edge, Eftekharzadeh, Eisenstein, Elliott, Ereza, Escoffier, Fagrelius, Fan, Fanning, Fawcett, Ferraro, Flaugher, Font-Ribera, Forero-Romero, Forero-S{\'{a}}nchez, Frenk, G{\"{a}}nsicke, Garc{\'{i}}a, Garc{\'{i}}a-Bellido, Garcia-Quintero, Garrison,
  Gil-Mar{\'{i}}n, Golden-Marx, Gontcho A~Gontcho, Gonzalez-Morales, Gonzalez-Perez, Gordon, Graur, Green, Gruen, Guy, Hadzhiyska, Hahn, Han, Hanif, Herrera-Alcantar, Honscheid, Hou, Howlett, Huterer, Ir{\v{s}}i{\v{c}}, Ishak, Jacques, Jana, Jiang, Jimenez, Jing, Joudaki, Joyce, Jullo, Juneau, Kara{\c{c}}aylı, Karim, Kehoe, Kent, Khederlarian, Kim, Kirkby, Kisner, Kitaura, Kizhuprakkat, Kneib, Koposov, Kov{\'{a}}cs, Kremin, Krolewski, L’Huillier, Lahav, Lambert, Lamman, Lan, Landriau, Lang, Lange, Lasker, Leauthaud, Le~Guillou, Levi, Li, Linder, Lyons, Magneville, Manera, Manser, Margala, Martini, McDonald, Medina, Medina-Varela, Meisner, Mena-Fern{\'{a}}ndez, Meneses-Rizo, Mezcua, Miquel, Montero-Camacho, Moon, Moore, Moustakas, Mueller, Mundet, Mu{\~{n}}oz-Guti{\'{e}}rrez, Myers, Nadathur, Napolitano, Neveux, Newman, Nie, Nikutta, Niz, Norberg, Noriega, Paillas, Palanque-Delabrouille, Palmese, Pan, Parkinson, Penmetsa, Percival, P{\'{e}}rez-Fern{\'{a}}ndez, P{\'{e}}rez-R{\`{a}}fols, Pieri, Poppett,
  Porredon, Pothier, Prada, Pucha, Raichoor, Ram{\'{i}}rez-P{\'{e}}rez, Ramirez-Solano, Rashkovetskyi, Ravoux, Rocher, Rockosi, Ross, Rossi, Ruggeri, Ruhlmann-Kleider, Sabiu, Said, Saintonge, Samushia, Sanchez, Saulder, Schaan, Schlafly, Schlegel, Scholte, Schubnell, Seo, Shafieloo, Sharples, Sheu, Silber, Sinigaglia, Siudek, Slepian, Smith, Soumagnac, Sprayberry, Stephey, Su{\'{a}}rez-P{\'{e}}rez, Sun, Tan, Tarl{\'{e}}, Tojeiro, Ure{\~{n}}a-L{\'{o}}pez, Vaisakh, Valcin, Valdes, Valluri, Vargas-Maga{\~{n}}a, Variu, Verde, Walther, Wang, Wang, Weaver, Weaverdyck, Wechsler, White, Xie, Yang, Y{\`{e}}che, Yu, Yuan, Zhang, Zhang, Zhao, Zheng, Zhou, Zhou, Zou, Zou, \& Zu}]{Adame2024TheInstrument}
Adame, A.~G., Aguilar, J., Ahlen, S., {et~al.} 2024, The Astronomical Journal, 168, 58, \dodoi{10.3847/1538-3881/ad3217}

\bibitem[{Ahumada {et~al.}(2020)Ahumada, Prieto, Almeida, Anders, Anderson, Andrews, Anguiano, Arcodia, Armengaud, Aubert, Avila, Avila-Reese, Badenes, Balland, Barger, Barrera-Ballesteros, Basu, Bautista, Beaton, Beers, Benavides, Bender, Bernardi, Bershady, Beutler, Bidin, Bird, Bizyaev, Blanc, Blanton, Boquien, Borissova, Bovy, Brandt, Brinkmann, Brownstein, Bundy, Bureau, Burgasser, Burtin, Cano-D{\'{i}}az, Capasso, Cappellari, Carrera, Chabanier, Chaplin, Chapman, Cherinka, Chiappini, Doohyun~Choi, Chojnowski, Chung, Clerc, Coffey, Comerford, Comparat, da~Costa, Cousinou, Covey, Crane, Cunha, Ilha, Dai~戴, Damsted, Darling, Davidson, Davies, Dawson, De, de~la Macorra, De~Lee, Queiroz, Deconto~Machado, de~la Torre, Dell’Agli, du~Mas~des Bourboux, Diamond-Stanic, Dillon, Donor, Drory, Duckworth, Dwelly, Ebelke, Eftekharzadeh, Davis~Eigenbrot, Elsworth, Eracleous, Erfanianfar, Escoffier, Fan, Farr, Fern{\'{a}}ndez-Trincado, Feuillet, Finoguenov, Fofie, Fraser-McKelvie, Frinchaboy, Fromenteau, Fu,
  Galbany, Garcia, Garc{\'{i}}a-Hern{\'{a}}ndez, Oehmichen, Ge, Maia, Geisler, Gelfand, Goddy, Gonzalez-Perez, Grabowski, Green, Grier, Guo, Guy, Harding, Hasselquist, Hawken, Hayes, Hearty, Hekker, Hogg, Holtzman, Horta, Hou, Hsieh, Huber, Hunt, Chitham, Imig, Jaber, Angel, Johnson, Jones, J{\"{o}}nsson, Jullo, Kim, Kinemuchi, Kirkpatrick~IV, Kite, Klaene, Kneib, Kollmeier, Kong, Kounkel, Krishnarao, Lacerna, Lan, Lane, Law, Le~Goff, Leung, Lewis, Li, Lian, Lin~林俐, Long, Longa-Pe{\~{n}}a, Lundgren, Lyke, Ted~Mackereth, MacLeod, Majewski, Manchado, Maraston, Martini, Masseron, Masters~何凱, Mathur, McDermid, Merloni, Merrifield, M{\'{e}}sz{\'{a}}ros, Miglio, Minniti, Minsley, Miyaji, Mohammad, Mosser, Mueller, Muna, Mu{\~{n}}oz-Guti{\'{e}}rrez, Myers, Nadathur, Nair, Nandra, do~Nascimento, Nevin, Newman, Nidever, Nitschelm, Noterdaeme, O’Connell, Olmstead, Oravetz, Oravetz, Osorio, Pace, Padilla, Palanque-Delabrouille, Palicio, Pan, Pan, Parker, Paviot, Peirani, Ramŕez, Penny, Percival,
  Perez-Fournon, P{\'{e}}rez-R{\`{a}}fols, Petitjean, Pieri, Pinsonneault, Poovelil, Povick, Prakash, Price-Whelan, Raddick, Raichoor, Ray, Rembold, Rezaie, Riffel, Riffel, Rix, Robin, Roman-Lopes, Rom{\'{a}}n-Z{\'{u}}{\~{n}}iga, Rose, Ross, Rossi, Rowlands, Rubin, Salvato, S{\'{a}}nchez, S{\'{a}}nchez-Menguiano, S{\'{a}}nchez-Gallego, Sayres, Schaefer, Schiavon, Schimoia, Schlafly, Schlegel, Schneider, Schultheis, Schwope, Seo, Serenelli, Shafieloo, Shamsi, Shao, Shen, Shetrone, Shirley, Aguirre, Simon, Skrutskie, Slosar, Smethurst, Sobeck, Sodi, Souto, Stark, Stassun, Steinmetz, Stello, Stermer, Storchi-Bergmann, Streblyanska, Stringfellow, Stutz, Su{\'{a}}rez, Sun, Taghizadeh-Popp, Talbot, Tayar, Thakar, Theriault, Thomas, Thomas, Tinker, Tojeiro, Toledo, Tremonti, Troup, Tuttle, Unda-Sanzana, Valentini, Vargas-Gonz{\'{a}}lez, Vargas-Maga{\~{n}}a, V{\'{a}}zquez-Mata, Vivek, Wake, Wang, Weaver, Weijmans, Wild, Wilson, Wilson, Wolthuis, Wood-Vasey, Yan, Yang, Y{\`{e}}che, Zamora, Zarrouk, Zasowski, Zhang,
  Zhao, Zhao, Zheng, Zheng, Zhu, \& Zou}]{Ahumada2020TheSpectra}
Ahumada, R., Prieto, C.~A., Almeida, A., {et~al.} 2020, The Astrophysical Journal Supplement Series, 249, 3, \dodoi{10.3847/1538-4365/ab929e}

\bibitem[{{Ajello} {et~al.}(2022){Ajello}, {Baldini}, {Ballet}, {Bastieri}, {Becerra Gonzalez}, {Bellazzini}, {Berretta}, {Bissaldi}, {Bonino}, {Brill}, {Bruel}, {Buson}, {Caputo}, {Caraveo}, {Cheung}, {Chiaro}, {Cibrario}, {Ciprini}, {Crnogorcevic}, {Cutini}, {D'Ammando}, {De Gaetano}, {Di Lalla}, {Di Venere}, {Dom{\'\i}nguez}, {Ramazani}, {Ferrara}, {Fiori}, {Fukazawa}, {Funk}, {Fusco}, {Gammaldi}, {Gargano}, {Garrappa}, {Gasparrini}, {Giglietto}, {Giordano}, {Giroletti}, {Green}, {Grenier}, {Guiriec}, {Horan}, {Hou}, {Kayanoki}, {Kuss}, {Larsson}, {Latronico}, {Lewis}, {Li}, {Liodakis}, {Longo}, {Loparco}, {Lott}, {Lovellette}, {Lubrano}, {Madejski}, {Maldera}, {Manfreda}, {Mart{\'\i}-Devesa}, {Mazziotta}, {Mereu}, {Michelson}, {Mirabal}, {Mitthumsiri}, {Mizuno}, {Monzani}, {Morselli}, {Moskalenko}, {Negro}, {Ojha}, {Orienti}, {Orlando}, {Ormes}, {Pei}, {Pe{\~n}a-Herazo}, {Persic}, {Pesce-Rollins}, {Petrosian}, {Pillera}, {Poon}, {Porter}, {Principe}, {Rain{\`o}}, {Rando}, {Rani}, {Razzano}, {Razzaque},
  {Reimer}, {Reimer}, {Scotton}, {Serini}, {Sgr{\`o}}, {Siskind}, {Spandre}, {Spinelli}, {Suson}, {Tajima}, {Torres}, {Valverde}, {Yassin}, \& {Zaharijas}}]{Fermi2022}
{Ajello}, M., {Baldini}, L., {Ballet}, J., {et~al.} 2022, \apjs, 263, 24, \dodoi{10.3847/1538-4365/ac9523}

\bibitem[{{Amirkhanian}(1985)}]{Amirkhanian1985}
{Amirkhanian}, V.~R. 1985, \apss, 108, 125, \dodoi{10.1007/BF00650124}

\bibitem[{Bellm {et~al.}(2019)Bellm, Kulkarni, Graham, Dekany, Smith, Riddle, Masci, Helou, Prince, Adams, Barbarino, Barlow, Bauer, Beck, Belicki, Biswas, Blagorodnova, Bodewits, Bolin, Brinnel, Brooke, Bue, Bulla, Burruss, Cenko, Chang, Connolly, Coughlin, Cromer, Cunningham, De, Delacroix, Desai, Duev, Eadie, Farnham, Feeney, Feindt, Flynn, Franckowiak, Frederick, Fremling, Gal-Yam, Gezari, Giomi, Goldstein, Golkhou, Goobar, Groom, Hacopians, Hale, Henning, Ho, Hover, Howell, Hung, Huppenkothen, Imel, Ip, Ivezi{\'{c}}, Jackson, Jones, Juric, Kasliwal, Kaspi, Kaye, Kelley, Kowalski, Kramer, Kupfer, Landry, Laher, Lee, Lin, Lin, Lunnan, Giomi, Mahabal, Mao, Miller, Monkewitz, Murphy, Ngeow, Nordin, Nugent, Ofek, Patterson, Penprase, Porter, Rauch, Rebbapragada, Reiley, Rigault, Rodriguez, Roestel, Rusholme, Santen, Schulze, Shupe, Singer, Soumagnac, Stein, Surace, Sollerman, Szkody, Taddia, Terek, Van~Sistine, van Velzen, Vestrand, Walters, Ward, Ye, Yu, Yan, \& Zolkower}]{Bellm2019}
Bellm, E.~C., Kulkarni, S.~R., Graham, M.~J., {et~al.} 2019, Publications of the Astronomical Society of the Pacific, 131, 018002, \dodoi{10.1088/1538-3873/aaecbe}

\bibitem[{{D'Abrusco} {et~al.}(2014){D'Abrusco}, {Massaro}, {Paggi}, {Smith}, {Masetti}, {Landoni}, \& {Tosti}}]{DAbrusco2014ApJS}
{D'Abrusco}, R., {Massaro}, F., {Paggi}, A., {et~al.} 2014, \apjs, 215, 14, \dodoi{10.1088/0067-0049/215/1/14}

\bibitem[{Dekany {et~al.}(2020)Dekany, Smith, Riddle, Feeney, Porter, Hale, Zolkower, Belicki, Kaye, Henning, Walters, Cromer, Delacroix, Rodriguez, Reiley, Mao, Hover, Murphy, Burruss, Baker, Kowalski, Reif, Mueller, Bellm, Graham, \& Kulkarni}]{Dekany2020TheSystem}
Dekany, R., Smith, R.~M., Riddle, R., {et~al.} 2020, Publications of the Astronomical Society of the Pacific, 132, 038001, \dodoi{10.1088/1538-3873/ab4ca2}

\bibitem[{{Denney} {et~al.}(2014){Denney}, {De Rosa}, {Croxall}, {Gupta}, {Bentz}, {Fausnaugh}, {Grier}, {Martini}, {Mathur}, {Peterson}, {Pogge}, \& {Shappee}}]{Denney2014}
{Denney}, K.~D., {De Rosa}, G., {Croxall}, K., {et~al.} 2014, \apj, 796, 134, \dodoi{10.1088/0004-637X/796/2/134}

\bibitem[{Denney {et~al.}(2014)Denney, De~Rosa, Croxall, Gupta, Bentz, Fausnaugh, Grier, Martini, Mathur, Peterson, Pogge, \& Shappee}]{Denney2014TheRole}
Denney, K.~D., De~Rosa, G., Croxall, K., {et~al.} 2014, The Astrophysical Journal, 796, 134, \dodoi{10.1088/0004-637X/796/2/134}

\bibitem[{{Dodd} {et~al.}(2025){Dodd}, {Huang}, {Davis}, \& {Ramirez-Ruiz}}]{Dodd2025}
{Dodd}, S.~A., {Huang}, X., {Davis}, S.~W., \& {Ramirez-Ruiz}, E. 2025, arXiv e-prints, arXiv:2506.19900, \dodoi{10.48550/arXiv.2506.19900}

\bibitem[{Elitzur {et~al.}(2014)Elitzur, Ho, \& Trump}]{Elitzur2014EvolutionNuclei}
Elitzur, M., Ho, L.~C., \& Trump, J.~R. 2014, Monthly Notices of the Royal Astronomical Society, 438, 3340, \dodoi{10.1093/mnras/stt2445}

\bibitem[{Fernandez {et~al.}(2022)Fernandez, Secrest, Johnson, Schmitt, Fischer, Cigan, \& Dorland}]{Fernandez2022FRAMEx.2992}
Fernandez, L.~C., Secrest, N.~J., Johnson, M.~C., {et~al.} 2022, The Astrophysical Journal, 927, 18, \dodoi{10.3847/1538-4357/ac4b5f}

\bibitem[{Ferrarese \& Ford(2005)}]{Ferrarese2005}
Ferrarese, L., \& Ford, H. 2005, Space Science Reviews, 116, 523, \dodoi{10.1007/s11214-005-3947-6}

\bibitem[{Flewelling {et~al.}(2020)Flewelling, Magnier, Chambers, Heasley, Holmberg, Huber, Sweeney, Waters, Calamida, Casertano, Chen, Farrow, Hasinger, Henderson, Long, Metcalfe, Narayan, Nieto-Santisteban, Norberg, Rest, Saglia, Szalay, Thakar, Tonry, Valenti, Werner, White, Denneau, Draper, Hodapp, Jedicke, Kaiser, Kudritzki, Price, Wainscoat, Chastel, McLean, Postman, \& Shiao}]{Flewelling2020TheProducts}
Flewelling, H.~A., Magnier, E.~A., Chambers, K.~C., {et~al.} 2020, The Astrophysical Journal Supplement Series, 251, 7, \dodoi{10.3847/1538-4365/abb82d}

\bibitem[{Frederick {et~al.}(2019)Frederick, Gezari, Graham, Cenko, van Velzen, Stern, Blagorodnova, Kulkarni, Yan, De, Fremling, Hung, Kara, Shupe, Ward, Bellm, Dekany, Duev, Feindt, Giomi, Kupfer, Laher, Masci, Miller, Neill, Ngeow, Patterson, Porter, Rusholme, Sollerman, \& Walters}]{Frederick2019ALINERs}
Frederick, S., Gezari, S., Graham, M.~J., {et~al.} 2019, The Astrophysical Journal, 883, 31, \dodoi{10.3847/1538-4357/ab3a38}

\bibitem[{Graham {et~al.}(2019)Graham, Kulkarni, Bellm, Adams, Barbarino, Blagorodnova, Bodewits, Bolin, Brady, Cenko, Chang, Coughlin, De, Eadie, Farnham, Feindt, Franckowiak, Fremling, Gezari, Ghosh, Goldstein, Golkhou, Goobar, Ho, Huppenkothen, Ivezi{\'{c}}, Jones, Juric, Kaplan, Kasliwal, Kelley, Kupfer, Lee, Lin, Lunnan, Mahabal, Miller, Ngeow, Nugent, Ofek, Prince, Rauch, Roestel, Schulze, Singer, Sollerman, Taddia, Yan, Ye, Yu, Barlow, Bauer, Beck, Belicki, Biswas, Brinnel, Brooke, Bue, Bulla, Burruss, Connolly, Cromer, Cunningham, Dekany, Delacroix, Desai, Duev, Feeney, Flynn, Frederick, Gal-Yam, Giomi, Groom, Hacopians, Hale, Helou, Henning, Hover, Hillenbrand, Howell, Hung, Imel, Ip, Jackson, Kaspi, Kaye, Kowalski, Kramer, Kuhn, Landry, Laher, Mao, Masci, Monkewitz, Murphy, Nordin, Patterson, Penprase, Porter, Rebbapragada, Reiley, Riddle, Rigault, Rodriguez, Rusholme, Santen, Shupe, Smith, Soumagnac, Stein, Surace, Szkody, Terek, Sistine, Velzen, Vestrand, Walters, Ward, Zhang, \&
  Zolkower}]{Graham2019}
Graham, M.~J., Kulkarni, S.~R., Bellm, E.~C., {et~al.} 2019, Publications of the Astronomical Society of the Pacific, 131, 078001, \dodoi{10.1088/1538-3873/ab006c}

\bibitem[{Graham {et~al.}(2020)Graham, Ross, Stern, Drake, McKernan, Ford, Djorgovski, Mahabal, Glikman, Larson, \& Christensen}]{Graham2020UnderstandingQuasars}
Graham, M.~J., Ross, N.~P., Stern, D., {et~al.} 2020, Monthly Notices of the Royal Astronomical Society, 491, 4925, \dodoi{10.1093/mnras/stz3244}

\bibitem[{Green {et~al.}(2022)Green, Pulgarin-Duque, Anderson, MacLeod, Eracleous, Ruan, Runnoe, Graham, Roulston, Schneider, Ahlf, Bizyaev, Brownstein, del Casal, Dodd, Hoover, Matt, Merloni, Pan, Ramirez, Ridder, \& Moseley}]{Green2022TheSDSS-IV}
Green, P.~J., Pulgarin-Duque, L., Anderson, S.~F., {et~al.} 2022, The Astrophysical Journal, 933, 180, \dodoi{10.3847/1538-4357/ac743f}

\bibitem[{Guo {et~al.}(2024)Guo, Zou, Fawcett, Canning, Juneau, Davis, Alexander, Jiang, Aguilar, Ahlen, Brooks, Claybaugh, de~la Macorra, Doel, Fanning, Forero-Romero, Gontcho A~Gontcho, Honscheid, Kisner, Kremin, Landriau, Meisner, Miquel, Moustakas, Nie, Pan, Poppett, Prada, Rezaie, Rossi, Siudek, Sanchez, Schubnell, Seo, Sui, Tarl{\'{e}}, \& Zhou}]{Guo2024Changing-lookData}
Guo, W.-J., Zou, H., Fawcett, V.~A., {et~al.} 2024, The Astrophysical Journal Supplement Series, 270, 26, \dodoi{10.3847/1538-4365/ad118a}

\bibitem[{Guolo {et~al.}(2021)Guolo, Ruschel-Dutra, Grupe, Peterson, Storchi-Bergmann, Schimoia, Nemmen, \& Robinson}]{Guolo2021The2992}
Guolo, M., Ruschel-Dutra, D., Grupe, D., {et~al.} 2021, Monthly Notices of the Royal Astronomical Society, 508, 144, \dodoi{10.1093/mnras/stab2550}

\bibitem[{{Hallinan} {et~al.}(2019){Hallinan}, {Ravi}, {Weinreb}, {Kocz}, {Huang}, {Woody}, {Lamb}, {D'Addario}, {Catha}, {Law}, {Kulkarni}, {Phinney}, {Eastwood}, {Bouman}, {McLaughlin}, {Ransom}, {Siemens}, {Cordes}, {Lynch}, {Kaplan}, {Brazier}, {Bhatnagar}, {Myers}, {Walter}, \& {Gaensler}}]{Hallinan2019}
{Hallinan}, G., {Ravi}, V., {Weinreb}, S., {et~al.} 2019, in Bulletin of the American Astronomical Society, Vol.~51, 255, \dodoi{10.48550/arXiv.1907.07648}

\bibitem[{{Hon} {et~al.}(2022){Hon}, {Wolf}, {Onken}, {Webster}, \& {Auchettl}}]{Hon2022}
{Hon}, W.~J., {Wolf}, C., {Onken}, C.~A., {Webster}, R., \& {Auchettl}, K. 2022, \mnras, 511, 54, \dodoi{10.1093/mnras/stab3694}

\bibitem[{Ivezi{\'{c}} {et~al.}(2019)Ivezi{\'{c}}, Kahn, Tyson, Abel, Acosta, Allsman, Alonso, AlSayyad, Anderson, Andrew, P.~Angel, Angeli, Ansari, Antilogus, Araujo, Armstrong, Arndt, Astier, Aubourg, Auza, Axelrod, Bard, Barr, Barrau, Bartlett, Bauer, Bauman, Baumont, Bechtol, Bechtol, Becker, Becla, Beldica, Bellavia, Bianco, Biswas, Blanc, Blazek, Blandford, Bloom, Bogart, Bond, Booth, Borgland, Borne, Bosch, Boutigny, Brackett, Bradshaw, Brandt, Brown, Bullock, Burchat, Burke, Cagnoli, Calabrese, Callahan, Callen, Carlin, Carlson, Chandrasekharan, Charles-Emerson, Chesley, Cheu, Chiang, Chiang, Chirino, Chow, Ciardi, Claver, Cohen-Tanugi, Cockrum, Coles, Connolly, Cook, Cooray, Covey, Cribbs, Cui, Cutri, Daly, Daniel, Daruich, Daubard, Daues, Dawson, Delgado, Dellapenna, Peyster, Val-Borro, Digel, Doherty, Dubois, Dubois-Felsmann, Durech, Economou, Eifler, Eracleous, Emmons, Neto, Ferguson, Figueroa, Fisher-Levine, Focke, Foss, Frank, Freemon, Gangler, Gawiser, Geary, Gee, Geha, Gessner, Gibson,
  Gilmore, Glanzman, Glick, Goldina, Goldstein, Goodenow, Graham, Gressler, Gris, Guy, Guyonnet, Haller, Harris, Hascall, Haupt, Hernandez, Herrmann, Hileman, Hoblitt, Hodgson, Hogan, Howard, Huang, Huffer, Ingraham, Innes, Jacoby, Jain, Jammes, Jee, Jenness, Jernigan, Jevremovi{\'{c}}, Johns, Johnson, Johnson, Jones, Juramy-Gilles, Juri{\'{c}}, Kalirai, Kallivayalil, Kalmbach, Kantor, Karst, Kasliwal, Kelly, Kessler, Kinnison, Kirkby, Knox, Kotov, Krabbendam, Krughoff, Kub{\'{a}}nek, Kuczewski, Kulkarni, Ku, Kurita, Lage, Lambert, Lange, Langton, Guillou, Levine, Liang, Lim, Lintott, Long, Lopez, Lotz, Lupton, Lust, MacArthur, Mahabal, Mandelbaum, Markiewicz, Marsh, Marshall, Marshall, May, McKercher, McQueen, Meyers, Migliore, Miller, Mills, Miraval, Moeyens, Moolekamp, Monet, Moniez, Monkewitz, Montgomery, Morrison, Mueller, Muller, Arancibia, Neill, Newbry, Nief, Nomerotski, Nordby, O’Connor, Oliver, Olivier, Olsen, O’Mullane, Ortiz, Osier, Owen, Pain, Palecek, Parejko, Parsons, Pease, Peterson,
  Peterson, Petravick, Petrick, Petry, Pierfederici, Pietrowicz, Pike, Pinto, Plante, Plate, Plutchak, Price, Prouza, Radeka, Rajagopal, Rasmussen, Regnault, Reil, Reiss, Reuter, Ridgway, Riot, Ritz, Robinson, Roby, Roodman, Rosing, Roucelle, Rumore, Russo, Saha, Sassolas, Schalk, Schellart, Schindler, Schmidt, Schneider, Schneider, Schoening, Schumacher, Schwamb, Sebag, Selvy, Sembroski, Seppala, Serio, Serrano, Shaw, Shipsey, Sick, Silvestri, Slater, Smith, Smith, Sobhani, Soldahl, Storrie-Lombardi, Stover, Strauss, Street, Stubbs, Sullivan, Sweeney, Swinbank, Szalay, Takacs, Tether, Thaler, Thayer, Thomas, Thornton, Thukral, Tice, Trilling, Turri, Berg, Berk, Vetter, Virieux, Vucina, Wahl, Walkowicz, Walsh, Walter, Wang, Wang, Warner, Wiecha, Willman, Winters, Wittman, Wolff, Wood-Vasey, Wu, Xin, Yoachim, \& Zhan}]{Ivezic2019LSST:Products}
Ivezi{\'{c}}, v., Kahn, S.~M., Tyson, J.~A., {et~al.} 2019, The Astrophysical Journal, 873, 111, \dodoi{10.3847/1538-4357/ab042c}

\bibitem[{{Jana} {et~al.}(2025){Jana}, {Ricci}, {Venselaar}, {Chang}, {Liao}, {Inoue}, {Kawamuro}, {Bauer}, {Shablovinskaya}, {Trakhtenbrot}, {Elford}, \& {Koss}}]{Arghajit1566}
{Jana}, A., {Ricci}, C., {Venselaar}, S.~M., {et~al.} 2025, arXiv e-prints, arXiv:2505.13242, \dodoi{10.48550/arXiv.2505.13242}

\bibitem[{{Kaaz} {et~al.}(2025){Kaaz}, {Lithwick}, {Liska}, \& {Tchekhovskoy}}]{Kaaz2025}
{Kaaz}, N., {Lithwick}, Y., {Liska}, M., \& {Tchekhovskoy}, A. 2025, \apj, 979, 192, \dodoi{10.3847/1538-4357/ad9a85}

\bibitem[{{Kellermann} \& {Pauliny-Toth}(1969)}]{Kellermann1969}
{Kellermann}, K.~I., \& {Pauliny-Toth}, I.~I.~K. 1969, \apjl, 155, L71, \dodoi{10.1086/180305}

\bibitem[{Koay {et~al.}(2016)Koay, Vestergaard, Bignall, Reynolds, \& Peterson}]{Koay2016Parsec-scale590}
Koay, J.~Y., Vestergaard, M., Bignall, H.~E., Reynolds, C., \& Peterson, B.~M. 2016, Monthly Notices of the Royal Astronomical Society, 460, 304, \dodoi{10.1093/mnras/stw975}

\bibitem[{Kormendy \& Richstone(1995)}]{Kormendy1995InwardNuclei}
Kormendy, J., \& Richstone, D. 1995, Annual Review of Astronomy and Astrophysics, 33, 581, \dodoi{10.1146/annurev.aa.33.090195.003053}

\bibitem[{{Koz{\l}owski}(2017)}]{Kozlowski2017}
{Koz{\l}owski}, S. 2017, \apjs, 228, 9, \dodoi{10.3847/1538-4365/228/1/9}

\bibitem[{Lacy {et~al.}(2020)Lacy, Baum, Chandler, Chatterjee, Clarke, Deustua, English, Farnes, Gaensler, Gugliucci, Hallinan, Kent, Kimball, Law, Lazio, Marvil, Mao, Medlin, Mooley, Murphy, Myers, Osten, Richards, Rosolowsky, Rudnick, Schinzel, Sivakoff, Sjouwerman, Taylor, White, Wrobel, Andernach, Beasley, Berger, Bhatnager, Birkinshaw, Bower, Brandt, Brown, Burke-Spolaor, Butler, Comerford, Demorest, Fu, Giacintucci, Golap, G{\"{u}}th, Hales, Hiriart, Hodge, Horesh, {Ivezi{\'{c}}}, Jarvis, Kamble, Kassim, Liu, Loinard, Lyons, Masters, Mezcua, Moellenbrock, Mroczkowski, Nyland, O’dea, O’sullivan, Peters, Radford, Rao, Robnett, Salcido, Shen, Sobotka, Witz, Vaccari, van Weeren, Vargas, Williams, \& Yoon}]{Lacy2020TheDesign}
Lacy, M., Baum, S.~A., Chandler, C.~J., {et~al.} 2020, Publications of the Astronomical Society of the Pacific, 132, \dodoi{10.1088/1538-3873/ab63eb}

\bibitem[{{Liu} {et~al.}(2019){Liu}, {Liu}, {Dong}, {Zhou}, {Wang}, {Lu}, \& {Yuan}}]{YangSMBH2019}
{Liu}, H.-Y., {Liu}, W.-J., {Dong}, X.-B., {et~al.} 2019, \apjs, 243, 21, \dodoi{10.3847/1538-4365/ab298b}

\bibitem[{Marinucci {et~al.}(2018)Marinucci, Bianchi, Braito, Matt, Nardini, \& Reeves}]{Marinucci2018TrackingStates}
Marinucci, A., Bianchi, S., Braito, V., {et~al.} 2018, Monthly Notices of the Royal Astronomical Society, 478, 5651, \dodoi{10.1093/mnras/sty1436}

\bibitem[{Masci {et~al.}(2019)Masci, Laher, Rusholme, Shupe, Groom, Surace, Jackson, Monkewitz, Beck, Flynn, Terek, Landry, Hacopians, Desai, Howell, Brooke, Imel, Wachter, Ye, Lin, Cenko, Cunningham, Rebbapragada, Bue, Miller, Mahabal, Bellm, Patterson, Juri{\'{c}}, Golkhou, Ofek, Walters, Graham, Kasliwal, Dekany, Kupfer, Burdge, Cannella, Barlow, Van~Sistine, Giomi, Fremling, Blagorodnova, Levitan, Riddle, Smith, Helou, Prince, \& Kulkarni}]{Masci2019}
Masci, F.~J., Laher, R.~R., Rusholme, B., {et~al.} 2019, Publications of the Astronomical Society of the Pacific, 131, 1, \dodoi{10.1088/1538-3873/aae8ac}

\bibitem[{{Massaro} {et~al.}(2015){Massaro}, {Maselli}, {Leto}, {Marchegiani}, {Perri}, {Giommi}, \& {Piranomonte}}]{BZCAT2015}
{Massaro}, E., {Maselli}, A., {Leto}, C., {et~al.} 2015, \apss, 357, 75, \dodoi{10.1007/s10509-015-2254-2}

\bibitem[{Mathur {et~al.}(2018)Mathur, Denney, Gupta, Vestergaard, De~Rosa, Krongold, Nicastro, Collinson, Goad, Korista, Pogge, \& Peterson}]{Mathur2018TheAwakening}
Mathur, S., Denney, K.~D., Gupta, A., {et~al.} 2018, The Astrophysical Journal, 866, 123, \dodoi{10.3847/1538-4357/aadd91}

\bibitem[{{McConnell} {et~al.}(2020){McConnell}, {Hale}, {Lenc}, {Banfield}, {Heald}, {Hotan}, {Leung}, {Moss}, {Murphy}, {O'Brien}, {Pritchard}, {Raja}, {Sadler}, {Stewart}, {Thomson}, {Whiting}, {Allison}, {Amy}, {Anderson}, {Ball}, {Bannister}, {Bell}, {Bock}, {Bolton}, {Bunton}, {Chippendale}, {Collier}, {Cooray}, {Cornwell}, {Diamond}, {Edwards}, {Gupta}, {Hayman}, {Heywood}, {Jackson}, {Koribalski}, {Lee-Waddell}, {McClure-Griffiths}, {Ng}, {Norris}, {Phillips}, {Reynolds}, {Roxby}, {Schinckel}, {Shields}, {Tremblay}, {Tzioumis}, {Voronkov}, \& {Westmeier}}]{McConnell2020}
{McConnell}, D., {Hale}, C.~L., {Lenc}, E., {et~al.} 2020, \pasa, 37, e048, \dodoi{10.1017/pasa.2020.41}

\bibitem[{{Meyer} {et~al.}(2025){Meyer}, {Laha}, {Shuvo}, {Roychowdhury}, {Green}, {Rhodes}, {Hankla}, {Philippov}, {Mbarek}, {laor}, {Begelman}, {Sadaula}, {Ghosh}, {Bruni}, {Panessa}, {Guainazzi}, {Behar}, {Masterson}, {Zhang}, {Yang}, {Gurwell}, {Keating}, {Williams-Baldwin}, {Bray}, {Bempong-Manful}, {Wrigley}, {Bianchi}, {Ricci}, {La Franca}, {Kara}, {Georganopoulos}, {Oates}, {Nicholl}, {Pal}, \& {Cenko}}]{Meyer2025}
{Meyer}, E.~T., {Laha}, S., {Shuvo}, O.~I., {et~al.} 2025, \apjl, 979, L2, \dodoi{10.3847/2041-8213/ad8651}

\bibitem[{Murphy {et~al.}(2013)Murphy, Chatterjee, Kaplan, Banyer, Bell, Bignall, Bower, Cameron, Coward, Cordes, Croft, Curran, Djorgovski, Farrell, Frail, Gaensler, Galloway, Gendre, Green, Hancock, Johnston, Kamble, Law, Lazio, Lo, MacQuart, Rea, Rebbapragada, Reynolds, Ryder, Schmidt, Soria, Stairs, Tingay, Torkelsson, Wagstaff, Walker, Wayth, \& Williams}]{Murphy2013VAST:Transients}
Murphy, T., Chatterjee, S., Kaplan, D.~L., {et~al.} 2013, {VAST: An ASKAP survey for variables and slow transients}, \dodoi{10.1017/pasa.2012.006}

\bibitem[{Murphy {et~al.}(2021)Murphy, Kaplan, Stewart, O’Brien, Lenc, Pintaldi, Pritchard, Dobie, Fox, Leung, An, Bell, Broderick, Chatterjee, Dai, d’Antonio, Doyle, Gaensler, Heald, Horesh, Jones, McConnell, Moss, Raja, Ramsay, Ryder, Sadler, Sivakoff, Wang, Wang, Wheatland, Whiting, Allison, Anderson, Ball, Bannister, Bock, Bolton, Bunton, Chekkala, Chippendale, Cooray, Gupta, Hayman, Jeganathan, Koribalski, Lee-Waddell, Mahony, Marvil, McClure-Griffiths, Mirtschin, Ng, Pearce, Phillips, \& Voronkov}]{Murphy2021TheSurvey}
Murphy, T., Kaplan, D.~L., Stewart, A.~J., {et~al.} 2021, Publications of the Astronomical Society of Australia, 38, e054, \dodoi{10.1017/pasa.2021.44}

\bibitem[{Nyland {et~al.}(2020)Nyland, Dong, Patil, Lacy, van Velzen, Kimball, Sarbadhicary, Hallinan, Baldassare, Clarke, Goulding, Greene, Hughes, Kassim, Kunert-Bajraszewska, Maccarone, Mooley, Mukherjee, Peters, Petrov, Polisensky, Rujopakarn, Whittle, \& Vaccari}]{Nyland2020QuasarsFIRST}
Nyland, K., Dong, D.~Z., Patil, P., {et~al.} 2020, The Astrophysical Journal, 905, 74, \dodoi{10.3847/1538-4357/abc341}

\bibitem[{{Oknyansky} {et~al.}(2019){Oknyansky}, {Winkler}, {Tsygankov}, {Lipunov}, {Gorbovskoy}, {van Wyk}, {Buckley}, \& {Tyurina}}]{Oknyansky2019}
{Oknyansky}, V.~L., {Winkler}, H., {Tsygankov}, S.~S., {et~al.} 2019, \mnras, 483, 558, \dodoi{10.1093/mnras/sty3133}

\bibitem[{Pacucci {et~al.}(2021)Pacucci, Mezcua, \& Regan}]{Pacucci2021TheGalaxies}
Pacucci, F., Mezcua, M., \& Regan, J.~A. 2021, The Astrophysical Journal, 920, 134, \dodoi{10.3847/1538-4357/ac1595}

\bibitem[{{Parker} {et~al.}(2019){Parker}, {Schartel}, {Grupe}, {Komossa}, {Harrison}, {Kollatschny}, {Mikula}, {Santos-Lle{\'o}}, \& {Tom{\'a}s}}]{Parker2019}
{Parker}, M.~L., {Schartel}, N., {Grupe}, D., {et~al.} 2019, \mnras, 483, L88, \dodoi{10.1093/mnrasl/sly224}

\bibitem[{{Plavin} {et~al.}(2022){Plavin}, {Kovalev}, \& {Pushkarev}}]{Plavin2022}
{Plavin}, A.~V., {Kovalev}, Y.~Y., \& {Pushkarev}, A.~B. 2022, \apjs, 260, 4, \dodoi{10.3847/1538-4365/ac6352}

\bibitem[{{Ricci} \& {Trakhtenbrot}(2023)}]{Ricci2022Changing-lookNuclei}
{Ricci}, C., \& {Trakhtenbrot}, B. 2023, Nature Astronomy, 7, 1282, \dodoi{10.1038/s41550-023-02108-4}

\bibitem[{{Saha} {et~al.}(2023){Saha}, {Markowitz}, {Homan}, {Krumpe}, {Haemmerich}, {Czerny}, {Graham}, {Frederick}, {Gromadzki}, {Gezari}, {Winkler}, {Buckley}, {Brink}, {Naddaf}, {Rau}, {Wilms}, {Gokus}, {Liu}, \& {Grotova}}]{Saha2023MultiwavelengthSeyfert}
{Saha}, T., {Markowitz}, A., {Homan}, D., {et~al.} 2023, arXiv e-prints, arXiv:2309.08956, \dodoi{10.48550/arXiv.2309.08956}

\bibitem[{Schawinski {et~al.}(2015)Schawinski, Koss, Berney, \& Sartori}]{Schawinski2015ActiveYr}
Schawinski, K., Koss, M., Berney, S., \& Sartori, L.~F. 2015, Monthly Notices of the Royal Astronomical Society, 451, 2517, \dodoi{10.1093/mnras/stv1136}

\bibitem[{Somalwar {et~al.}(2022)Somalwar, Ravi, Dong, Graham, Hallinan, Law, Lu, \& Myers}]{Somalwar2022TheChange}
Somalwar, J.~J., Ravi, V., Dong, D., {et~al.} 2022, The Astrophysical Journal, 929, 184, \dodoi{10.3847/1538-4357/ac5e29}

\bibitem[{Stewart {et~al.}(2025)Stewart, Dobie, O'Brien, \& Kaplan}]{vasttools}
Stewart, A., Dobie, D., O'Brien, A., \& Kaplan, D. 2025, askap-vast/vast-tools: v3.2.0, v3.2.0,  Zenodo, \dodoi{10.5281/zenodo.8365236}

\bibitem[{{The Astropy Collaboration} {et~al.}(2022){The Astropy Collaboration}, Price-Whelan, Lim, Earl, Starkman, Bradley, Shupe, Patil, Corrales, Brasseur, N{\"{o}}the, Donath, Tollerud, Morris, Ginsburg, Vaher, Weaver, Tocknell, Jamieson, van Kerkwijk, Robitaille, Merry, Bachetti, G{\"{u}}nther, Aldcroft, Alvarado-Montes, Archibald, B{\'{o}}di, Bapat, Barentsen, Baz{\'{a}}n, Biswas, Boquien, Burke, Cara, Cara, Conroy, Conseil, Craig, Cross, Cruz, D’Eugenio, Dencheva, Devillepoix, Dietrich, Eigenbrot, Erben, Ferreira, Foreman-Mackey, Fox, Freij, Garg, Geda, Glattly, Gondhalekar, Gordon, Grant, Greenfield, Groener, Guest, Gurovich, Handberg, Hart, Hatfield-Dodds, Homeier, Hosseinzadeh, Jenness, Jones, Joseph, Kalmbach, Karamehmetoglu, Ka{\l}uszy{\'{n}}ski, Kelley, Kern, Kerzendorf, Koch, Kulumani, Lee, Ly, Ma, MacBride, Maljaars, Muna, Murphy, Norman, O’Steen, Oman, Pacifici, Pascual, Pascual-Granado, Patil, Perren, Pickering, Rastogi, Roulston, Ryan, Rykoff, Sabater, Sakurikar, Salgado, Sanghi,
  Saunders, Savchenko, Schwardt, Seifert-Eckert, Shih, Jain, Shukla, Sick, Simpson, Singanamalla, Singer, Singhal, Sinha, Sip{\H{o}}cz, Spitler, Stansby, Streicher, {\v{S}}umak, Swinbank, Taranu, Tewary, Tremblay, Val-Borro, Van~Kooten, Vasovi{\'{c}}, Verma, de~Miranda~Cardoso, Williams, Wilson, Winkel, Wood-Vasey, Xue, Yoachim, Zhang, \& Zonca}]{TheAstropyCollaboration2022}
{The Astropy Collaboration}, Price-Whelan, A.~M., Lim, P.~L., {et~al.} 2022, The Astrophysical Journal, 935, 167, \dodoi{10.3847/1538-4357/ac7c74}

\bibitem[{Trakhtenbrot {et~al.}(2019)Trakhtenbrot, Arcavi, MacLeod, Ricci, Kara, Graham, Stern, Harrison, Burke, Hiramatsu, Hosseinzadeh, Howell, Smartt, Rest, Prieto, Shappee, Holoien, Bersier, Filippenko, Brink, Zheng, Li, Remillard, \& Loewenstein}]{Trakhtenbrot20191ESMonths}
Trakhtenbrot, B., Arcavi, I., MacLeod, C.~L., {et~al.} 2019, The Astrophysical Journal, 883, 94, \dodoi{10.3847/1538-4357/ab39e4}

\bibitem[{{Tripathi} \& {Dewangan}(2022)}]{Tripathi2022}
{Tripathi}, P., \& {Dewangan}, G.~C. 2022, \apj, 930, 117, \dodoi{10.3847/1538-4357/ac610f}

\bibitem[{{Vagnetti} {et~al.}(2011){Vagnetti}, {Turriziani}, \& {Trevese}}]{Vagnetti2011}
{Vagnetti}, F., {Turriziani}, S., \& {Trevese}, D. 2011, \aap, 536, A84, \dodoi{10.1051/0004-6361/201118072}

\bibitem[{{Walker}(1998)}]{Walker1998}
{Walker}, M.~A. 1998, \mnras, 294, 307, \dodoi{10.1046/j.1365-8711.1998.01238.x10.1111/j.1365-8711.1998.01238.x}

\bibitem[{Wo{\l}owska {et~al.}(2021)Wo{\l}owska, Kunert-Bajraszewska, Mooley, Siemiginowska, Kharb, Ishwara-Chandra, Hallinan, Gromadzki, \& Kozie{\l}-Wierzbowska}]{Woowska2021Caltech-NRAOState}
Wo{\l}owska, A., Kunert-Bajraszewska, M., Mooley, K.~P., {et~al.} 2021, The Astrophysical Journal, 914, 22, \dodoi{10.3847/1538-4357/abe62d}

\bibitem[{Xu {et~al.}(2024)}]{Xu2024}
Xu, D.~W., {et~al.} 2024, Universe, 10, 61, \dodoi{10.3390/universe10020061}

\bibitem[{Yang {et~al.}(2021{\natexlab{a}})Yang, van Bemmel, Paragi, Komossa, Yuan, Yang, An, Koay, Reynolds, Oonk, Liu, \& Wu}]{Yang2021AMrk590}
Yang, J., van Bemmel, I., Paragi, Z., {et~al.} 2021{\natexlab{a}}, Monthly Notices of the Royal Astronomical Society: Letters, 502, L61, \dodoi{10.1093/mnrasl/slab005}

\bibitem[{Yang {et~al.}(2021{\natexlab{b}})Yang, Paragi, Beswick, Chen, van Bemmel, Wu, An, Wu, Fan, Oonk, Liu, \& Wang}]{Yang2021A2617}
Yang, J., Paragi, Z., Beswick, R.~J., {et~al.} 2021{\natexlab{b}}, Monthly Notices of the Royal Astronomical Society, 503, 3886, \dodoi{10.1093/mnras/stab706}

\bibitem[{Zeltyn {et~al.}(2024)Zeltyn, Trakhtenbrot, Eracleous, Yang, Green, Anderson, LaMassa, Runnoe, Assef, Bauer, Brandt, Davis, Frederick, Fries, Graham, Grogin, Guolo, Hern{\'{a}}ndez-Garc{\'{i}}a, Koekemoer, Krumpe, Liu, Mart{\'{i}}nez-Aldama, Ricci, Schneider, Shen, {\'{S}}niegowska, Temple, Trump, Xue, Brownstein, Dwelly, Morrison, Bizyaev, Pan, \& Kollmeier}]{Zeltyn2024ExploringResults}
Zeltyn, G., Trakhtenbrot, B., Eracleous, M., {et~al.} 2024, The Astrophysical Journal, 966, 85, \dodoi{10.3847/1538-4357/ad2f30}

\bibitem[{Zhang {et~al.}(2022)Zhang, Shu, Sun, Yang, Jiang, Dou, Wang, \& Wang}]{Zhang2022TransientSurveys}
Zhang, F., Shu, X., Sun, L., {et~al.} 2022, The Astrophysical Journal, 938, 43, \dodoi{10.3847/1538-4357/ac8a9a}

\end{thebibliography}
\bibliographystyle{aasjournal}

\end{document}